\documentclass[]{amsart}
\usepackage{amsfonts}
\usepackage{amsmath}
\usepackage{amssymb}
\usepackage{graphicx}
\usepackage{float}
\usepackage{hyperref}
\usepackage{geometry}
\usepackage{xcolor}
\numberwithin{equation}{section}

\newcommand{\C}{\mathbb{C}}
\newcommand{\vect}[1]{\mathbf{#1}}
\newcommand{\R}{\mathbb{R}}
\newcommand{\Z}{\mathbb{Z}}
\newcommand{\N}{\mathbb{N}}
\newcommand{\I}{\mathrm{i}}

\newcommand{\ernst}{\mathcal{E}}

\newcommand{\thetapq}{\vartheta_{\mathrm{pq}}}

\newcommand{\B}{\tau}
\renewcommand{\L}{\mathcal{L}}
\newcommand{\p}{\mathrm{p}}
\newcommand{\q}{\mathrm{q}}
\newcommand*{\ii}{\mathrm{i}}
\newcommand*{\ee}{\mathrm{e}}

\begin{document}
\title[Ray tracing in the toron spacetime]{Gravitational lensing and shadows in the toron solution of Einstein's equations using ray tracing methods}
\author{E. de Leon}
    \email{eddybrandon11@hotmail.com}
\address{Institut de Math\'ematiques de Bourgogne,
		Universit\'e de Bourgogne-Europe, 9 avenue Alain Savary, 21078 Dijon
		Cedex, France}
\author[C. Klein]{C. Klein}
\address{Institut de Math\'ematiques de Bourgogne,  UMR 5584;\\
Institut Universitaire de France \\
Universit\'e de Bourgogne-Europe, 9 avenue Alain Savary, 21078 Dijon
                Cedex, France} 
\email{Christian.Klein@u-bourgogne.fr}
\author{D. Korotkin}
\address{Department of Mathematics and Statistics;\\
Concordia University\\  1455 De Maisonneuve Blvd. W.
Montreal, QC  H3G 1M8
Canada
} 
\email{dmitry.korotkin@concordia.ca}
\date{\today} 
\thanks{This work was partially supported by the EIPHI Graduate 
School (contract ANR-17-EURE-0002), the Bourgogne
Franche-Comt\'e Region and the ANR project ISAAC-ANR-23-CE40-0015-01. 
DK acknowledges support by an NSERC Discovery grant}
\begin{abstract}
We present a numerical and analytical study of the so-called `toron'  
solution of the stationary axisymmetric Einstein equations in vacuum 
expressed in terms of elliptic functions.
The asymptotic behavior of this solution coincides with the one of 
the NUT solution, i.e., it has a  `gravimagnetic' mass known as the 
NUT parameter while the ordinary mass vanishes. The physical 
properties of this spacetime are studied via  ray tracing. The 
results are compared to known 
 geodesic flows in Schwarzschild, Kerr and NUT 
spacetimes to   
discuss similarities and differences, with a particular emphasis on 
the comparison of NUT and toron spacetimes.  

\end{abstract}

\maketitle

\section{Introduction} 
The physical discussion of relativistic spacetimes in comparison to 
Newtonian gravity is best based on 
the tracing of light rays that are null geodesics in a relativistic 
setting but straight lines in a Newton theory. As already Einstein 
discussed in \cite{einstein}, light passing near strong gravitational 
sources on its way to an observer is subject to \emph{gravitational 
lensing} which can lead to an apparent 
displacement of an object on the celestial sphere, the appearance of 
several copies of 
the same object and to an object directly behind the light bending 
source appearing as a ring-like structure called \emph{Einstein 
ring}.  It is also possible that no light from certain regions of the 
spacetime reaches an observer which leads to a \emph{shadow}, see 
Synge \cite{synge} for the Schwarzschild and Bardeen \cite{bardeen} for the Kerr 
black hole. This can be due to photons actually orbiting the black 
hole along trajectories (which are circular orbits in the case of 
Schwarzschild, but can have a  complicated shape in the case of Kerr \cite{Teo})  restricted to a shell of spherical topology  called the \emph{photon sphere} (see the recent reviews \cite{PT,MHC}).

Ray tracing approaches are especially fruitful in asymptotically flat stationary 
vacuum spacetimes since the virtual camera can be placed at any point 
in the spacetime  at some distance from the strongly gravitating 
object. In the stationary axisymmetric case there is a large class of 
known exact solutions since the Einstein equations in vacuum can be 
reduced to the Ernst equation \cite{ernst1968a,ernst1968b} which were 
shown to be completely integrable in \cite{BZ,Mai,Neu}. In \cite{K88} 
a large class of solutions was constructed in terms of theta 
functions on hyperelliptic Riemann surfaces. They contain the Kerr 
solution as a limiting case. The simplest solution corresponding to 
an elliptic curve was 
discussed in detail in  \cite{K91}; it was called {\it toron} due to 
the toroidal shape of the ergosphere. In \cite{NM} a special class of solutions corresponding to genus two algebraic curves  
was found to satisfy the boundary conditions of a rigidly rotating thin dust disk; in \cite{KRPRL} these results were extended to the case when the disk consists of two counter-rotating components.
see also \cite{KR}. In \cite{DFK} ray tracing in these spacetimes was 
studied in detail using  the numerical approach of \cite{FK}.

It is the goal of this paper to apply the numerical ray tracing approach developed in 
\cite{DFK} to study the  toron solution of  \cite{K91} in more detail.  The asymptotical behavior of the toron coincides with 
the one of the NUT solution with zero real mass, i.e., the standard mass of this solution is 
replaced by a so-called NUT parameter also called a `gravimagnetic mass'. Therefore, near infinity the
geodesic flow of the toron is similar to the one of the NUT solution with zero real mass; the 
significant difference is observed near the object itself. 

To begin with we visualize known 
properties on the null geodesics in NUT and Kerr spacetimes and 
present a detailed comparison to null geodesics in the toron 
spacetime. 

The paper is organized as follows: In section \ref{sec:kerr-nut-metric} we collect known 
facts on the Kerr and the NUT solution. In section 
\ref{sec:elliptic_metric} we discuss the toron solution of 
\cite{K91}. The numerical approach of \cite{DFK} to the visualization of  geodesics 
is briefly summarized in section \ref{geodesic_section}. It is  illustrated in section \ref{sec:nut_geodesics} for 
the Kerr metric and the NUT metric, similar to the spacetimes studied 
in \cite{MHC}. In section \ref{toron_geod} we 
discuss geodesics in the gravitational field of the toron. We add some concluding remarks in section \ref{conclusion_sec}.

\section{Kerr and NUT spacetimes} \label{sec:kerr-nut-metric}

In this section we collect  known facts on the  Kerr and the NUT solution. 

\subsection{Ernst equation}

Throughout the entire paper, we consider spacetimes in Weyl 
coordinates. The Weyl-Lewis-Papapetrou form of the metric is, see 
\cite{Exact}, 
\begin{equation}
    ds^2 = -f (dt +A d\phi)^2 + f^{-1} [e^{2k} (d\rho^2 +dz^2) + \rho^2 d\phi^2 ].
	\label{eq:wlp}
\end{equation}
The functions $f$, $A$ and $k$ are only dependent on the coordinates 
$\rho$ and $z$. 

Complex notation leads to a simplification of the Einstein equations 
in this case: we put 
$\xi:=z-\ii\rho$ and define the \emph{twist potential} $b$ via
\begin{equation}
  b_{\xi}:=-\frac{\ii}{\rho}A_{\xi}f^{2}
  \label{bxi}.
\end{equation}
Partial derivatives are denoted via subscripts here.
For given $A$, the potential $b$ is determined up to a constant via a 
quadrature. In asymptotically flat spacetimes this constant is chosen such that $b$ 
vanishes at infinity. The Ernst potential $\mathcal{E}=f+\ii b$ was 
introduced in \cite{ernst1968a,ernst1968b}. 

The vacuum Einstein equations imply the  Ernst equation for $ \mathcal{E}$:
\begin{equation}
    \mathcal{E}_{\xi\bar{\xi}}-\frac{1}{2(\bar{\xi}-\xi)}(
    \mathcal{E}_{\bar{\xi}}-\mathcal{E}_{\xi}) = 
    \frac{2}{\mathcal{E}+
    \bar{\mathcal{E}}}\mathcal{E}_{\xi}\mathcal{E}_{\bar{\xi}}\;
    \label{eq:ernstcom}.
\end{equation}

The metric functions 
(\ref{eq:wlp}) can be obtained from a given Ernst potential 
$\mathcal{E}$ via the formula  $f=\Re \mathcal{E}$ and quadratures
\begin{equation}
  A_{\xi}=2\rho\frac{(\mathcal{E}-\bar{\mathcal{E}})_{\xi}}{
    (\mathcal{E}+\bar{\mathcal{E}})^{2}}
  \label{axi},
\end{equation}
and
\begin{equation}
  k_{\xi}=(\xi-\bar{\xi})
  \frac{\mathcal{E}_{\xi}\bar{\mathcal{E}}_{\xi}}{
    (\mathcal{E}+\bar{\mathcal{E}})^{2}}\;.
  \label{kxi}
\end{equation}

It was shown in \cite{Mai,BZ,Neu} that the Ernst equation is completely integrable in the sense   of soliton theory: it possesses the $U-V$ pair with variable spectral parameter which makes it possible to
construct multisoliton and algebro-geometric solutions, see \cite{BZ,K91}.

\subsection{Kerr spacetimes}

The most prominent solution of the Ernst equation is the Kerr 
solution (a special multi-soliton solution from the perspective of the theory of integrable systems), see \cite{Exact},  which can be interpreted as a rotating black hole. It 
depends on two physical parameters,  the mass parameter $m>0$ and 
another parameter $\varphi\in[0,\pi/2]$  (related to the angular 
momentum of the black hole via  the relation $J=m^2 \sin \varphi$):

\begin{equation}
    \mathcal{E}=\frac{\cos \varphi \,X - \ii\sin \varphi\, Y-1}{
    \cos \varphi \,X - \ii\sin \varphi\, Y+1}
    \label{kerr5a},
\end{equation}
where 
$$X=\frac{r_+ + r_-}{2m\cos\varphi}\;\;,\hskip0.6cm Y=\frac{r_+ - 
r_-}{2m\cos\varphi}$$ 
with $r_\pm^2 = (z\pm m\cos\varphi)^2 + 
\rho^2$ being prolate ellipsoidal coordinates.  The Schwarzschild solution corresponds to  $\varphi=0$. 

The metric coefficients of the Kerr metric which corresponds to the Ernst potential (\ref{kerr5a}) read
\begin{equation} \label{eq:Kerr_metric}
    \begin{split}
    f &= \frac{\cos^2\varphi X^2 + \sin^2\varphi Y^2 - 1}{(\cos\varphi X + 1)^2 + \sin^2 Y^2}, \\
    e^{2k} &= \frac{m^2}{r_+ r_-} \left( \sin^2\varphi (Y^2-1) + \cos^2\varphi (X^2 - 1) \right) ,\\
    A &=  \frac{2m \sin\varphi (1-Y^2)(1+X\cos\varphi)}{\cos^2\varphi 
	X^2 + \sin^2\varphi Y^2 - 1}.
    \end{split}
\end{equation}
The metric is equatorially symmetric which means that all functions 
(\ref{eq:Kerr_metric}) are even in $z$. 

\begin{figure}[!htb]
  \includegraphics[width=0.49\hsize]{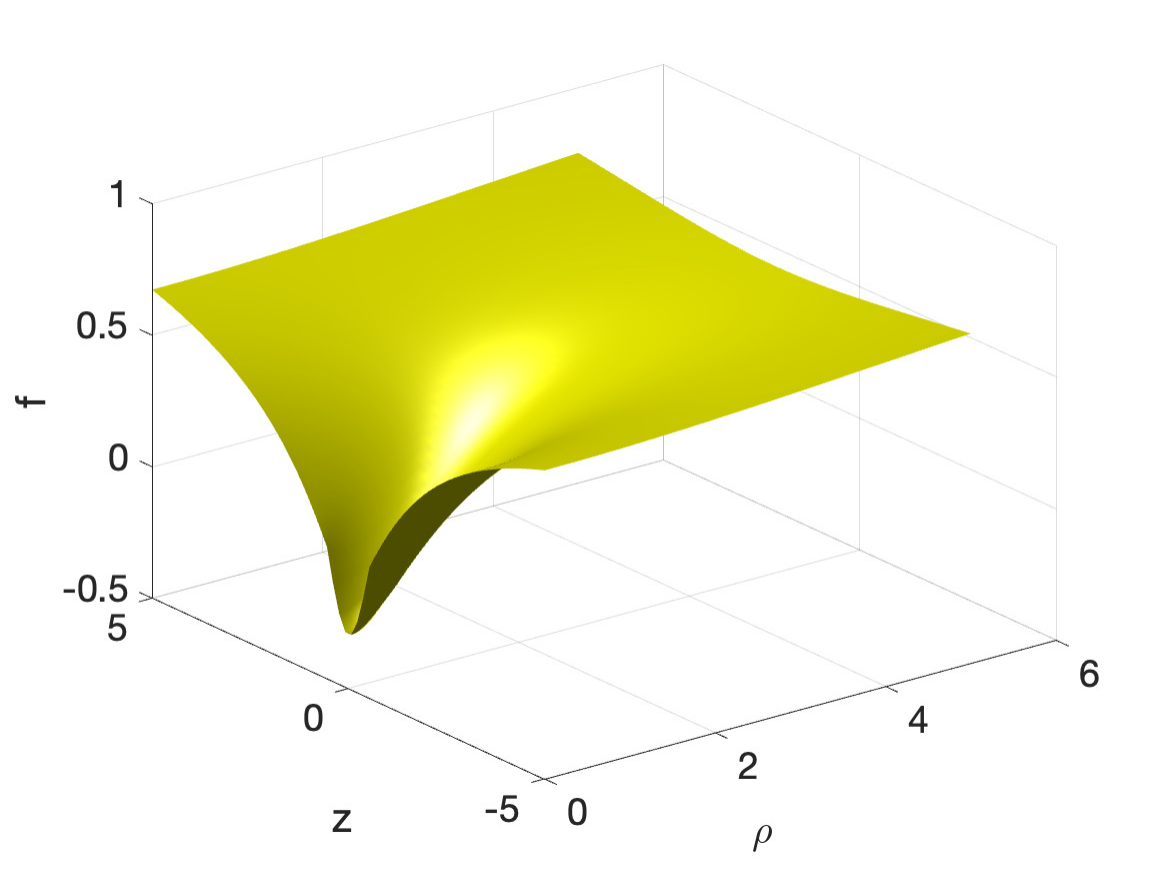}
  \includegraphics[width=0.49\hsize]{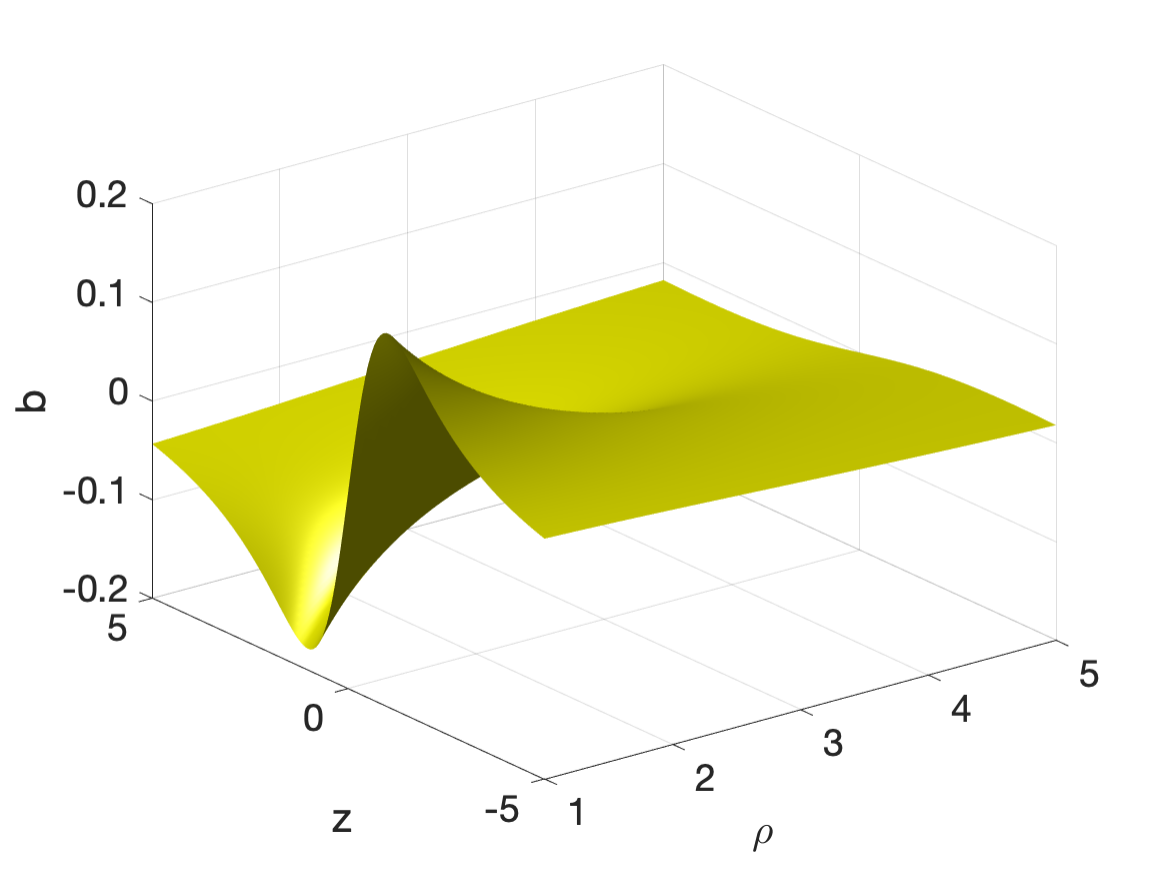}\\
  \includegraphics[width=0.49\hsize]{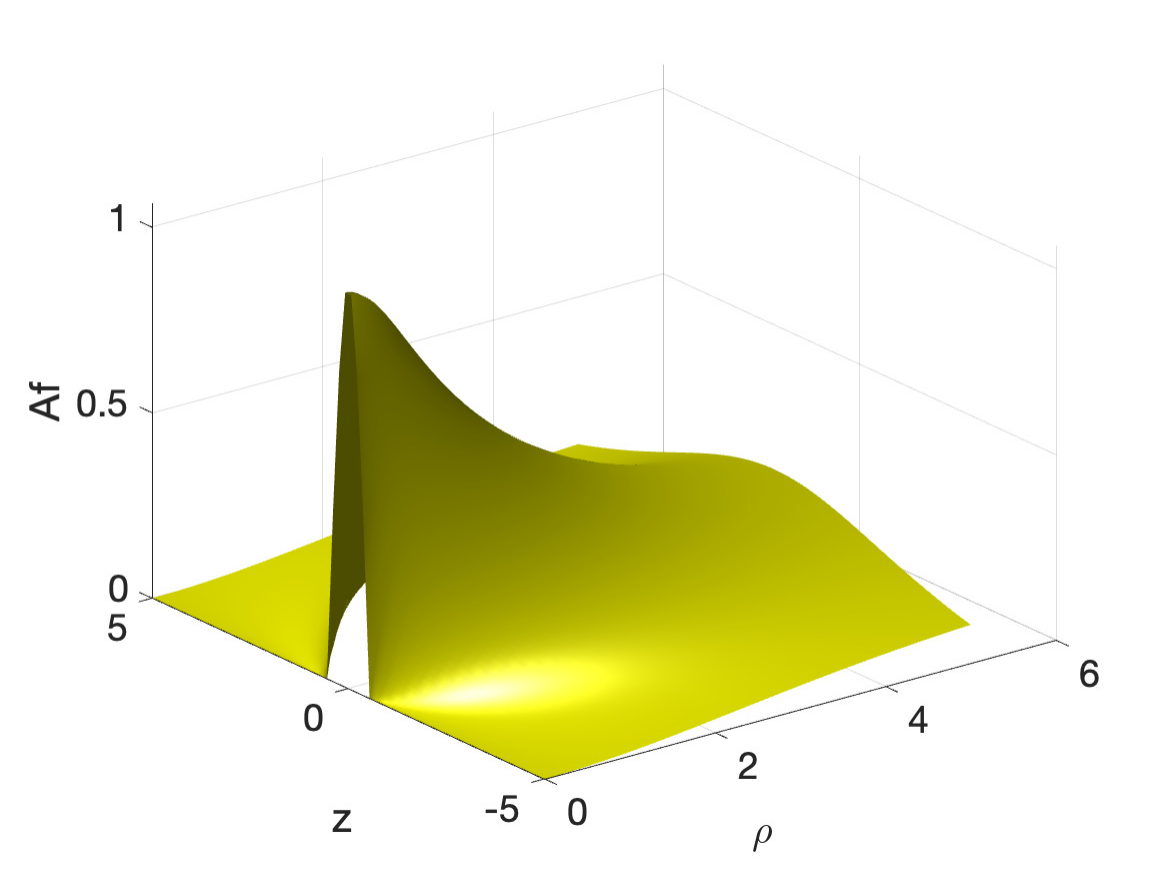} 
\includegraphics[width=0.49\hsize]{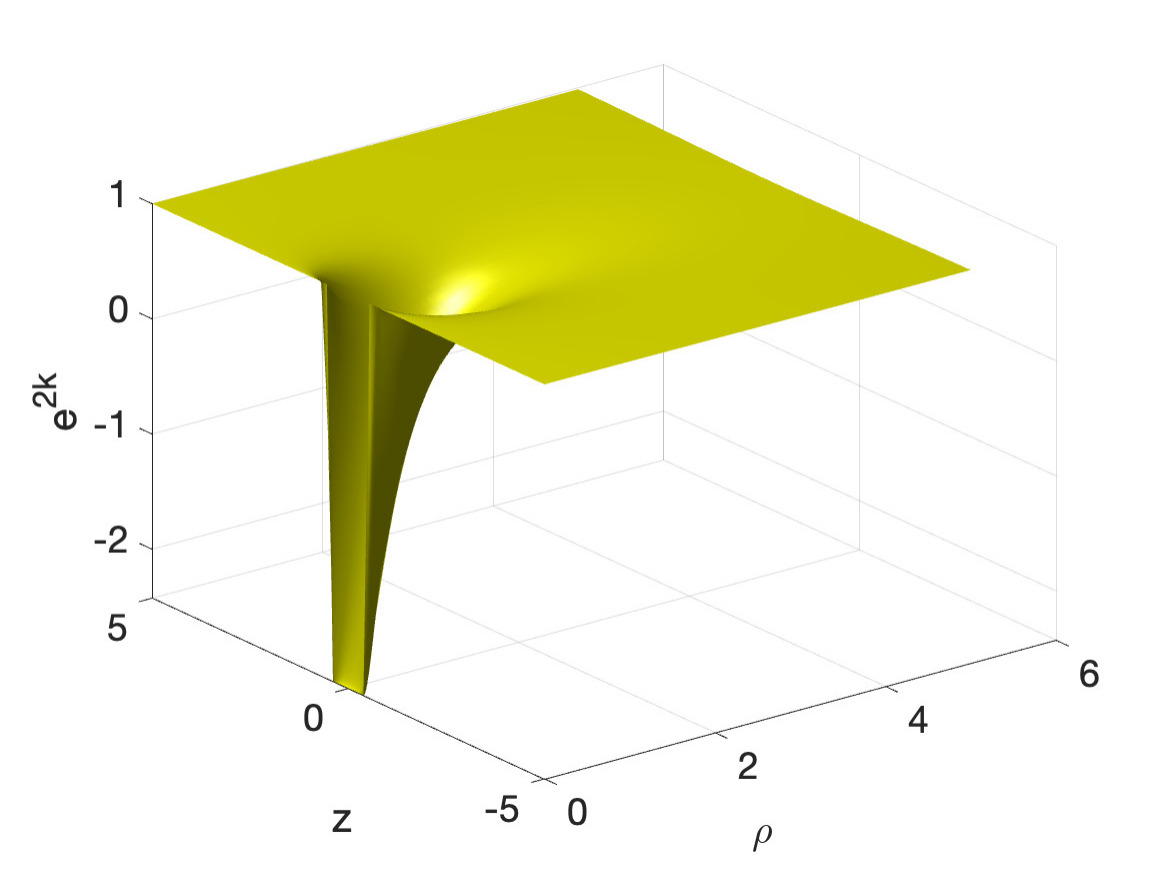}
  \caption{Ernst potential and metric functions for the  Kerr solution for $m=1$ and $\varphi=1$.}
  \label{kerrernst}
\end{figure}

The metric functions are shown for $m=1$ and $\varphi=1$   
in 
Fig.\ref{kerrernst}.     

In Weyl coordinates the horizon of the Kerr black hole is located on 
the axis for $z\in[-m\cos\varphi,m\cos\varphi]$. 
The  
\emph{ergoregion}, 
where $g_{tt}\geq 0$, i.e. $f\leq 0$, is bounded by the \emph{ergosphere} defined by the equation
$$
\Re   \mathcal{E} =0,
$$
where an 
observer stationary at infinity will note an infinite redshift.  The 
ergosphere of the Kerr solution
touches the horizon at the poles of the event horizon, i.e. at the 
points $z=\pm m \cos \varphi$ of  the axis.
 The functions $A$ and $k$ vanish on 
the regular part of the axis. On the horizon they take the constant 
values
\begin{equation}
  \ee^{2k}=-\tan^{2}\varphi,\quad A=-1/\Omega_{BH}
  \label{kerr8},
\end{equation}
where $\Omega_{BH}$ is the angular velocity of the horizon, given by $1/\Omega_{BH}=2m\cot\frac{\varphi}{2}$.

For $z>m\cos\varphi$, the Ernst potential on the axis reads
$$
	\mathcal{E}=\frac{z/m-\I\sin\varphi-1}{z/m-\I\sin\varphi+1},
$$
i.e., for $z\gg1$ one has 
$$\mathcal{E}=1-\frac{2m}{z}+\frac{2m^{2}(1-\I \sin\varphi)}{z^{2}}+\mathcal{O}(1/z^{3})\;.$$

Thus the $1/z$
term in the asymptotic expansion of the Ernst potential in the case of the Kerr solution is real (which is not the case for the 
NUT solution considered below).
The angular momentum $J=m^{2}\sin\varphi$ appears in order $1/z^{2}$ 
in the expansion of the imaginary part of the Ernst potential. 

\subsection{NUT spacetimes}
The Ernst potential of the Newman-Unti-Tambourini 
(NUT) solution  depends on the two parameters, $m\geq 0$ and $\ell\in\R$,  (see \cite{MR} and Sec. 20 of \cite{Exact}):
\begin{equation} \label{eq:nut}
    \mathcal{E} = \frac{\tilde{X} - m - \I\ell}{\tilde{X} + m + \I\ell},
\end{equation}
where $\tilde{X} = (\tilde{r}_+ + \tilde{r}_-)/2$ and $\tilde{Y}= (\tilde{r}_+ - \tilde{r}_-)/2$, with $\tilde{r}_{\pm}^2 = \rho^2 + (z\pm \sqrt{m^2+\ell^2})^2$. The NUT metric functions are 
\begin{equation} \label{eq:NUT_metric}
    \begin{split}
    f &= \frac{\tilde{X}^2-(m^2+\ell^2)}{(\tilde{X}+m)^2+\ell^2},\\
    e^{2k} &= \frac{\tilde{X}^2-(m^2+\ell^2)}{\tilde{X}^2-\tilde{Y}^2},\\
    A &= \frac{2\ell}{\sqrt{m^2+\ell^2}} \tilde{Y}.
    \end{split}
\end{equation}
This solution can be interpreted as a black hole with real mass $m$ and gravimagnetic mass $\ell$. The latter is often referred to as the NUT parameter or gravimagnetic momentum in the literature, following the interpretation of \cite{bonor} in which $\ell$ is interpreted as a massless source of angular momentum.

\begin{figure}[!htb]
  \includegraphics[width=0.49\hsize]{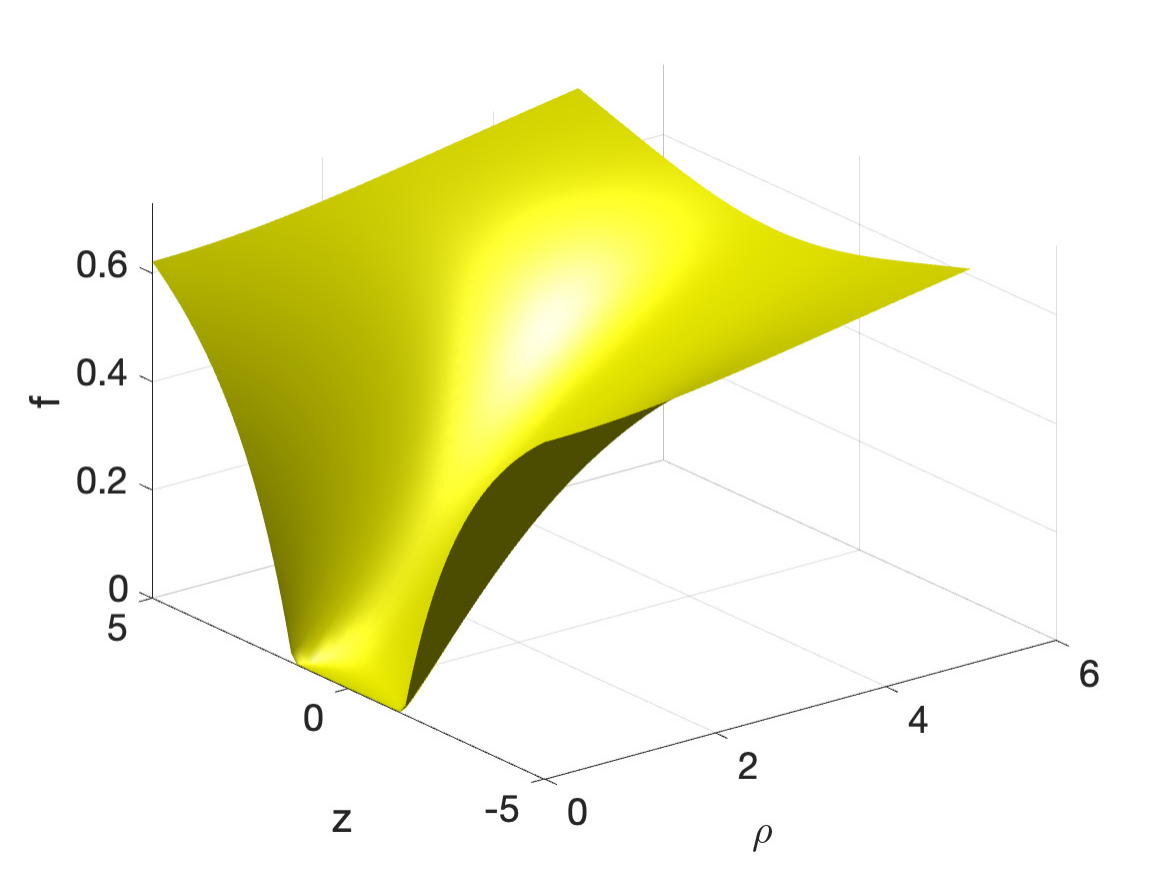}
  \includegraphics[width=0.49\hsize]{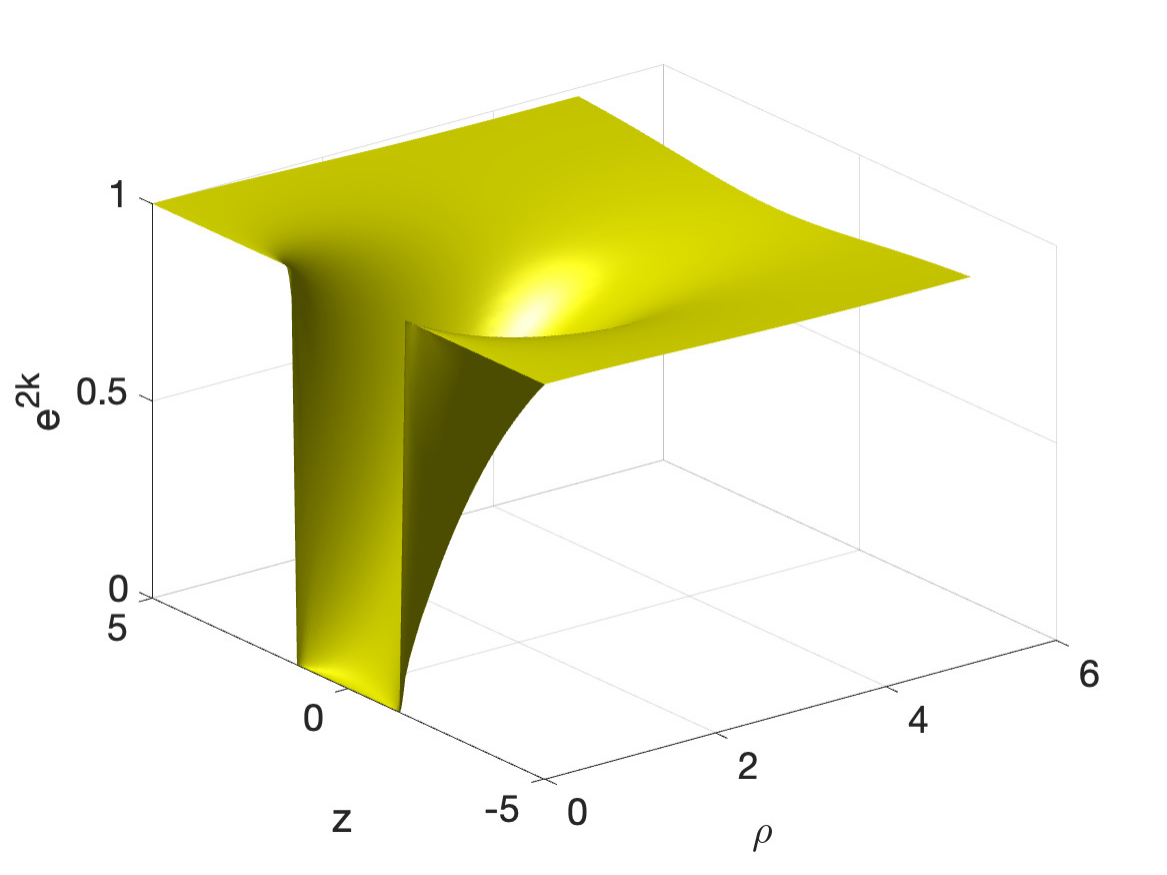}
  \caption{Metric functions for the NUT solutions 
  (\ref{eq:NUT_metric}) for $m=\ell=1$, on the left $f$, on the right 
  $e^{2k}$.}
  \label{figNUTfk}
\end{figure}

We plot the metric functions $f$ and $e^{2k}$ for $m=\ell=1$ in 
Fig.~\ref{figNUTfk}.

\begin{figure}[!htb]
  \includegraphics[width=0.49\hsize]{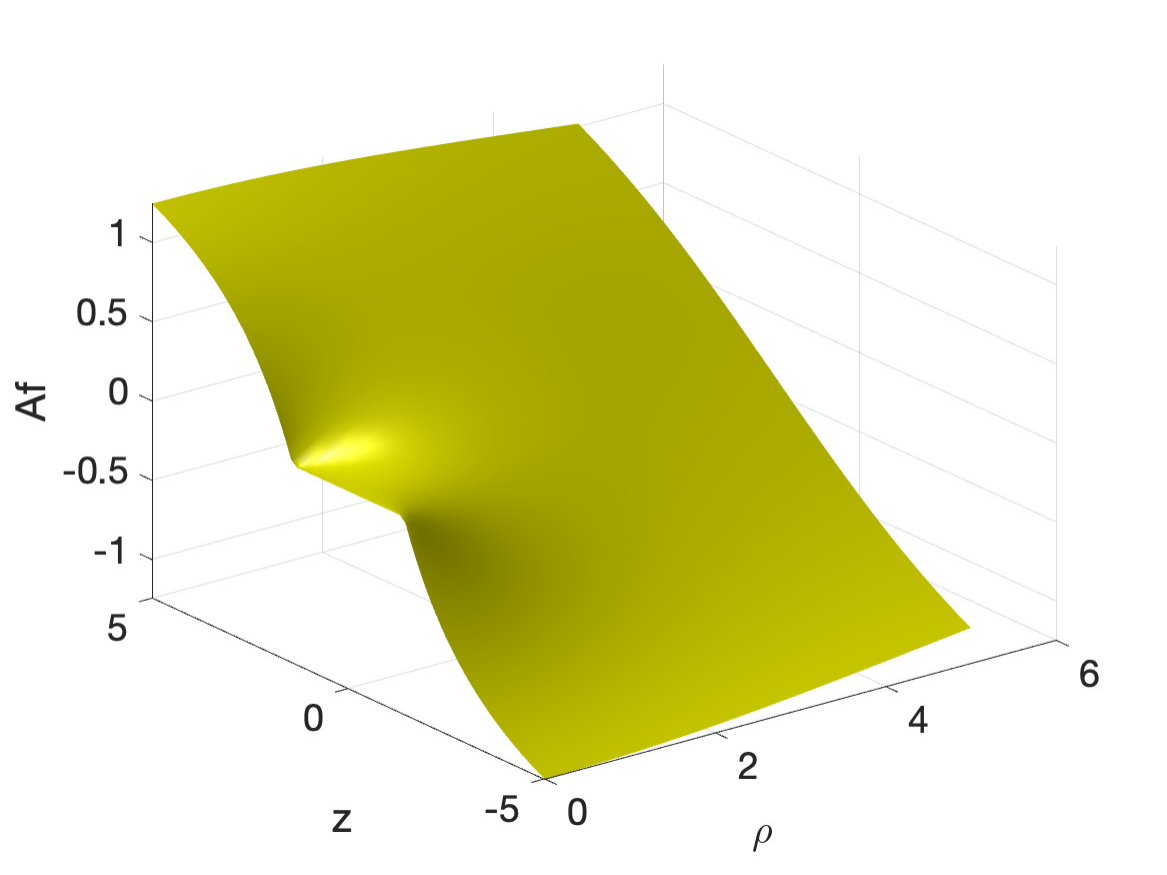}
  \includegraphics[width=0.49\hsize]{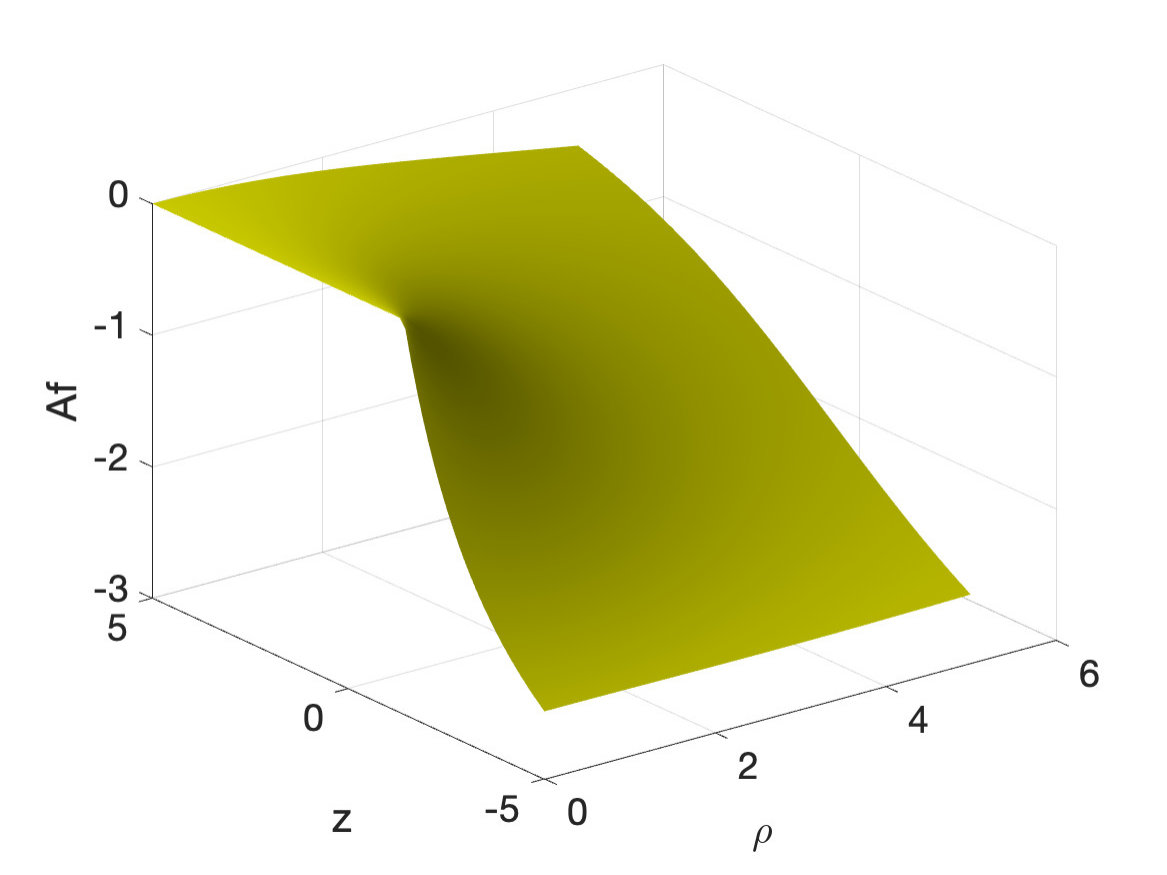}
  \caption{Metric function $Af$ for the NUT solutions 
  (\ref{eq:NUT_metric}) for $m=\ell=1$, on the left  for $A$ odd, on the right 
  such that it vanishes for $z>\sqrt{m^{2}+\ell^{2}}$.}
  \label{figNUTa}
\end{figure}

Similarly to the Kerr solution, there is an event  horizon of spherical topology occupying the interval  of the symmetry axis
$z\in[-\sqrt{m^{2}+\ell^{2}},\sqrt{m^{2}+\ell^{2}}]$. 
The  metric functions $f$ and  $e^{2k}$  plotted in Fig.~\ref{figNUTfk} are  even in the
$z$ variable, similarly to the case of the Kerr solution. This is no longer the case for the function $A$ given by (\ref{eq:NUT_metric}) which is odd in 
$z$ (in contrast to the function $A$ for the Kerr solution given by (\ref{eq:Kerr_metric}) which is also even in $z$), see the left pane  of Fig.~\ref{figNUTa}. However, 
$A$ can be alternatively chosen to vanish on the positive (respectively negative) regular 
part of the axis by replacing $A$ by  $A-2\ell$ (respectively $A+2\ell$), which would 
correspond to the choice of a rotating coordinate system. 
It is not possible in NUT spacetimes to choose $A$ to vanish on the 
whole regular part of the $z$-axis. 

Since the metric function $A$ is at the origin 
of so-called gravimagnetic effects as frame dragging, there is a 
considerable difference between geodesics in NUT and Kerr spacetimes 
as discussed below in sections \ref{geodesic_section} and \ref{sec:nut_geodesics}. For large $r = \sqrt{\rho^{2}+z^{2}}$, we get 
$$A \sim \ell\frac{z}{r}$$ which shows again that $A$ does not tend to 0 for 
$r\to\infty$ whereas this is the case for the Kerr solution. 
We get the following expression for the  Ernst 
potential (\ref{eq:nut}) on the axis: 
$$\mathcal{E}=\frac{z-m-\ii \ell}{z+m+\ii \ell}$$
for $z>\sqrt{m^{2}+\ell^{2}}$. 

For large $z$ this 
leads to the asymptotics
$$\mathcal{E}=1-\frac{2m+\ii \ell}{z}+\mathcal{O}(1/z^{2})\;.$$
 In contrast to the Kerr solution,
 there is thus an imaginary contribution to the mass which is 
precisely the NUT parameter.
A special case is when the real mass $m$ vanishes, which happens also in the elliptic solutions of the Ernst equation.

One consequence of the form of the function $A$ in NUT spacetimes is that the frame of 
reference is dragged in opposite directions, see \cite{MR}, depending 
on the sign of $z$.
Regardless of whether we choose the function $A$ to vanish on part of 
the axis or not, one can observe that if $\ell>0$ then $\partial_z 
A(\rho,z)\geq 0$ for all $z,\rho$, where equality occurs on 
the part of the axis outside the horizon. In particular, 
$\partial_z A(\rho,z)$ is strictly positive on the equatorial 
plane (strictly negative if $\ell<0$). This will have an important effect on the geodesics, as we will see in section \ref{sec:nut_geodesics}.

%
%
\section{Elliptic solutions to the stationary axisymmetric Einstein equations} \label{sec:elliptic_metric}
In this section we summarize a few facts on the elliptic solution to 
the Ernst equation \cite{K88,K91}. 


This class of solutions is the simplest special case of the algebro-geometric solutions of the Ernst equation originally found in 
\cite{K88}. The first attempt to understand geometric properties of the elliptic solutions was made in \cite{K91,K91_1}
using analytical methods; elliptic solutions studied in these two papers differ by choice of some built-in parameters. Here we shall study solutions from \cite{K91} (called there ``torons" due to toroidal structure of the ergosphere) in more detail using a combination of numerical and analytical methods.

The metric functions describing the toron spacetimes are given in 
terms of theta functions and Abel maps on elliptic curves. In this 
section we shall simply present the metric, the reader is referred to 
Appendix \ref{sec:elliptic_curves} for definitions.
All elliptic functions  can be 
computed in standard way (it is implemented in Matlab as well as other environments).


We remark that the dependence of the Ernst potential on the physical 
coordinates is 
via the branch points $\xi$ and $\bar{\xi}$ of the family of elliptic 
curves (\ref{eq:curves_L_xi}). 

\begin{figure}[!htb]
  \includegraphics[width=0.49\hsize]{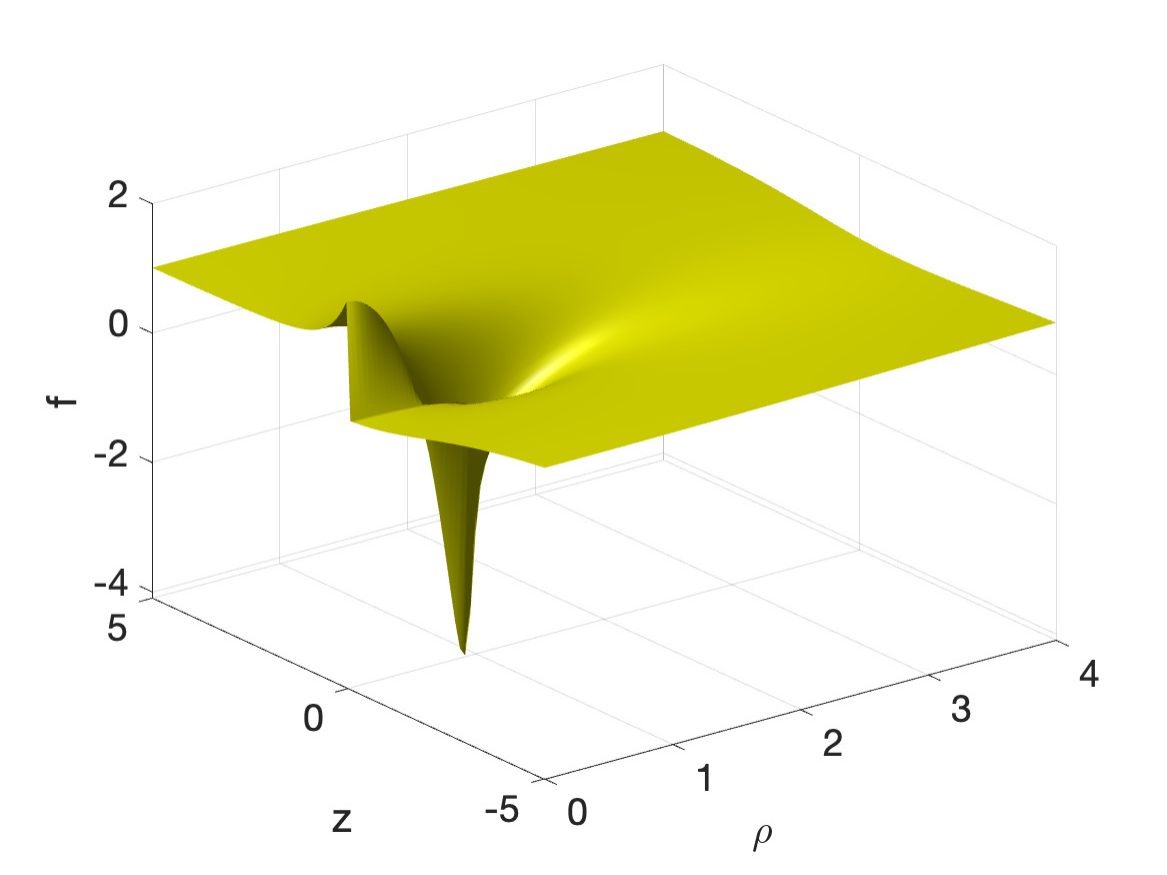}
  \includegraphics[width=0.49\hsize]{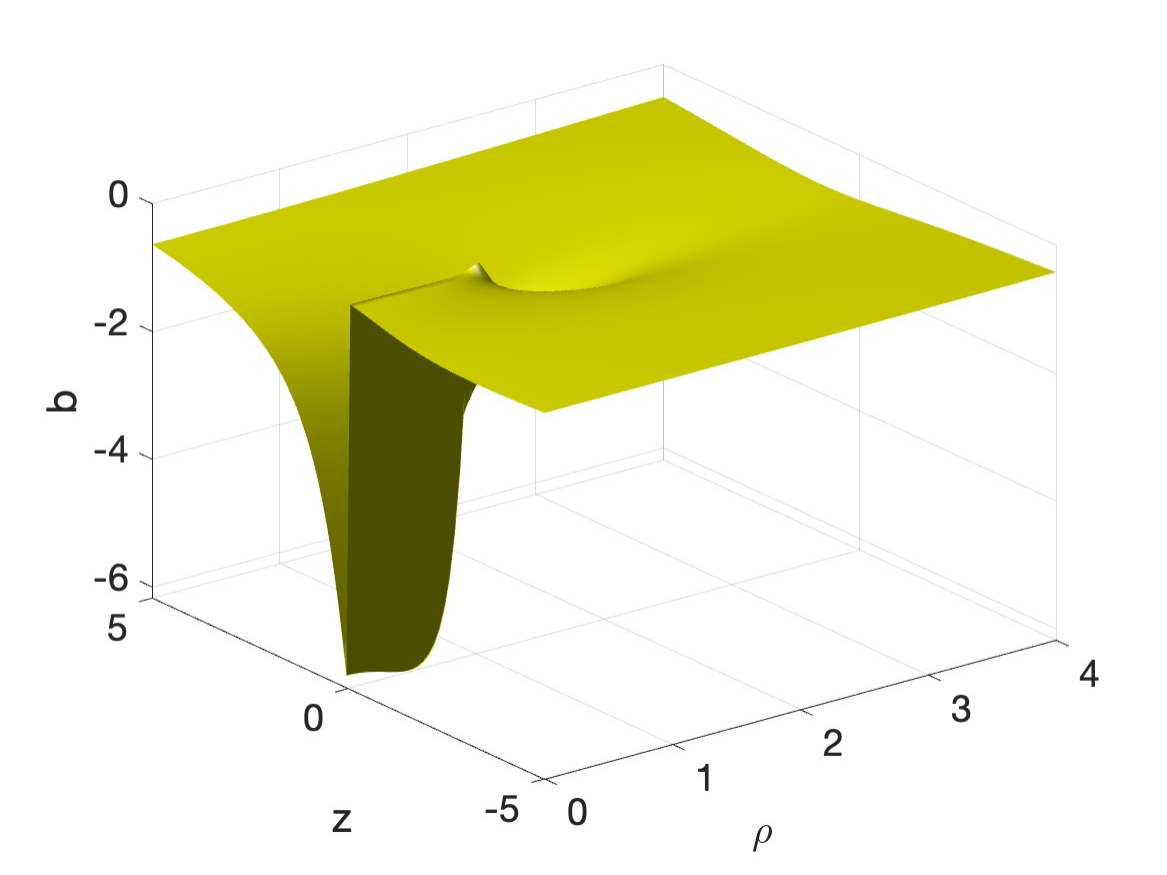}
  \caption{ Ernst potential (\ref{eq:ernst_potential}) for 
  $\alpha=0.3$, $\p=0$. On the left the real part, on the right the 
  imaginary part.}  \label{ellipticernst}
\end{figure}

It is important to note that the integral  $\int^{\infty^+}_\xi 
	 \omega$ is branched in the interval $0\leq \rho\leq 1$ of the equatorial plane, which 
implies that the Ernst potential is discontinuous for $0<\rho<1$ as can be 
seen in Fig.~\ref{ellipticernst}. In contrast to the disk discussed 
in \cite{KRPRL} the metric is not continuous at the disk which makes 
an interpretation as a delta-type distribution, for instance a dust 
disk, problematic since the energy-momentum tensor would be 
proportional to the derivative of a delta function in this case. One possible
 interpretation of this solution is thus that it is defined 
on an infinitely-sheeted spacetime whose slice for constant $\phi$ 
and $t$ is the Riemann surface in the $\xi$-coordinate whose ``branch ring''
 is located at $z=0$, $\rho=1$, see \cite{K91}. In this paper we shall be considering only the first ``sheet" of the spacetime ignoring its non-trivial topology
 and, therefore, 
we will treat 
the disk $0\leq \rho\leq 1$ as in \cite{DFK} as totally light absorbing. Thus we 
discuss the spacetime with 
a discontinuity 
at the interval $z=0$,   $0\leq \rho\leq 1$. 
The Ernst potential for the toron solution reads
\begin{equation}
 \ernst(\xi,\bar{\xi}) =\frac{\vartheta(\int^{\infty^+}_\xi 
	 \omega+\ii\alpha ,\B)}{\vartheta(\int^{\infty^-}_\xi \omega+\ii\alpha,\B)}\;.
\label{toron}
\end{equation}
We show an example in Fig.~\ref{ellipticernst}.

\begin{figure}[!htb]
  \includegraphics[width=0.49\hsize]{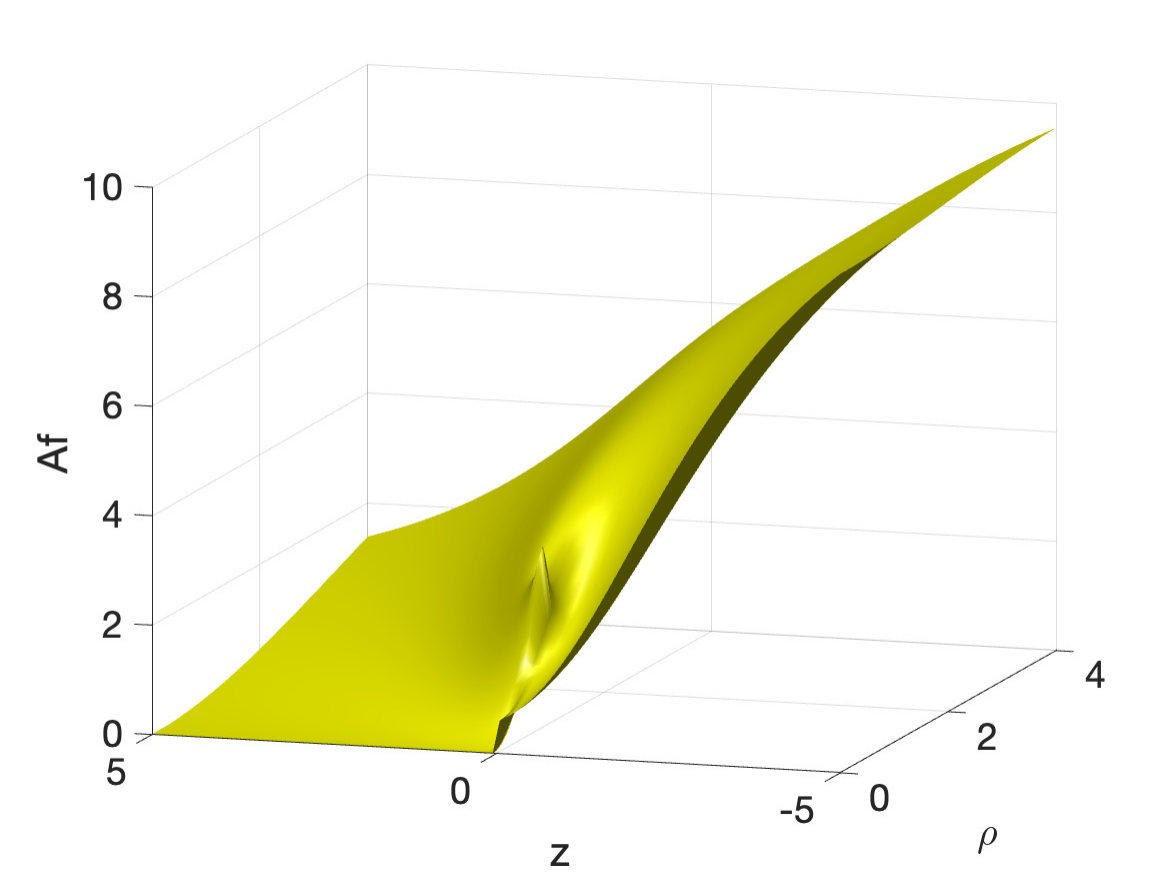}
  \includegraphics[width=0.49\hsize]{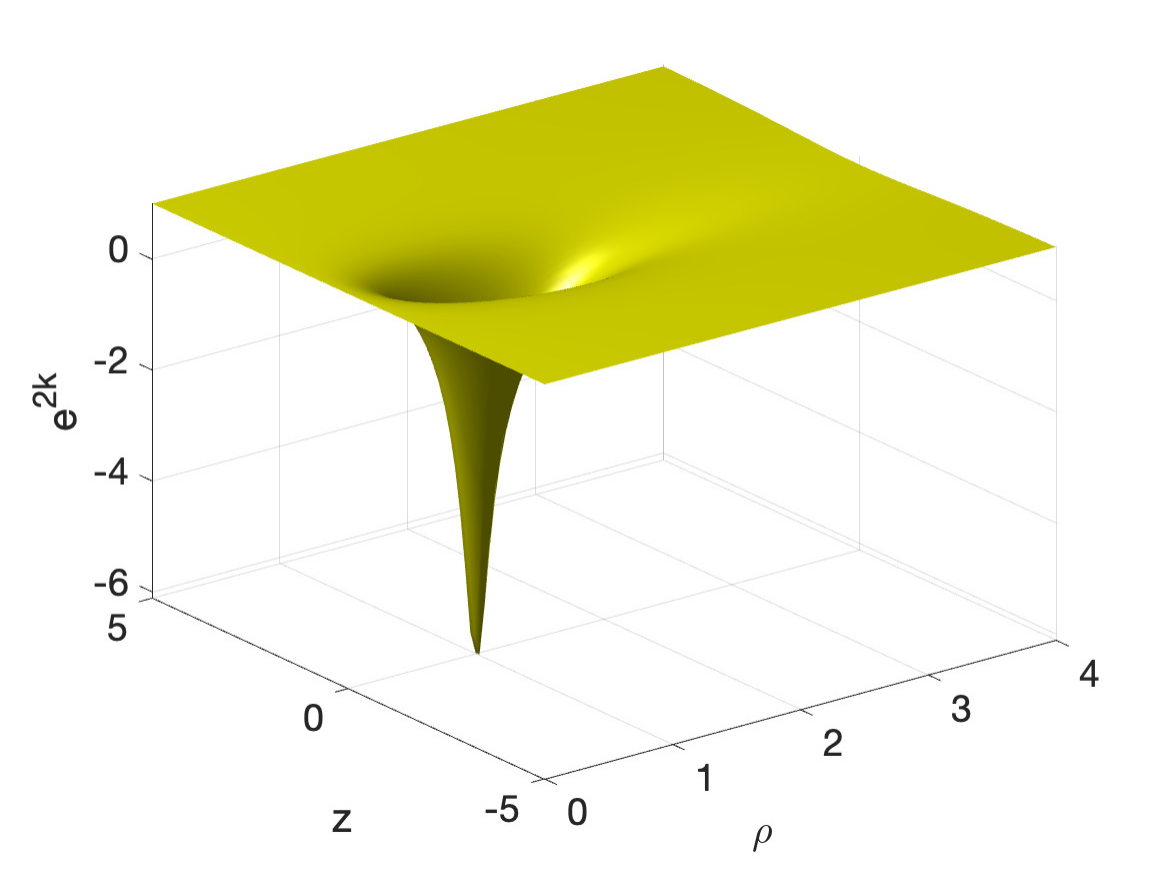}
  \caption{Metric function $Af$ on the left and $e^{2k}$ on the right 
  for the Ernst potential shown in Fig.~\ref{ellipticernst}. }
  \label{ellipticmetric}
\end{figure}

The metric functions corresponding to the Ernst potential 
(\ref{toron}) can be given in closed form, see \cite{KKS}:
 \begin{equation}
     f = Q  \frac{\vartheta(\ii\alpha) \vartheta(\ii\alpha+1/2)}{\vartheta(\int^{\infty^-}_{\xi} \omega + \ii\alpha) \vartheta(\int^{\infty^-}_{\Bar{\xi}}\omega+\ii\alpha)},
     \label{fmet}
 \end{equation}
 \begin{equation}
     e^{2k} = \frac{\vartheta(\ii\alpha) \vartheta(\ii\alpha +1/2) }{ \vartheta(0) \vartheta(1/2 ) },
     \label{kmet}
 \end{equation}
 \begin{equation}
     (A-A_0)f = -\rho \left( \frac{1}{Q} \frac{\vartheta(\ii\alpha) \vartheta(\int^{\infty^-}_{\xi} \omega + \int^{\infty^-}_{\Bar{\xi}} \omega+\ii\alpha) }{ \vartheta(\int^{\infty^-}_{\xi} \omega+\ii\alpha ) \vartheta(\int^{\infty^-}_{\Bar{\xi}} \omega+\ii\alpha) } - 1 \right)
     \label{Amet},
 \end{equation}
where the function $Q$ is
 \begin{equation}
     Q =\frac{\vartheta(\int^{\infty^-}_{\xi} \omega ) \vartheta(\int^{\infty^-}_{\Bar{\xi}} \omega)}{\vartheta(0) \vartheta(1/2)}.
 \end{equation}
 The metric functions for the  Ernst potential shown in 
 Fig.~\ref{ellipticernst} are plotted in Fig.~\ref{ellipticmetric}. 
 Similarly    to the case of 
 the NUT solution, the function $A$ cannot be chosen to 
 vanish on the whole $z$-axis; $A$ can vanish either   for $z>0$ or for $z<0$. 
 The function $A$ chosen to vanish  for $z>0$ is shown on the left pane of 
 Fig.~\ref{ellipticmetric}. The corresponding function $e^{2k}$ can be 
 seen on the right  pane. It is important to remark here that only the metric function $e^{2k}$ is symmetric with respect to the equatorial plane, while the functions $f$ and $A$ are neither even nor odd.

\subsection{Physical properties}

We are interested in providing a physical interpretation of the 
parameter $\alpha \in \R_+$, which is done by studying the 
asymptotic behavior of the metric and by studying light rays in 
various spacetimes in a later section. Before that, we describe the peculiar ergospheres in these spacetimes.

Let us recall that ergospheres are defined as the hypersurfaces satisfying the condition $g_{tt}=0$. Given a spacetime with $\alpha$, it is shown in Appendix \ref{sec:physical_prop} that this spacetime admits the infinite family of ergospheres described by

\begin{equation*}
    (\rho-\rho_n)^2 + z^2 = R_n^2,
\end{equation*}
for all $n\in\N$, 
where $\rho_n$ and $R_n$ are functions that increase monotonically with the magnitude of $\tau_n= \I\alpha /(n+\frac{1}{2})$, meaning that their value in dependence of $n$ is decreasing. The explicit form of both radii is given by \eqref{eq:E_n} in Appendix \ref{sec:physical_prop}.

Therefore, for each value of the parameter $\alpha$ (which 
parametrizes the class of elliptic solutions), there is an infinite family of ergospheres parametrized by $n\in\N$ and since the spacetime is stationary axisymmetric, the spatial projection of each of these surfaces is a torus with major radius $\rho_n$ and minor radius $R_n$. As mentioned above, both of these radii decrease monotonically in dependence of $n$; moreover, as $n$ tends to infinity we observe that $\rho_n \to 1$ and $R_n \to 0$, i.e., the limiting ergosphere is a ring of radius 1.

Numerically, we will only be interested in the largest ergosphere $E_0$ (which corresponds to $n=0$, i.e. $\tau_0=2\ii\alpha$)
 since the coordinate time of test particles 
and photons goes to infinity as they approach $E_0$.  The ergospheres $E_0$ for a family of spacetimes in dependence of $\alpha$ are described by the following data.

\begin{table}[H]
    \centering
    \begin{tabular}{ p{7mm}|p{7mm}|c|c|c|c|c|c|c}
   \hline \centering $\alpha$ & \centering 0 & 0.1 & 0.15 & 0.2 & 0.25 & 0.3 & 0.4 & 0.5 \\ \hline
    \centering $\rho_0$ & \centering 1 & 1.0000 & 1.0009 & 1.0125 & 1.0607 & 1.1778 & 1.7416 & 3.0000 \\ \hline
    \centering $R_0$ &\centering 0 & 0.0031 & 0.0426 & 0.1584 & 0.3536 & 0.6222 & 1.4259 & 2.8284 \\ \hline
    \end{tabular}
    \caption{First ergospheres $E_0$ in dependence of $\alpha$.} \label{table:ergo_q}
\end{table}
The size of the main ergosphere $E_0$ increases in dependence of $\alpha$, i.e. both radii $\rho_0$ and $R_0$ increase monotonically with $\alpha$ (the same can be said about the $n$th ergosphere, for every $n\in\N$), which can be deduced from the explicit formulas presented in Appendix \ref{sec:physical_prop}.
This monotonic increase can be observed in Table \ref{table:ergo_q} for $E_0$ for some values of $\alpha$.

\begin{figure}[H]
    \centering
    \includegraphics[width=6cm]{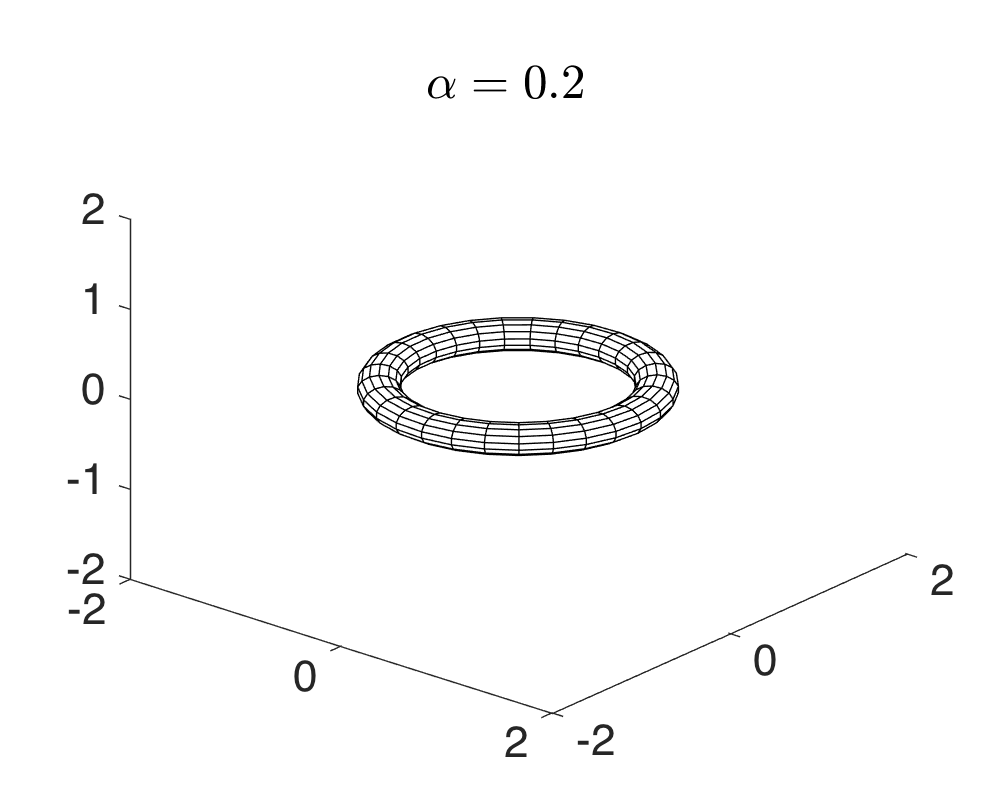} 
    \includegraphics[width=6cm]{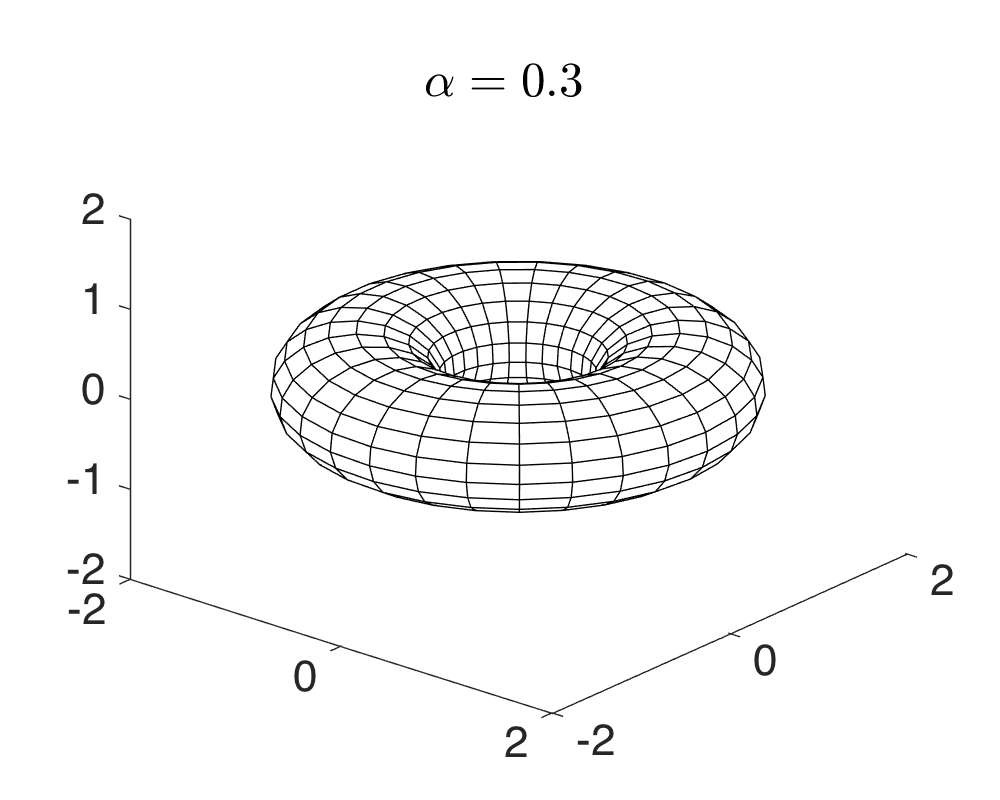}
    \caption{3D representation of the ergospheres of spacetimes with a fixed $\p=0$ and parameters $\alpha=0.2$ and $\alpha=0.3$.} \label{fig:ergo0}
\end{figure}

Fig. \ref{fig:ergo0} shows the three-dimensional representation of the ergospheres in Table \ref{table:ergo_q} with $\alpha=0.2$ and $\alpha=0.3$.

On the axis ($\rho\to 0 $) the branch cut $[\xi,\bar{\xi}]$ collapses 
to a point, and all quantities on the elliptic curves 
(\ref{eq:curves_L_xi}) can be expressed in terms of quantities 
defined on the genus 0 curve $\mu^{2}=\lambda^2+1$, see for 
instance \cite{KR} and references therein. For small $\rho$, the 
$b$-period has an expansion of the form $\B= 2\ln \rho 
+\sum_{n=0}^{\infty}c_{n}\rho^{n}$ with constants $c_{n}$, 
$n=0,1,\ldots$. 


If the parameter $\p\neq 0$ then there will 
be terms proportional to $\rho^{2\mathrm{p}}$ in the expansion of the 
Ernst potential near the axis, i.e., the axis will not be regular 
for $\p$ not equal to $0$ or $1/2$.
For this reason in this paper we  only consider the case $\p=0$ in 
detail. 

To study the asymptotic behavior for the elliptic Ernst potential, this value is computed analytically on the axis when the branch points $\xi$ and $\bar{\xi}$ coincide and the elliptic curve degenerates. The Ernst potential on the axis (the step-by-step computation is presented in Appendix \ref{sec:physical_prop}) is described by the elementary functions
\begin{equation}
\ernst(z,\rho=0)=\frac{z+\sqrt{z^2+\sigma^2}-\ii \sigma e^{-2\pi \alpha}}{z+\sqrt{z^2+\sigma^2}-\ii \sigma e^{2\pi \alpha}}\;,
\hskip0.7cm z < 0,
\label{zetal0}
\end{equation}
and
\begin{equation}
\ernst(z,\rho=0)=\frac{z-\sqrt{z^2+\sigma^2}-\ii \sigma e^{-2\pi \alpha}}{z-\sqrt{z^2+\sigma^2}-\ii \sigma e^{2\pi \alpha}}\;,
\hskip0.7cm z > 0.
\label{zetag0}
\end{equation}

The expressions (\ref{zetal0}) and (\ref{zetag0}) imply the expansion
\begin{equation}
	\mathcal{E}=1-\frac{\ii\sigma}{z}\sinh(2\pi\alpha)+\mathcal{O}(1/z^{2})\; 
	\label{mass1}
\end{equation}
around infinity. Therefore, the ``mass"  $\ii\sigma\sinh(2\pi\alpha)$  is purely imaginary in the elliptic solution. In the sequel, we refer to this solution as the toron spacetime. 
\section{Geodesics in stationary axisymmetric spacetimes}
\label{geodesic_section}
In this section we discuss ray  tracing in stationary axisymmetric 
vacuum spacetimes with the approach \cite{DFK}, see Appendix 
\ref{appgeodesics} for details. 
We describe the expected behavior of individual geodesics given some particular initial conditions
and then explain the process to simulate the apparent images of 
extended objects in the presence of a strongly gravitating body, such as a 
black hole or a gravitating disk. 

\subsection{Light rays on the equatorial plane and the plane passing through the symmetry axis}

We study the expected behavior of individual light rays on a plane with initial conditions of two types: those initially moving on the equatorial plane and those initially moving on a plane containing the $z$-axis.

To make sure the light rays initially stay on the equatorial plane, we  assume that the initial conditions satisfy
 $z_0=p^z_0=0$. 
Considering $s$ to be an affine parameter, as used in Appendix \ref{appgeodesics},
the deviation of the light rays from the equatorial plane is determined by the value of $d p^z/ds$: they will move in upwards (downwards) direction if this value is positive (negative). 
 For spacetimes whose metric functions $f$ and $e^{2k}$ are even in $z$, such as Kerr and NUT, their derivatives with respect to $z$ vanish identically on the equatorial plane. Therefore, for such spacetimes

\begin{equation} \label{eq:parallel_xy}
    \left.\frac{d p^z}{ds} \right|_{s=0} = - \left. \frac{f\cdot \partial_z A}{h}  \right|_{(\rho_0,z_0)} p^t_0 \, p^\phi_0.
\end{equation}
In Kerr spacetimes $\partial_z A$ also vanishes identically on the plane, and, therefore, light rays of this type are bound to the plane. 

On the other hand, in NUT spacetimes with $\ell>0$ , the derivative 
$\partial_z A$ is strictly positive on the equatorial plane (as 
observed in section \ref{sec:kerr-nut-metric}). Therefore, in the 
case of the NUT metric the light rays do not stay on the plane, unless they move in a purely radial direction. Otherwise, they will go in downwards (respectively upwards) direction if $p^\phi_0$ is positive (respectively negative). 
On the other hand, due to the lack of symmetry of the metric 
functions in toron spacetimes, the component $d p^z/ds$ of the geodesic equations cannot be simplified to the form \eqref{eq:parallel_xy}, and we must rely on numerical computations in order to analyze the behavior of geodesics with initial conditions of this type. 

For the light rays which initially stay on a the plane passing through the symmetry axis, we must consider initial conditions with
$$p_0^\phi=0\;.$$
Then the conserved energy and momentum  have the form 
$$E = f^{(0)} p_0^t \;, \hskip0.7cm L=-f^{(0)} A^{(0)} p^t_0\;,$$
 where the superscripts indicate the initial value of the metric functions. Hence, the geodesic equations (\ref{ode_geodesics}) imply
\begin{equation} \label{eq:pphi_eq}
  \frac{d\phi}{ds} = \frac{1}{\rho^2} f \cdot f^{(0)} \cdot (A-A^{(0)}) p^t_0 \;. 
\end{equation}
The component $p_0^t$ is positive for  light rays  chosen  to be moving forward in time. Moreover, we know that the function $f$ is strictly positive outside the ergosphere. Therefore, equation (\ref{eq:pphi_eq}) indicates that the clockwise or counterclockwise deviation of light rays of this type is controlled by the difference $(A-A^{(0)})$ for $s\geq 0$. For instance, if the light rays are initially moving in an inwards radial direction, they bend in counterclockwise (resp. clockwise) direction if $\partial_\rho A |_{s=0}$ is positive (resp. negative). This conclusion holds for any stationary axisymmetric spacetime.

\subsection{Photon spheres in Kerr and light rings in NUT}

For Kerr and NUT spacetimes there exist light-like geodesics traveling on spheres of fixed Boyer-Lindquist radii $r$ \cite{Wilkins,Teo,PT,MHC}, called the \textit{photon spheres}. 
Such geodesics are called the {\it fundamental photon orbits} in general stationary axisymmetric spacetimes if their trajectory is a closed path \cite{CH}.
If this trajectory is planar, then it is referred to as a \textit{light ring}.
In Kerr spacetime light rings lie in the equatorial plane, while in NUT spacetimes the light rings can be lifted from the equatorial plane.
In Kerr spacetimes there are two light rings: the prograde one (the direction of photons coincides with the direction of rotation) of smaller radius $r_{\min}$ and the retrograde one (the direction of photons is the opposite to  with the direction of rotation) of larger radius $r_{\max}$.
The radii of photon spheres where photons not constrained to the equatorial plane are moving, lie between $r_{\min}$ and $r_{\max}$.

A necessary condition for the photon to stay on the photon sphere  is $dr/ds=0$ for all $s\in\R$. Therefore, $d^2r/ds^2=0$ for all $s\in\R$ is also necessary. Thus, the possible photon sphere radii $r_p$ in a given spacetime in the presence of rotation depends on the initial condition of the null geodesic and can be found from the  system of equations $dr/ds = d^2r/ds^2=0$.
Such values have been computed explicitly for a family of multiparametric spacetimes, see \cite{GPL}, which include Kerr and NUT as particular examples. 
In Schwarzschild and NUT spacetimes there exists a unique radius 
satisfying the above condition.
Its radius 
depends on both the mass $m$ and the gravimagnetic mass $\ell$. From 
the conditions for the radii of the light ring given in \cite{GPL}, the Boyer-Lindquist radius of the equatorial light ring  is given explicitly by the real solution of the equation 
$$ 2r^3-6mr^2-6r\ell^2+m\ell^2=0,$$
which is 
\begin{equation} \label{eq:rad_photon_sphere}
    r_p = m + 2\sqrt{m^2+\ell^2} \cos\left(\frac{1}{3} \arctan(\ell/m) \right),
\end{equation}
if $m\neq 0$ (which includes the well-known radius $r_p=3m$ in the Schwarzschild spacetime) and $r_p=\sqrt{3}\ell$ if $m=0$. 



The explicit description of fundamental photon orbits not staying in the equatorial plane is not known analytically (it is known that they are light rings in NUT, but the explicit equation of the plane containing it is not).
It is even more  non-trivial to determine the
 explicit shape or even existence of the fundamental photon orbits (i.e. photon orbits staying in a neighborhood of a compact object) and corresponding photon surfaces  in   general stationary axisymmetric spacetimes (including the  toron
spacetimes); since no equivalent to the Carter 
constant is  known, the geodesic equations cannot be decoupled and thus, the Boyer-Lindquist coordinates are no longer natural.
In order to determine numerically the fundamental  photon orbits in NUT and an approximation thereof in toron spacetimes we use the ray tracing techniques discussed below.

\subsection{Relativistic ray tracing}
\begin{figure}[H]
    \centering
    \fbox{\includegraphics[width=6.5cm]{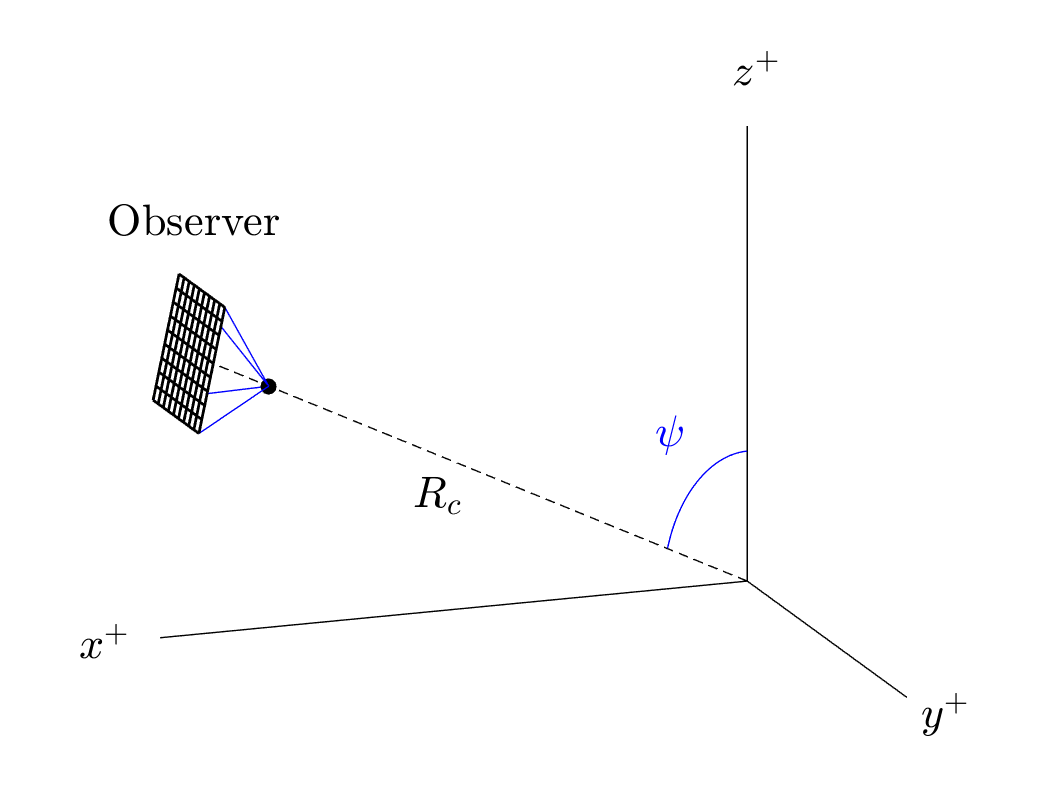}}\fbox{\includegraphics[width=6.5cm]{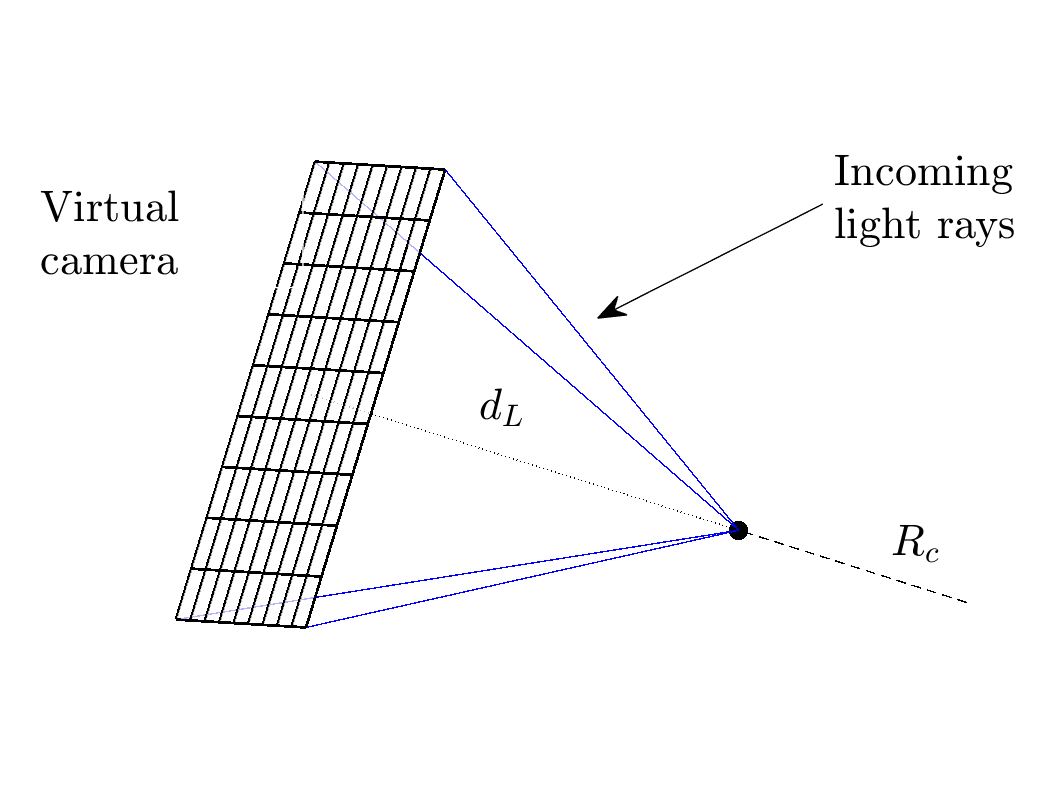}}
    \caption{Diagram of the virtual camera (not to scale for representation purposes) with respect to the black hole and an enlarged image of the screen with some incoming light rays.} \label{fig:camera_diagram} 
\end{figure}

In order to perform the simulation of the apparent image of extended 
objects in curved spacetimes
we use ray tracing techniques, as 
described in \cite{DFK} for general stationary axisymmetric 
spacetimes. This technique consists in studying the light rays 
reaching a virtual camera (which is used to simulate the apparent 
image seen by a distant observer), as shown in Fig. 
\ref{fig:camera_diagram}. The observer is located at a distance $R_c$ 
from the  gravitational source (such as a black hole or a gravitating 
disk) and its line of sight is at an angle $\psi$ with respect to the symmetry axis. The screen of the virtual camera has a width $d_H$, height $d_V$, a resolution of $I\times J$ pixels, with $I,J\in\mathbb{N}$, and a varying focal length $d_L$. 
To compute the horizontal and vertical angular apertures $\delta^c_H$ and $\delta^c_V$ of the camera in terms of these values we observe that the vertical and horizontal sections passing through both the center of the screen and the angular aperture are isosceles triangles, see Fig. \ref{fig:camera_diagram}. These triangles have a base $d_H$ (respectively $d_V$), height $d_L$ and the angle opposite to the base is in fact $\delta^c_H$ (respectively $\delta^c_V$). Thus, the horizontal and vertical angular apertures of the camera are
\begin{equation*}
    \delta^c_{H,V} = 2 \cdot \arctan\left( \frac{d_{H,V}}{2d_L} \right).
\end{equation*}

In this paper we choose fixed values of $d_H$ and $d_V$. Thus, the angular apertures and therefore, the field of vision of the observer will depend inversely on the value of the focal length $d_L$. 
Once the camera features are set, each pixel on the screen gives an initial condition for the initial value problem --IVP-- \eqref{eq:ivp}, which is then integrated backwards in time in order to determine the sources of light. 

The following are the chosen criteria to terminate the time integration of the geodesics, namely, the computations are stopped if:
\begin{itemize}
    \item[1.] The light ray hits an object.
    \item[2.] $\Re {\mathcal E}=f<\epsilon$, for a small $\epsilon>0$. This means that the photon approaches the first ergosphere.
    \item[3.] $r>R_{\infty}$ for some $R_\infty \gg 1$, where $r=\sqrt{\rho^2+z^2}$. This will indicate that the photon escapes to infinity.
    \item[4.] $s_n>S_{\max}$. This means that the integration time has surpassed a predefined maximum threshold.
\end{itemize}
1. If the photon hits an object, we record the color of the 
point hit by such a photon and associate that to the corresponding 
pixel. This condition is added as an event function in Matlab. \\
2. If the photon approaches the first ergosphere such that $f<\epsilon$, the color given 
to the corresponding pixel is black, which would indicate that a ray reaching that pixel cannot originate from any light source without passing through the ergosphere. 
The computations are stopped
since the coordinate time will go to infinity; therefore, the ergosphere
cannot be crossed by any light ray in Weyl coordinates. \\
3. The condition  $r>R_\infty$ indicates that the light ray has originated from a large sphere with radius $R_\infty$ concentric with the strongly gravitating object. In the sequel, we refer to such a sphere as the \textit{celestial sphere} in analogy with the terminology used in astronomy.
If a photon meets this condition, we assign a color to the corresponding 
pixel according to a predefined coloring of the celestial sphere. \\
4. Finally, if the computations are interrupted because $s_n>S_{\max}$ we color the pixel in black, since this means that the photon has made several turns around the gravitating body without falling towards the center nor escaping to infinity. We choose a value for $S_{\max}$ such that at least the secondary copy of the celestial sphere is visible, but one could choose a sufficiently large value in order to show subsequent copies.

\subsubsection{Angle of view and coloring of the celestial sphere}
\begin{figure}[htb!]
    \centering
    \includegraphics[width=5cm]{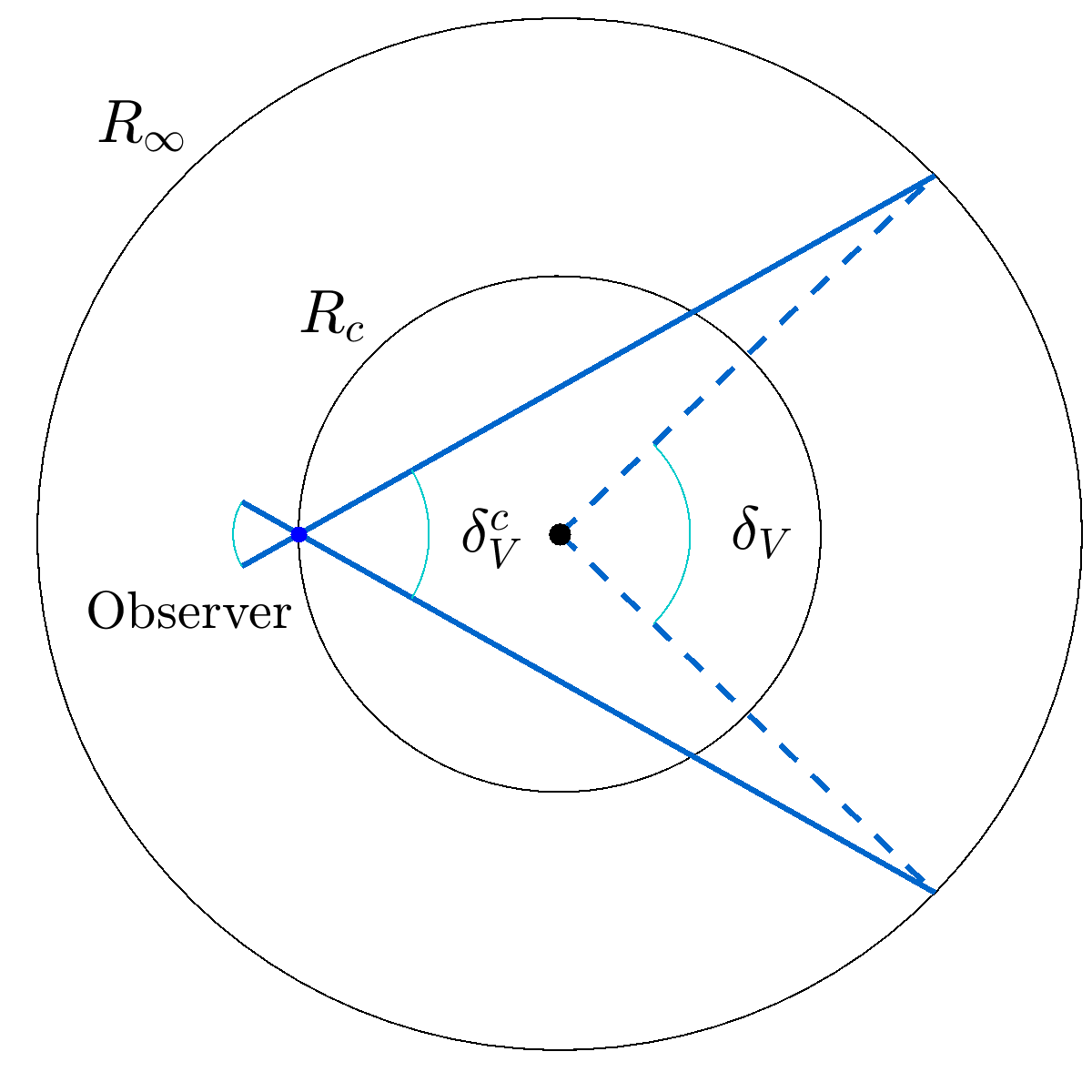}
    \caption{Vertical section of the celestial and observer's spheres, showing the relation between the angular aperture $\delta^c_V$ of the virtual camera and the effective angle of view of the celestial sphere $\delta_V$. The same relation applies to the angles $\delta^c_H$ and $\delta_H$. }  \label{fig:camera_angles}
\end{figure}

\begin{figure}[H]
    \centering
    \includegraphics[width=10cm]{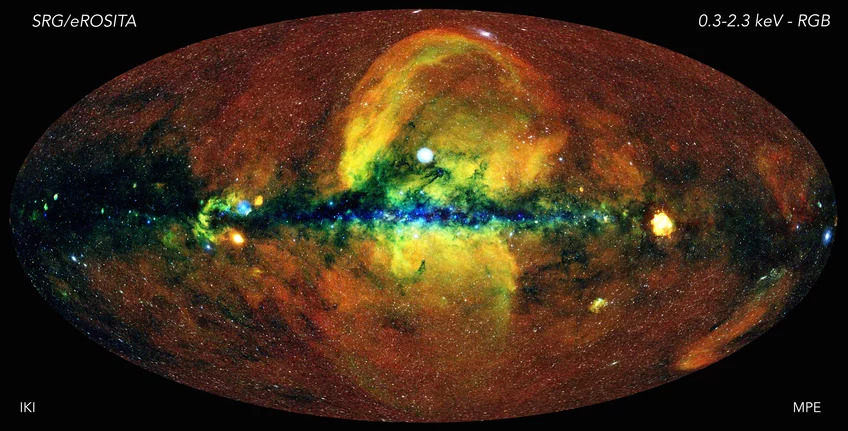}
    \caption{All-sky survey by the eROSITA X-ray telescope. The celestial sphere is projected onto an ellipse with the Aitoff projection. Credit: J. Sanders, H. Brunner, A. Merloni and the eSASS Team (MPE); E. Churazov, M. Gilfanov, R. Sunyaev (IKI). } \label{fig:celestial_sphere} 
\end{figure}

\begin{figure}[H]
    \centering
    \includegraphics[width=11cm]{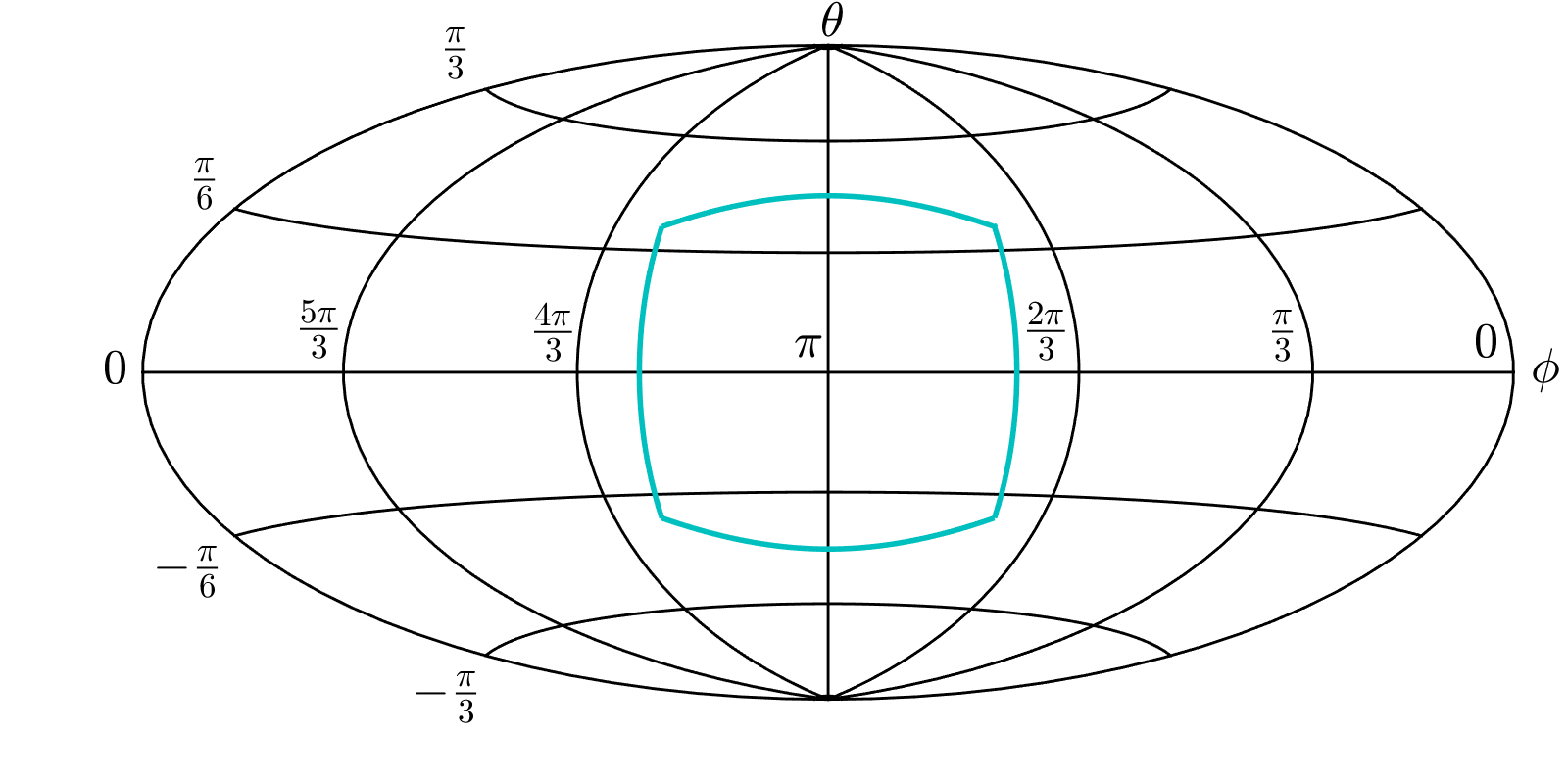}
    \caption{Aitoff projection of the celestial sphere with constant coordinates $\phi$ and $\theta=\arctan(z/\rho)$. The boundary of the recorded image in Minkowski spacetime with angles of view $\delta_H=\delta_V=90^\circ$ is indicated by the blue contour. } \label{fig:aitoff} 
\end{figure}

Since the spheres with radii $R_c$ and $R_\infty$ are chosen to be concentric (see Fig. \ref{fig:camera_angles}), the effective horizontal and vertical angles of view $\delta_{H}$ and $\delta_{V}$ for a given $R_\infty$ in Minkowski spacetime are given by the formula 
\begin{equation*}
    \delta_{H,V} = \delta^c_{H,V} + 2 \cdot \arcsin\left(\frac{R_c}{R_
    \infty} \sin\left(\delta^c_{H,V}/2 \right) \right).
\end{equation*}
Thus, given a camera with aperture angles $\delta^c_{H,V}$, the observed portion of the celestial sphere will be within angles $\delta_{H,V}$ (notice that the limit when $R_\infty$ tends to infinity is just $\delta_{H,V}=\delta^c_{H,V}$), see Fig. \ref{fig:aitoff} in which $\delta_{H}=\delta_{V}=90^\circ$.
However, this formula only holds in flat spacetimes; in gravitating spacetimes it will be an upper bound for the so-called primary image due to the bending of light, as we will observe later. Nevertheless, the deformation with respect to the contour shown in Fig. \ref{fig:aitoff} (which would only correspond to the contour of the primary image in gravitating spacetimes) will indicate the type of optical effect produced by a particular class of spacetimes.
Moreover, in order to observe additional properties, in 
this paper we add an artificial coloring to the celestial sphere following the colors shown 
by Fig. \ref{fig:celestial_sphere} in galactic coordinates, i.e., the 
celestial sphere is projected onto an ellipse using the Aitoff 
projection (see below). 
This is done in order to determine the influence of the various values parametrizing the classes of spacetimes considered here on the apparent images (for instance, $\ell$ for NUT spacetimes and $\alpha$ for toron spacetimes). Namely, we determine some of the optical effects of the gravitating object by comparing the deformation of the observed image with respect the image one would expect in a Minkowski spacetime. 

Fig. \ref{fig:aitoff} shows the Aitoff projection with some constant coordinates $\phi$ and $\theta$, where $\theta=\arccos(z/\rho)$. The line of sight of the observer is directed towards the gravitating object. In terms of the celestial sphere coordinates, the line of sight is directed towards $(\phi,\theta)=(\pi,0)$. Notice that we use the Weyl coordinate $\phi$ for the longitudes, while the usual choice is to consider this angle in the interval $[-\pi,\pi]$. The recorded image in Minkowski spacetime would correspond to the area enclosed by the blue contour in Fig. \ref{fig:aitoff}, whose size (i.e. field of view) will depend directly on the angular aperture of the virtual camera. In curved spacetimes, the boundary of the observed region will change, as we will see in the following sections, but this contour will serve as a reference to assess such deformations. 

\subsection{Shadows}
One important consequence of strongly gravitating objects is that if the line of sight of an observer is directed towards the object, then this object casts a shadow on the apparent image. This means that no light from the celestial sphere will reach the observer if the field of view is sufficiently small. 

Given a screen representing the apparent image seen by an observer in the presence of strongly gravitating body, see Fig. \ref{fig:camera_diagram}, some of the light rays traveling backwards in time (corresponding to each pixel of the screen) hit a predefined object (such as an accretion disk) while others escape to infinity (i.e., towards the celestial sphere) or cross the photon spheres. In the third situation the corresponding photons  fall towards the gravitating object, which is generally assumed not to emit light. If this is the case, then we conclude that these `light rays' did not originate from any light source, and thus the pixels corresponding to such light rays are not illuminated. The set of all such pixels is known as the \textit{shadow} of the gravitating object.  

We recall that in the Schwarzschild and NUT spacetimes there exists a unique 
photon sphere.
These black holes cast therefore a circular 
shadow on the images seen by a distant observer, see \cite{GPL}. However, photon spheres are not unique in Kerr spacetimes; their radii are in an interval $r_{\min}<r_p<r_{\max}$ for some positive $r_{\min}$, $r_{\max}$ with the value of $r_p$ depending on the initial conditions of the geodesics. Therefore, the shadow cast by the rotating black hole  is not circular since the radius of the photon sphere for which the photon will fall into the horizon once it is crossed  depends on the direction at which the photon approaches it with respect to the frame-dragging direction. Let us recall that in Kerr spacetimes, $r_{\min}$ corresponds to a photon in prograde motion and $r_{\max}$ to a photon in retrograde motion.  Moreover, the size of the photon spheres  directly determines the size of the shadow. For instance, it can be inferred from formula \eqref{eq:rad_photon_sphere} that the size of the shadow of a Schwarzschild black hole increases monotonically with its mass $m$, while the shadow of a NUT black hole (given a fixed $m$) increases monotonically with the magnitude of its gravimagnetic mass $\ell$. The computations for toron spacetimes are less straightforward, but their image simulations together with the knowledge on Kerr and NUT spacetimes will indicate the dependence of the shadow size and shape on the imaginary mass of the toron.


\section{Geodesics in Schwarzschild, Kerr and  NUT spacetimes} 
\label{sec:nut_geodesics}

In this section we study numerically  the behavior of individual light rays in NUT 
spacetimes and compare it with the geodesics in Schwarzschild and Kerr. We also  simulate  the apparent images of 
extended objects. The aim is to study the influence of the 
gravimagnetic mass $\ell$ on null geodesics and use these results as a reference for the study of similar phenomena in toron spacetimes.

\subsection{Initially-parallel light rays} \label{sec:parallel_nut}

We start by shooting a beam of parallel light rays in the $xy$-plane. Then we repeat the calculation by  shooting 
a beam of parallel light rays lying  in the  $xz$-plane.
With the usual relation between Weyl and Cartesian 
coordinates ($x=\rho \cos\phi$, $y=\rho\sin\phi$), the following initial conditions will produce initially-parallel light rays on the aforementioned planes. 

\begin{figure}[H]
\centering
\includegraphics[width=6cm]{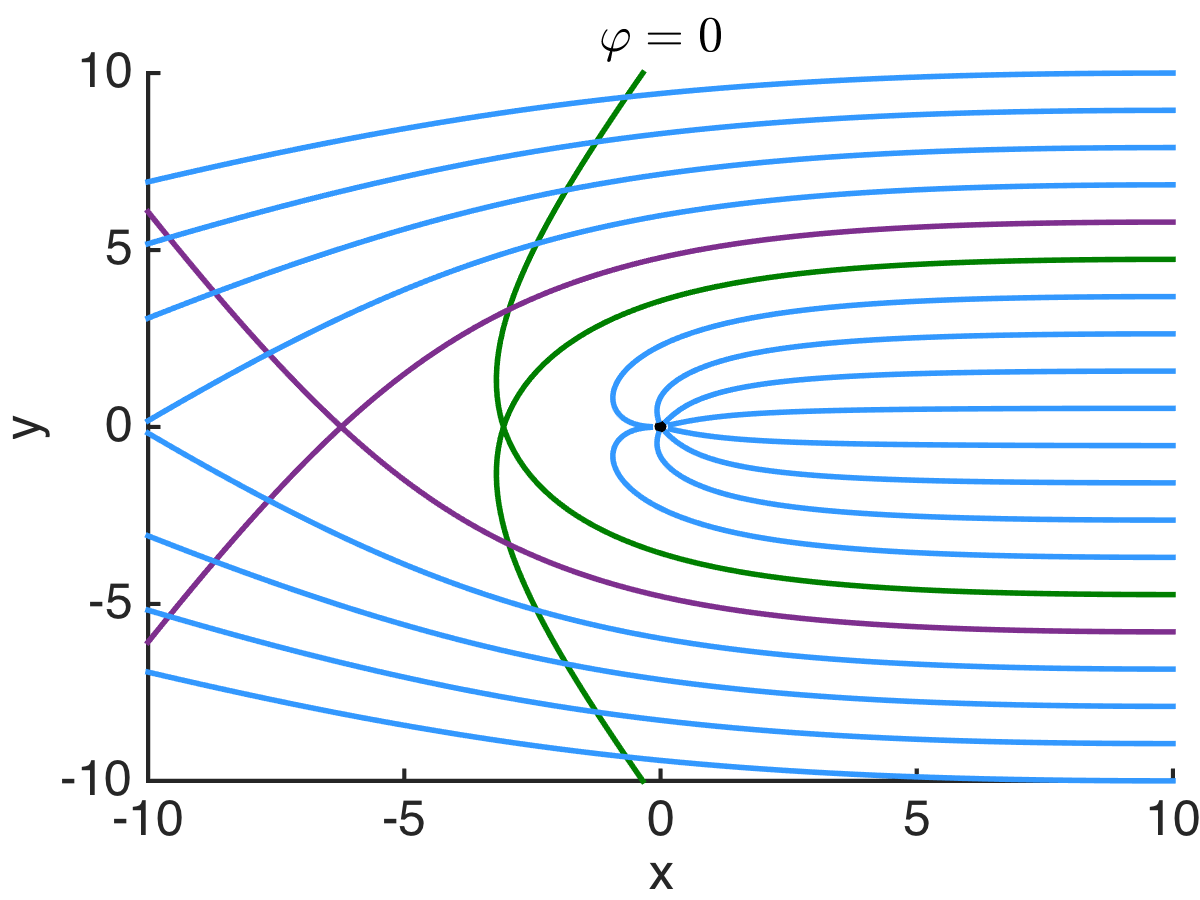} \includegraphics[width=6cm]{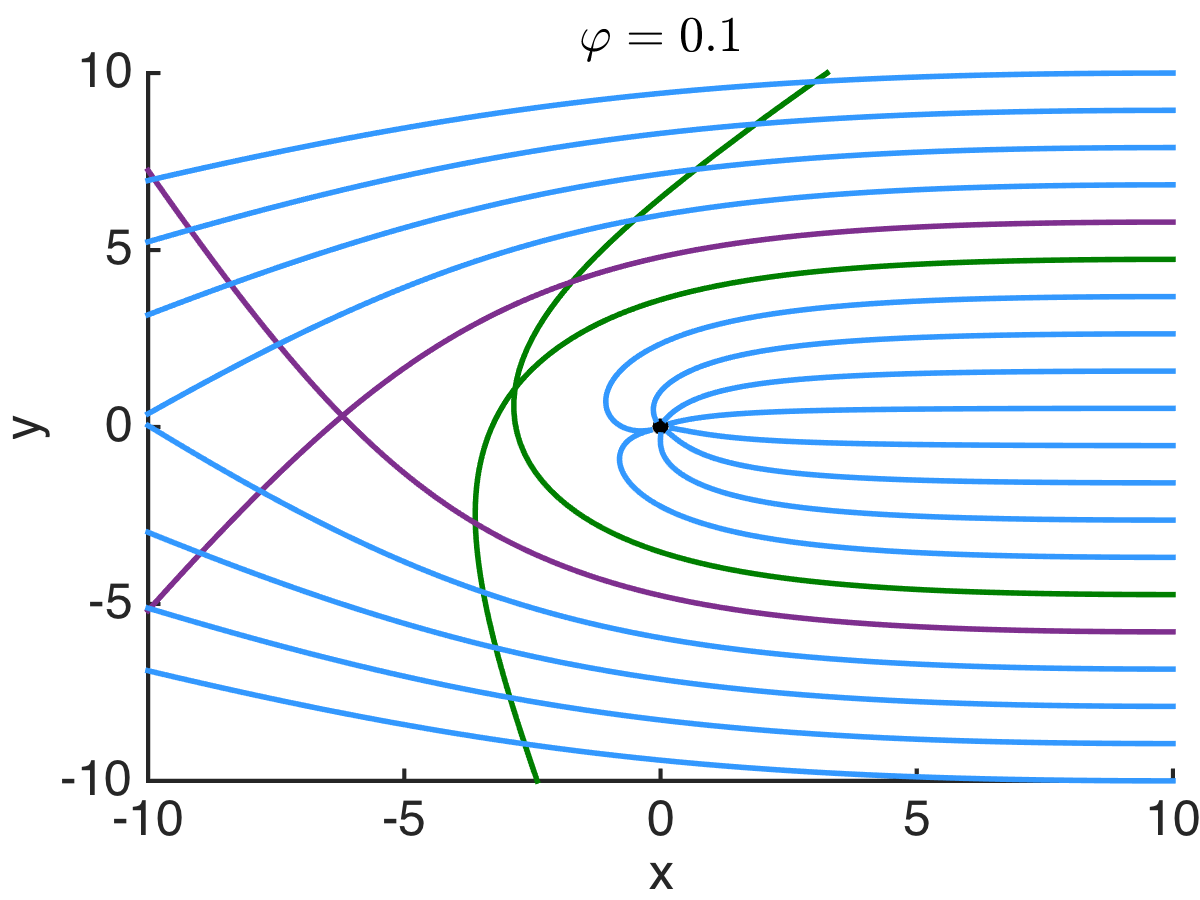} \includegraphics[width=6cm]{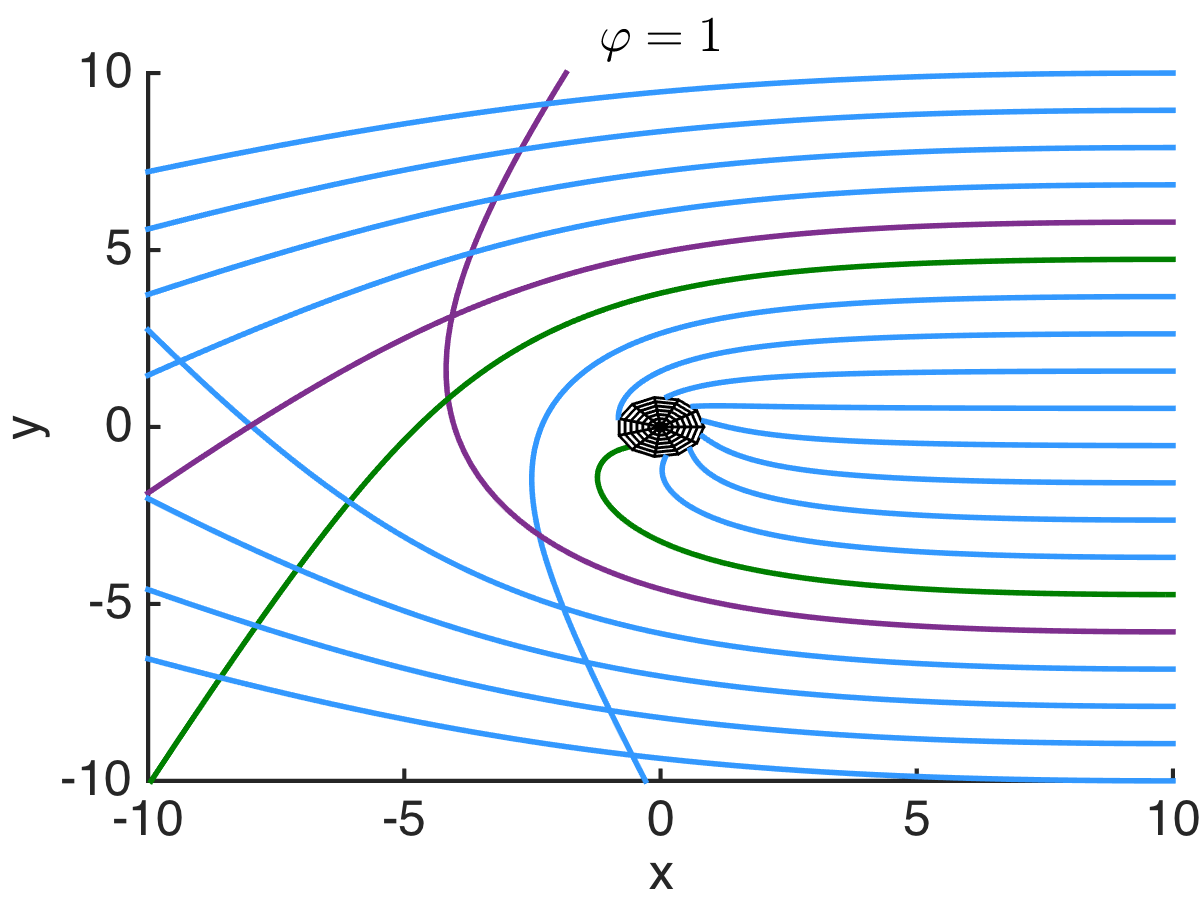} \includegraphics[width=6cm]{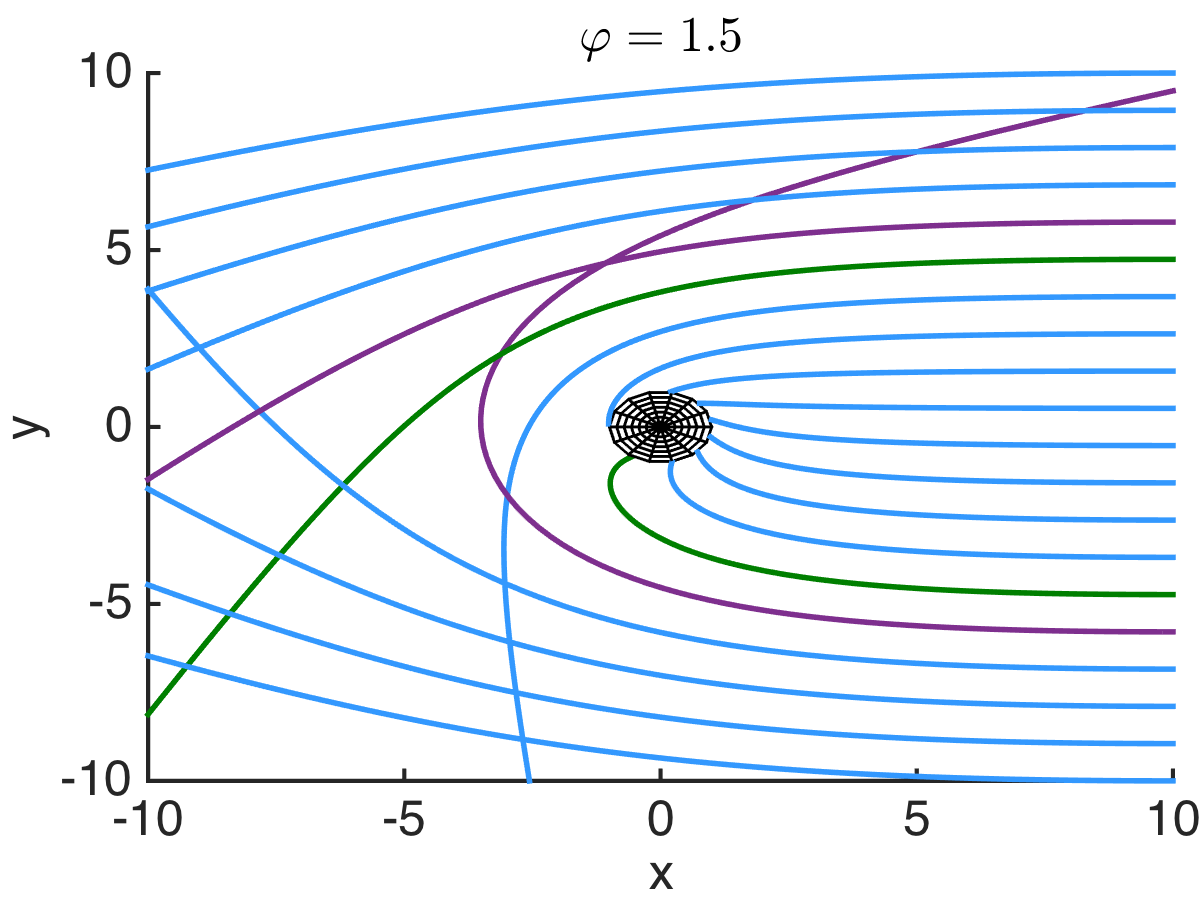} 
\caption{Initially-parallel light rays in the equatorial plane 
for Schwarzschild ($\varphi=0$) and various Kerr spacetimes. For each example, two pairs of light rays corresponding to the same initial conditions except for the $y$-component (we chose $y_0>0$ for one of the rays and $y'_0=-y_0$ for the other) are colored in purple and green. The surface in the middle is the ergosphere. } \label{fig:Kerr_xy}
\end{figure}

\begin{figure}[H]
   \centering
    \includegraphics[width=6cm]{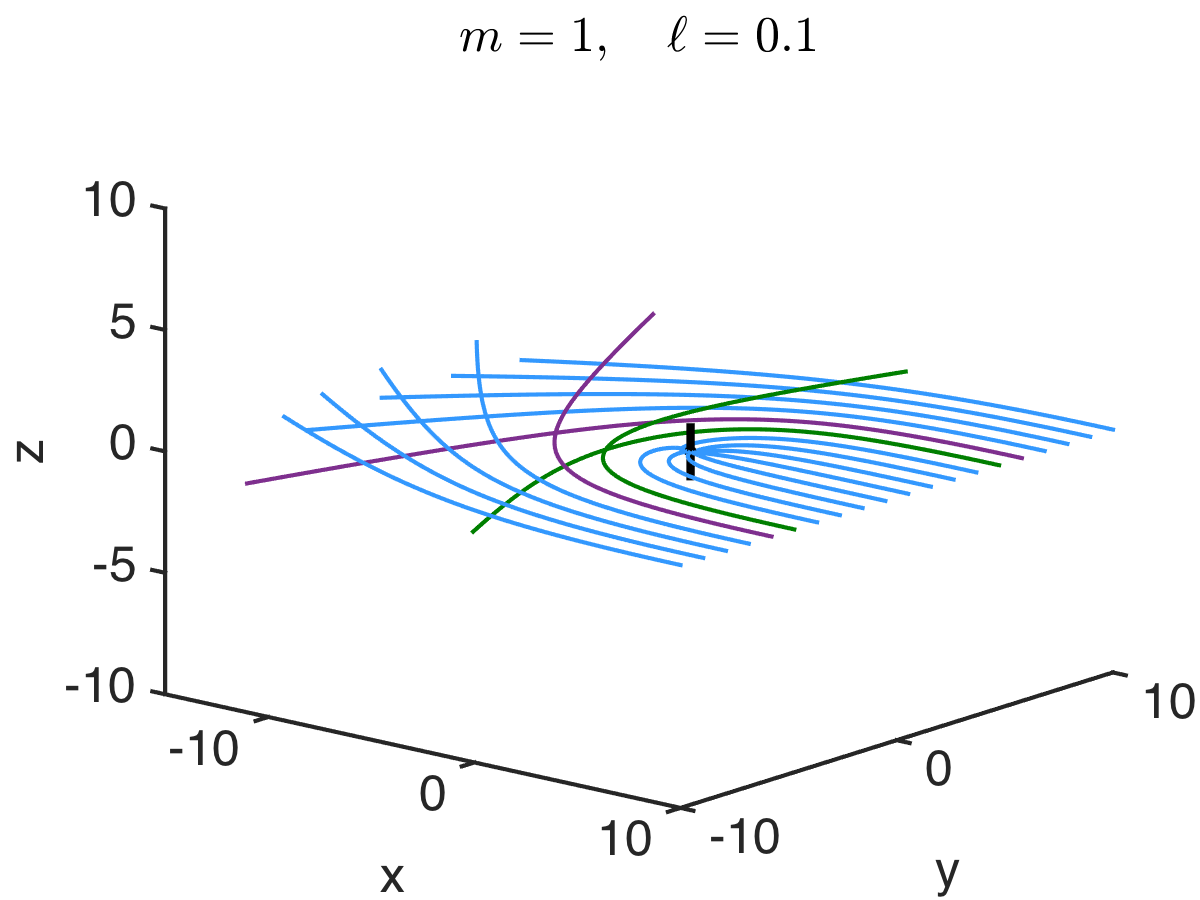}    
    \includegraphics[width=6cm]{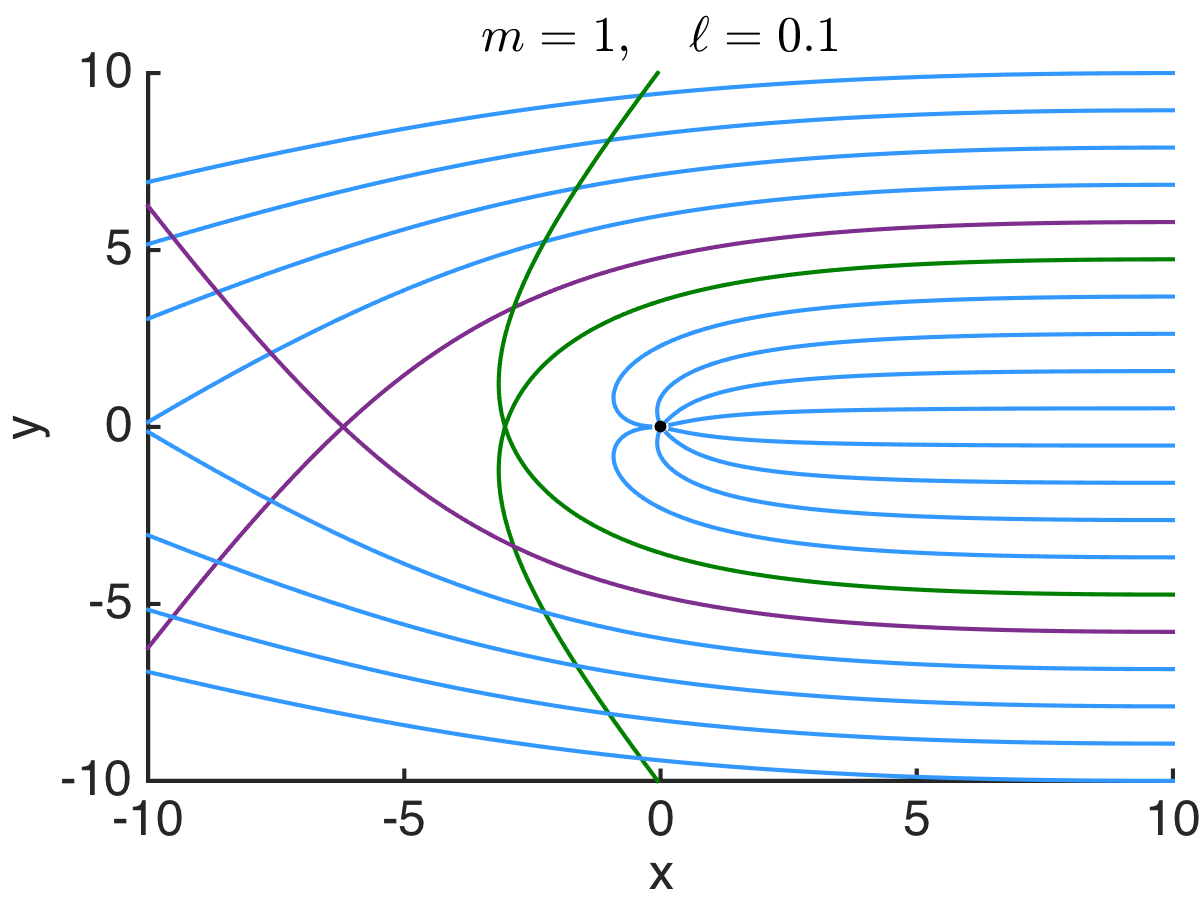}
    \includegraphics[width=6cm]{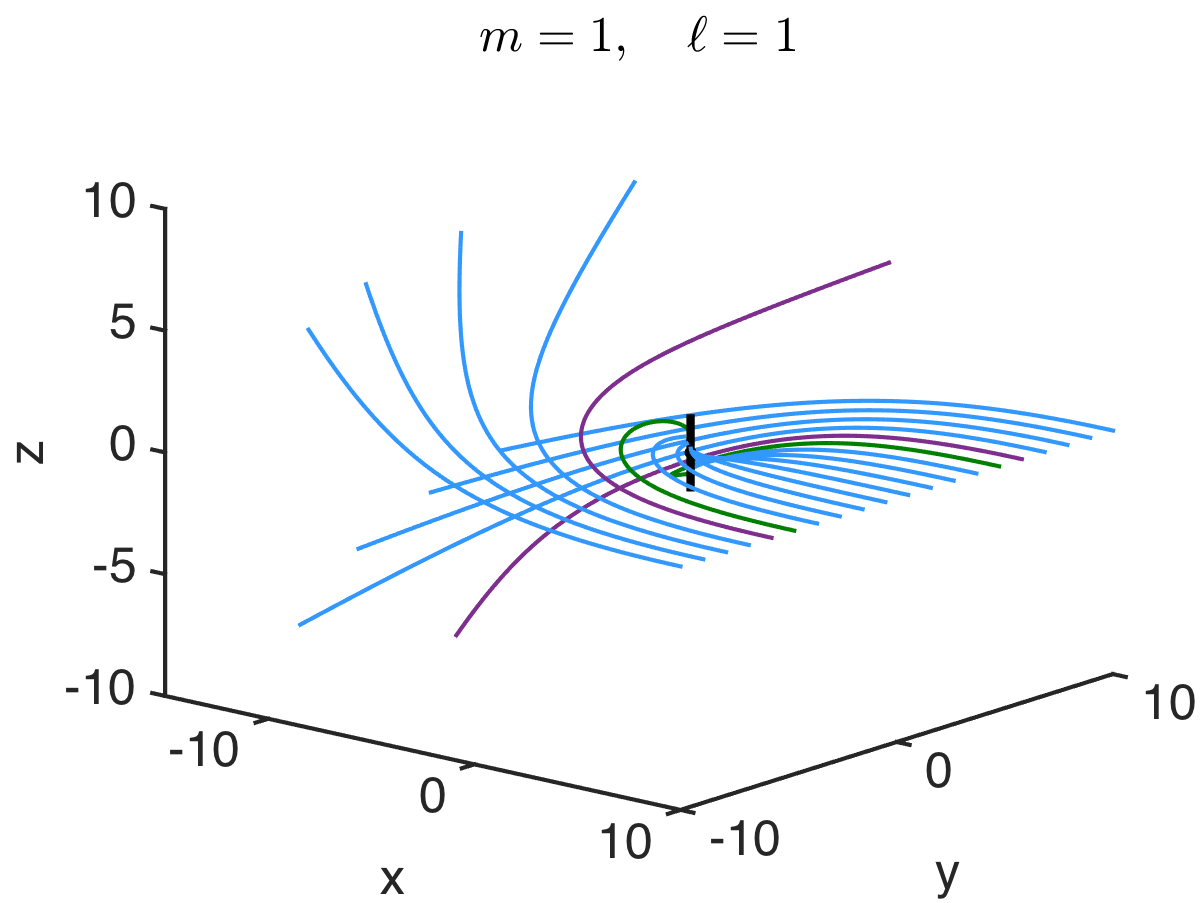}    
    \includegraphics[width=6cm]{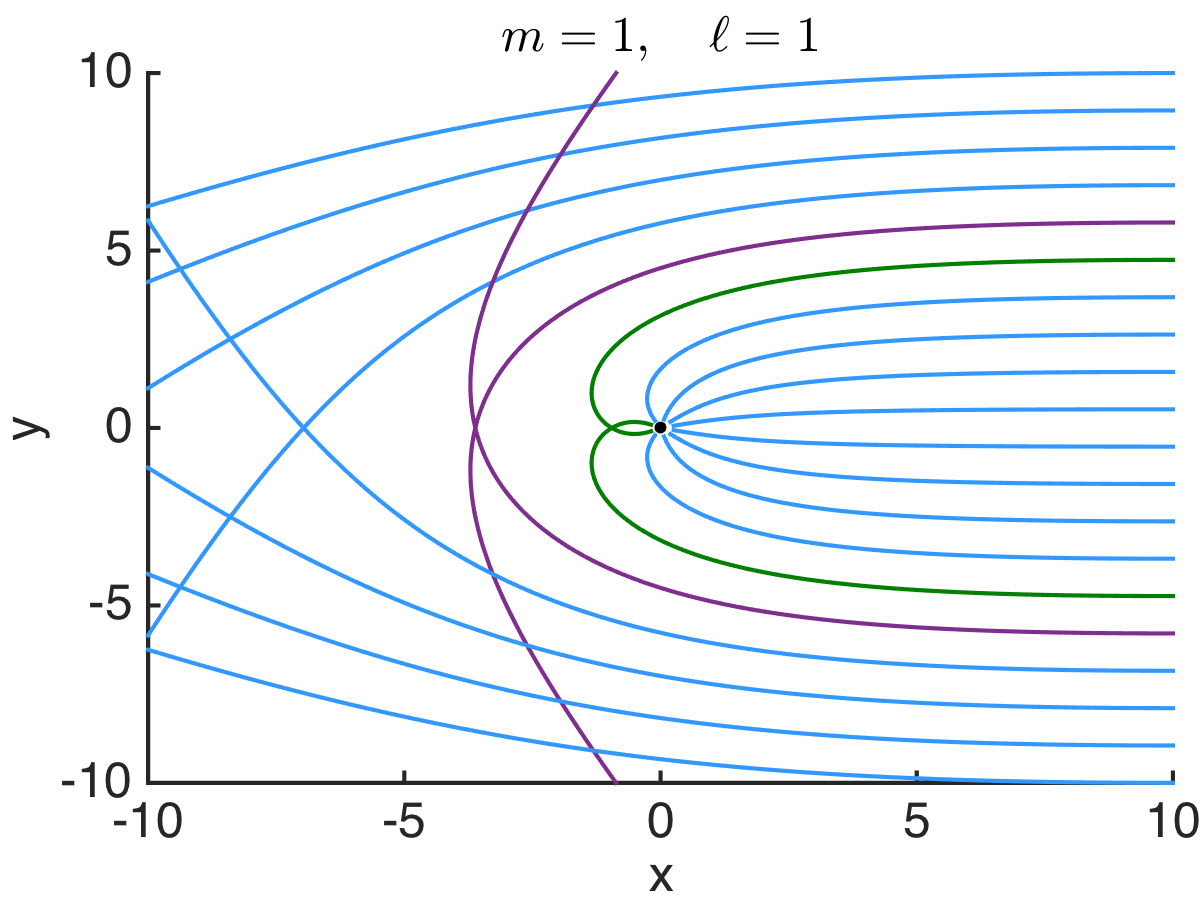}
    \includegraphics[width=6cm]{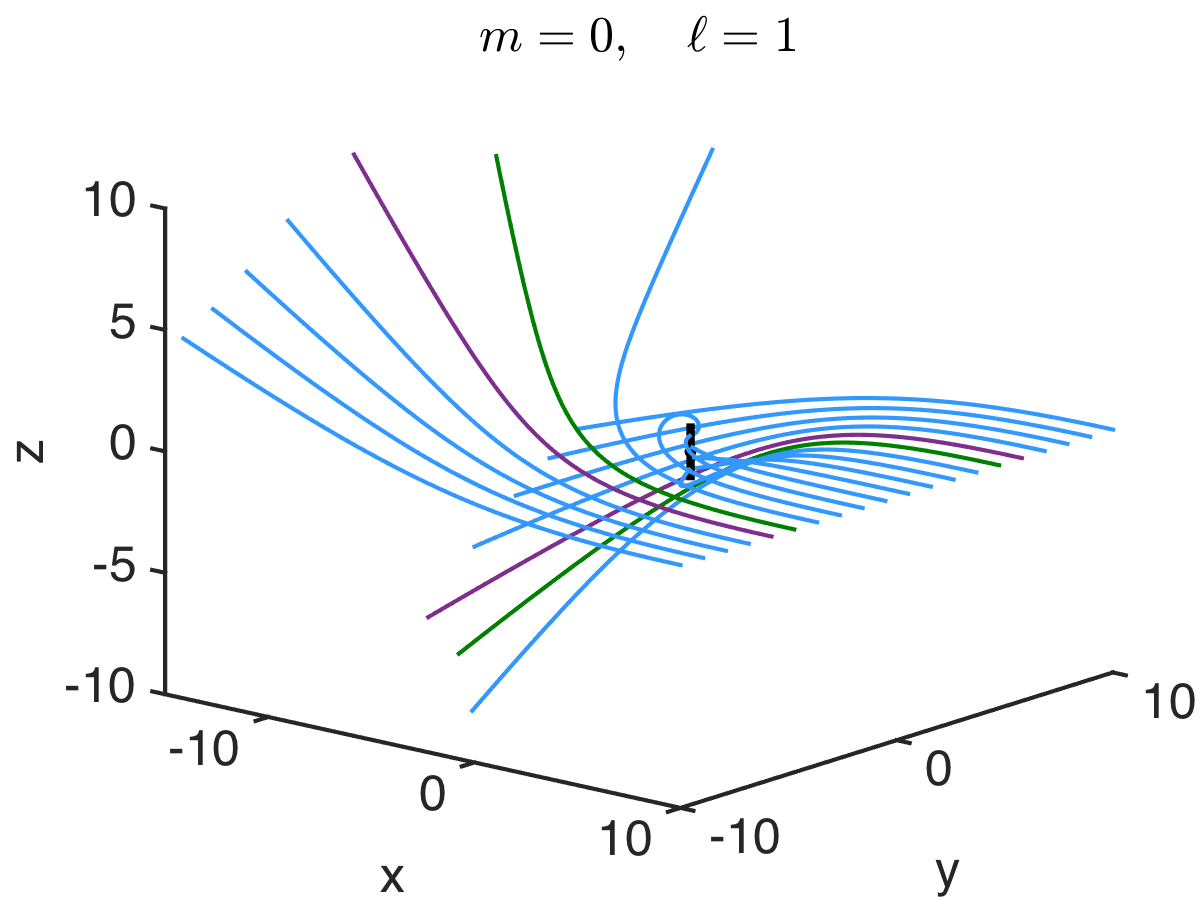}    
    \includegraphics[width=6cm]{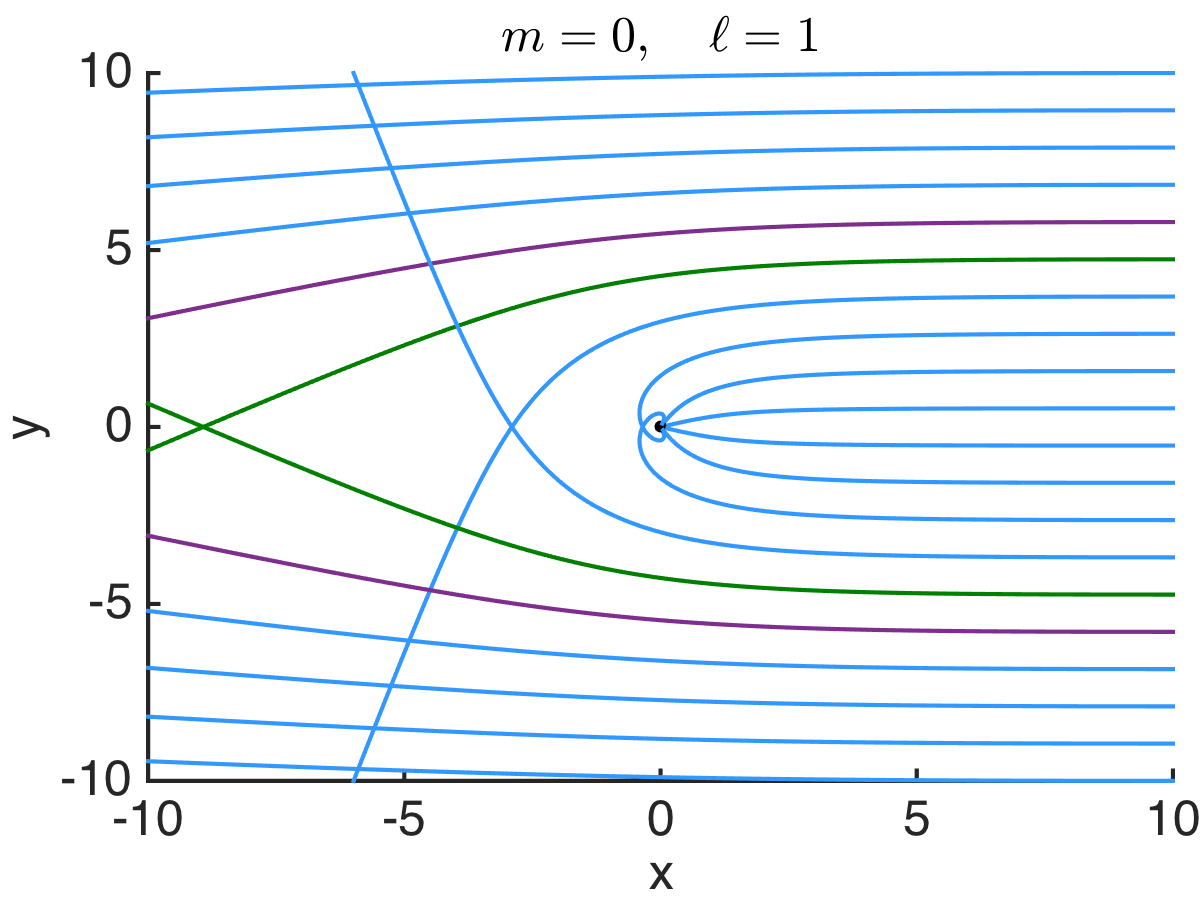}
    \caption{Initially-parallel light rays in the equatorial plane 
	for various NUT spacetimes and their projection on the $xy$-plane. For each example, two pairs of light rays corresponding to the same initial conditions except for the $y$-component (we chose $y_0>0$ for one of the rays and $y'_0=-y_0$ for the other) are colored in purple and green. The rod on the axis is the ergosphere.}  \label{fig:NUT_xy}
\end{figure}

\begin{figure}[H]
   \centering
    \includegraphics[width=6cm]{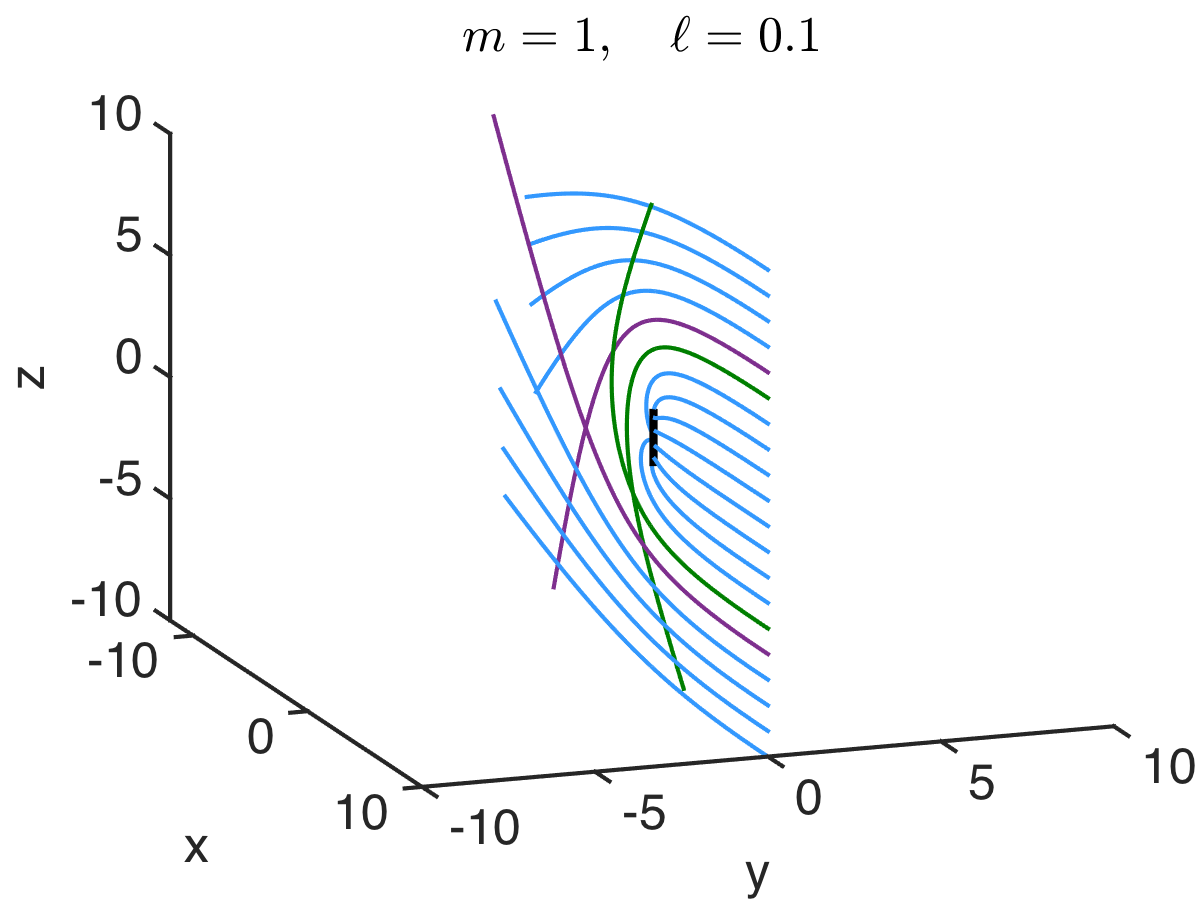}    
    \includegraphics[width=6cm]{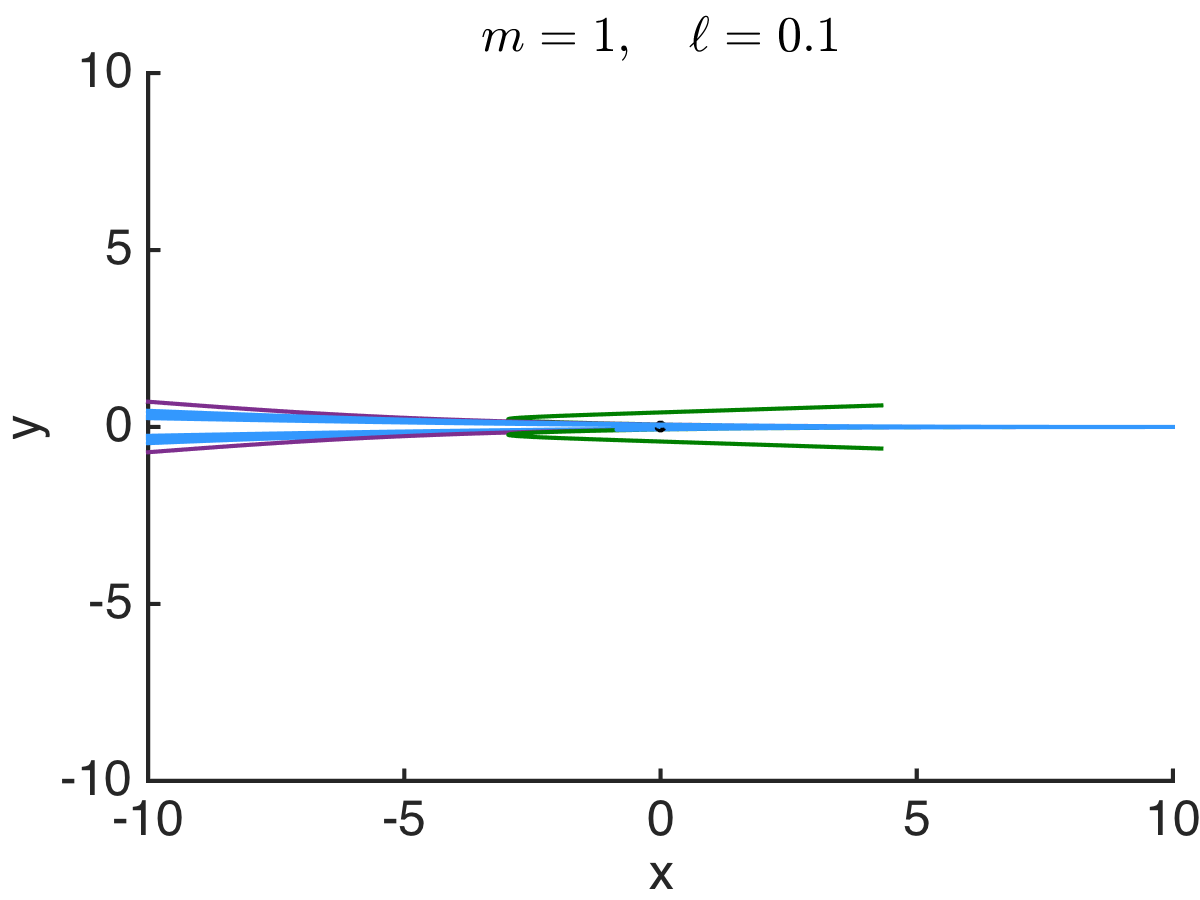}
    \includegraphics[width=6cm]{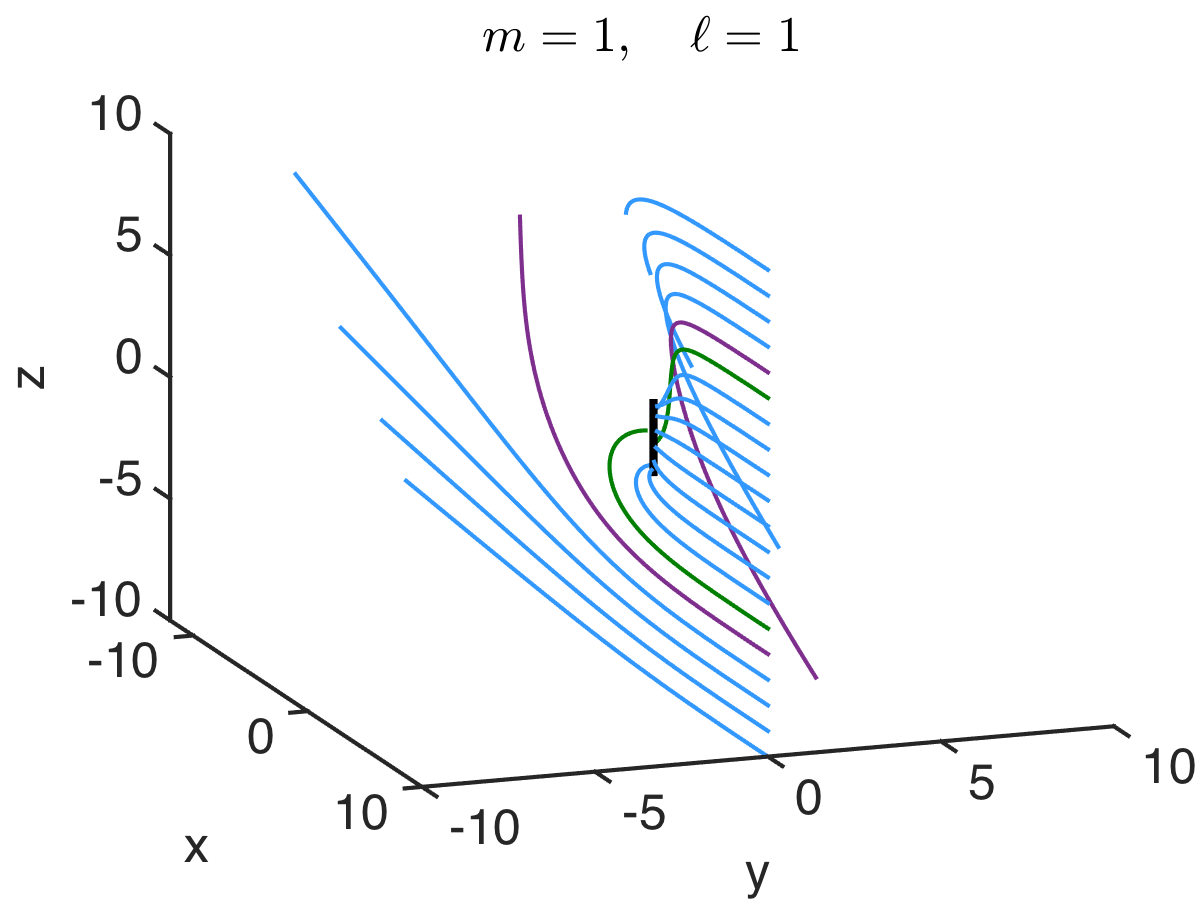}    
    \includegraphics[width=6cm]{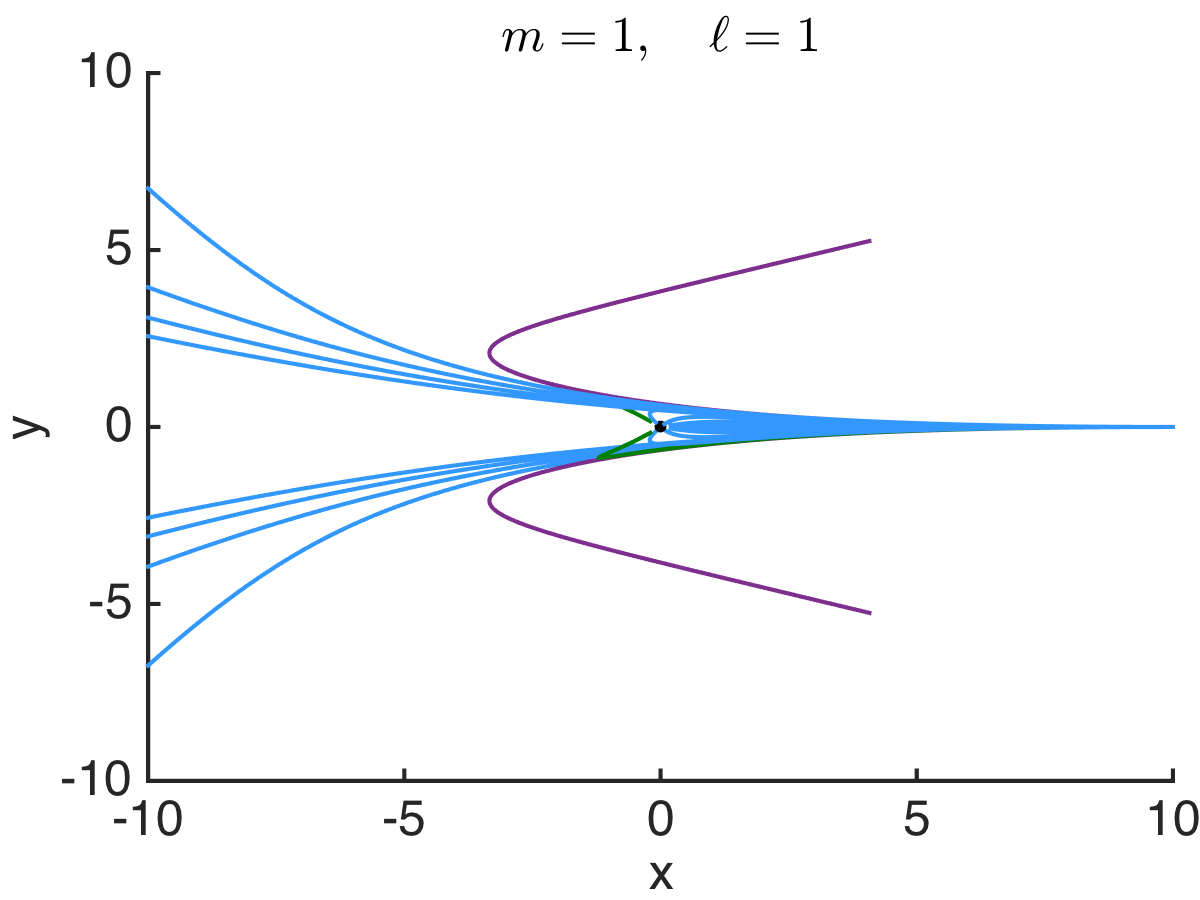}
    \includegraphics[width=6cm]{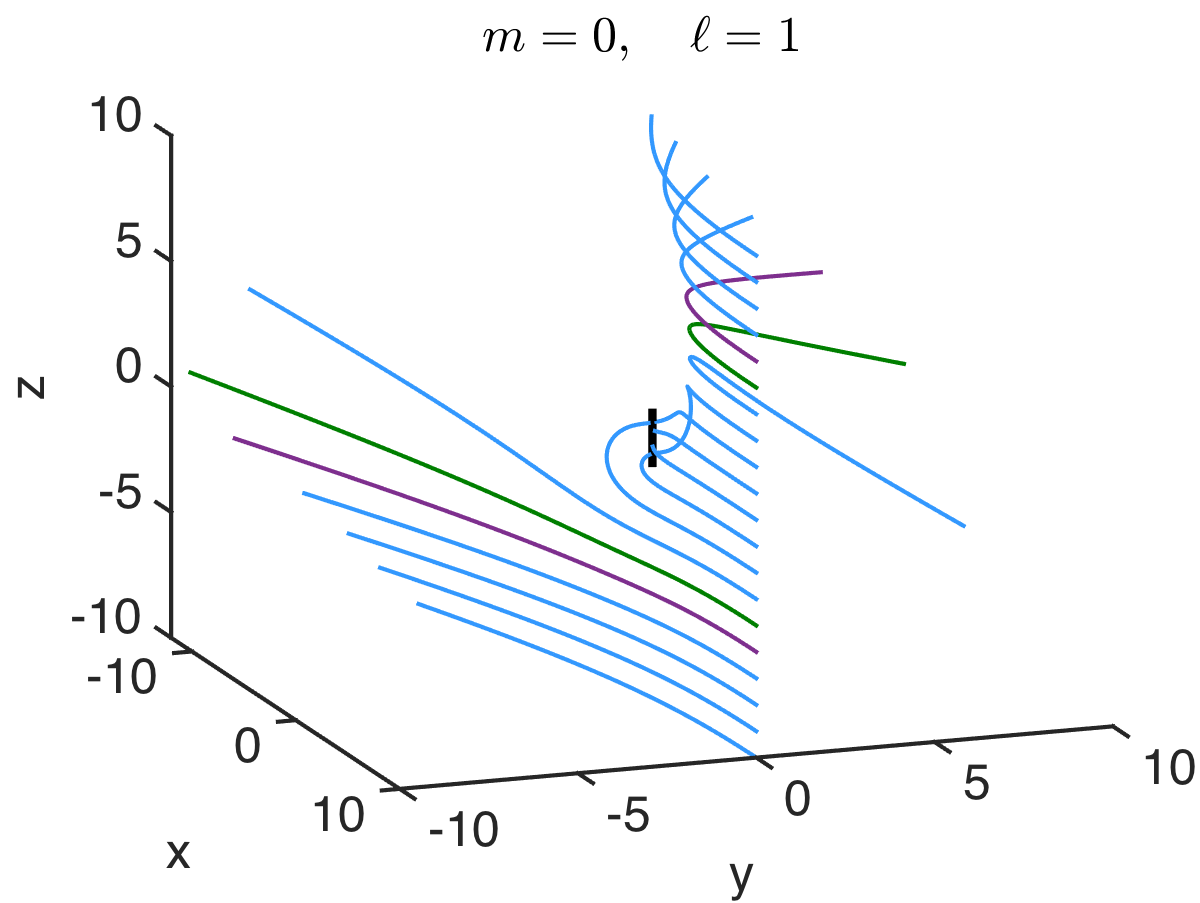}    
    \includegraphics[width=6cm]{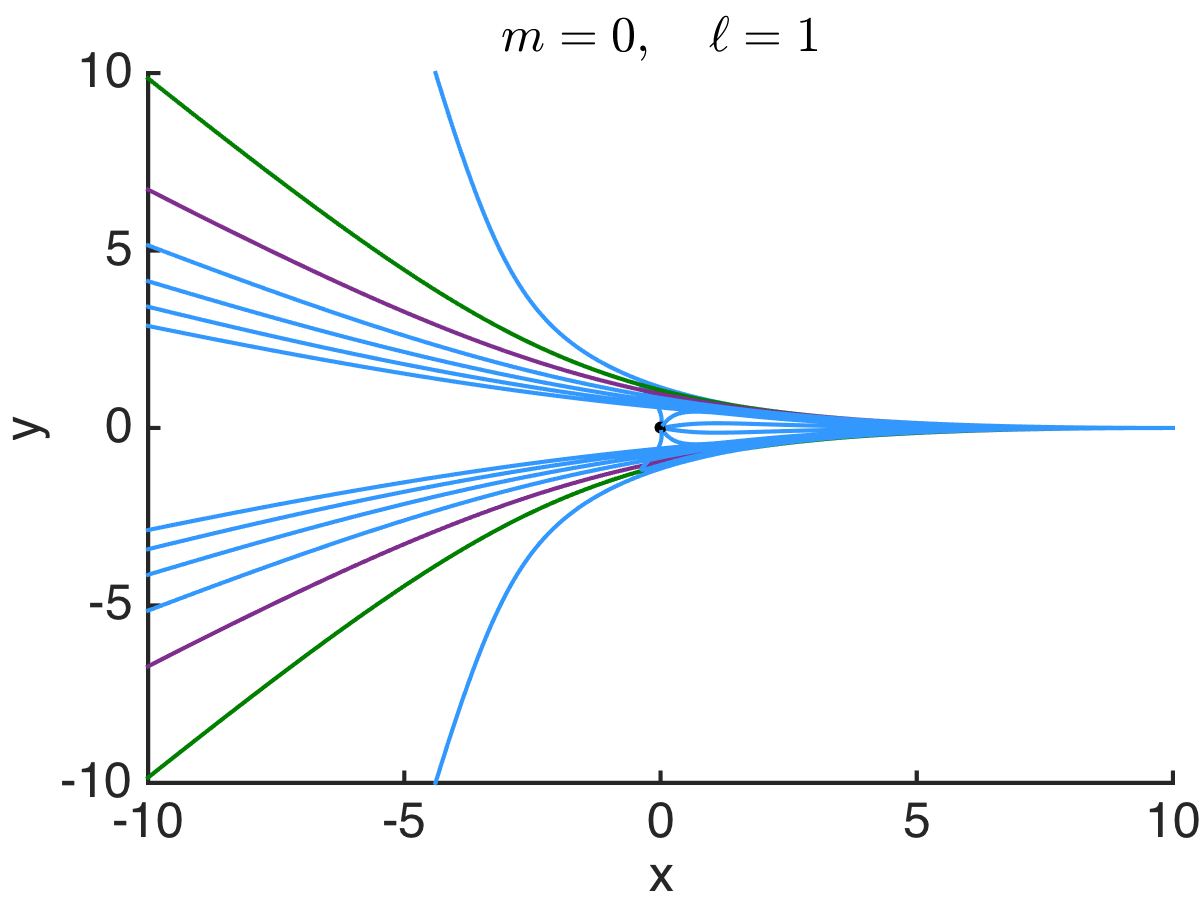}
   \caption{Initially-parallel light rays on the $xz$-plane for various NUT spacetimes and their projection on the $xy$-plane. For each example, two pairs of light rays corresponding to the same initial conditions except for the $z$-component (we chose $z_0>0$ for one of the rays and $z'_0=-z_0$ for the other) are colored in purple and green. The rod on the axis is the ergosphere.} \label{fig:NUT_xz}
\end{figure}

\begin{itemize}
\item[(i)] Photons in the equatorial plane directed towards the negative $x$-direction for the same $x_0$ and for several equispaced values for $y_0$. The initial conditions for these light rays have the form
  \begin{equation*}
    \left\{
      \begin{array}{llll}
        t_0 = 0, & \rho_0=\sqrt{x_0^2+y_0^2}, & z_0 = 0, & \phi_0=\arctan{(y_0/x_0)},\\
        p^t_0 , & p^\rho_0 = - \frac{x_0}{\rho_0} , & p^z_0 =0, &  p^\phi_0 = \frac{y_0}{\rho_0^2}.
      \end{array}
    \right.
  \end{equation*}
\item[(ii)] Photons on the $xz$-plane directed in the negative $x$-direction for the same $x_0$ and several equispaced values for $z_0$. The initial conditions for these light rays have the form
  \begin{equation*}
    \left\{
      \begin{array}{llll}
        t_0 = 0, & \rho_0=x_0, & z_0 = z_0, & \phi_0=0,\\
        p^t_0 , & p^\rho_0 = -1, & p^z_0 =0, &  p^\phi_0=0.
      \end{array}
    \right.
  \end{equation*}
\end{itemize}
The component $p^t$ is obtained as the positive solution of $g_{\mu\nu}p^\mu p^\nu=0$, since we are considering photons moving forward in time. We will refer to these conditions as initial conditions of type (i) and (ii), respectively.
Figs. \ref{fig:Kerr_xy} and \ref{fig:NUT_xy} show light rays corresponding to initial conditions of type (i) with equispaced values $y_0\in[-10,10]$, while Fig. \ref{fig:NUT_xz} shows light rays with conditions of type (ii) with equispaced values $z_0\in [-10,10]$. In both cases, we choose a fixed value $x_0=10$ and shoot a beam with 20 light rays.

In order to provide a physical explanation to observed effects in toron spacetimes which are not manifested  in the NUT solution, we briefly discuss first the behavior of light rays with initial conditions of type (i) in Kerr spacetimes. 
It is well known that a non-zero parameter $\varphi$ in (\ref{kerr5a}) gives rise to a frame-dragging effect in the direction of the black hole rotation, which increases with the magnitude of $\varphi$ (let us recall that the angular momentum is given by $J=m^2\sin\varphi$). In the Schwarzschild case, pairs of light rays with the same initial conditions except for the $y$-component, in which $y_0>0$ for one light ray and $y'_0=-y_0$ for the other (which induce initial conditions with $\phi'_0=-\phi_0$ and $p^{\phi'}_0 =-p^\phi_0$ in Weyl coordinates), are symmetric with respect to the $x$-axis. 
However, due to the frame-dragging effect produced by a non-zero $\varphi$, this symmetry is lost. 
Indeed, Fig. \ref{fig:Kerr_xy} shows that photons in prograde motion are accelerated, while those in retrograde motion are slowed down.

We continue with the same analysis in NUT spacetimes. As in
the previous example, we are interested in determining how a non-zero NUT parameter affects the symmetry of these light rays when compared to those in a Schwarzschild spacetime. Moreover, it is possible to choose a vanishing mass, and thus one can analyze the purely gravimagnetic effects. In the following examples, we always choose positive $\ell$.
From formula \eqref{eq:parallel_xy}, we expect that light rays with conditions of type (i)  go in downwards (respectively upwards) direction if $p^\phi_0$ is positive (respectively negative). This is indeed what is observed in the numerical integration of the corresponding geodesics shown Fig. \ref{fig:NUT_xy}. However, it is interesting to note that their $xy$-projection is still symmetric with respect to the $x$-axis.

This frame-dragging effect in opposite directions is also observed on light rays with initial conditions of type (ii).
The consequence of a non-vanishing metric function $A$ on light rays 
with initial conditions of this type is that $p^{\phi}$, which 
indicates the direction in which the light rays are deflected, will depend on the sign of $z_0$, as implied by the formula \eqref{eq:pphi_eq}. 
If $m\neq 0$ and $\ell=0$, which is Schwarzschild, then the function $A$ vanishes identically, implying that the rays will stay on the same plane. If $\ell > 0$, then the value of the derivative $\partial_\rho A |_{(\rho_0,z_0)}$ is positive if $z_0>0$ and negative if $z_0<0$. This explains the observed behavior in Fig. \ref{fig:NUT_xz}, in which the light rays deviate in clockwise (resp. counterclockwise) direction if $z_0>0$ (resp. $z_0<0$). The dragging effect in opposite directions increases in dependence of the gravimagnetic mass $\ell$ and it is particularly noticeable for the case $m=0$ with $\ell\neq 0$.
Moreover, the $xy$ projection of two light rays with initial 
conditions of this type, one with $z_0>0$ and the other with 
$z'_0=-z_0$, are symmetric with respect to the $x$-axis, 
meaning that such light rays are deflected with the same magnitude, but in opposite directions. This can be deduced from formula \eqref{eq:pphi_eq} and the parity of the NUT metric functions in $z$. 

\subsection{Light rings} 
\begin{figure}[H]
    \centering
    \includegraphics[width=5cm]{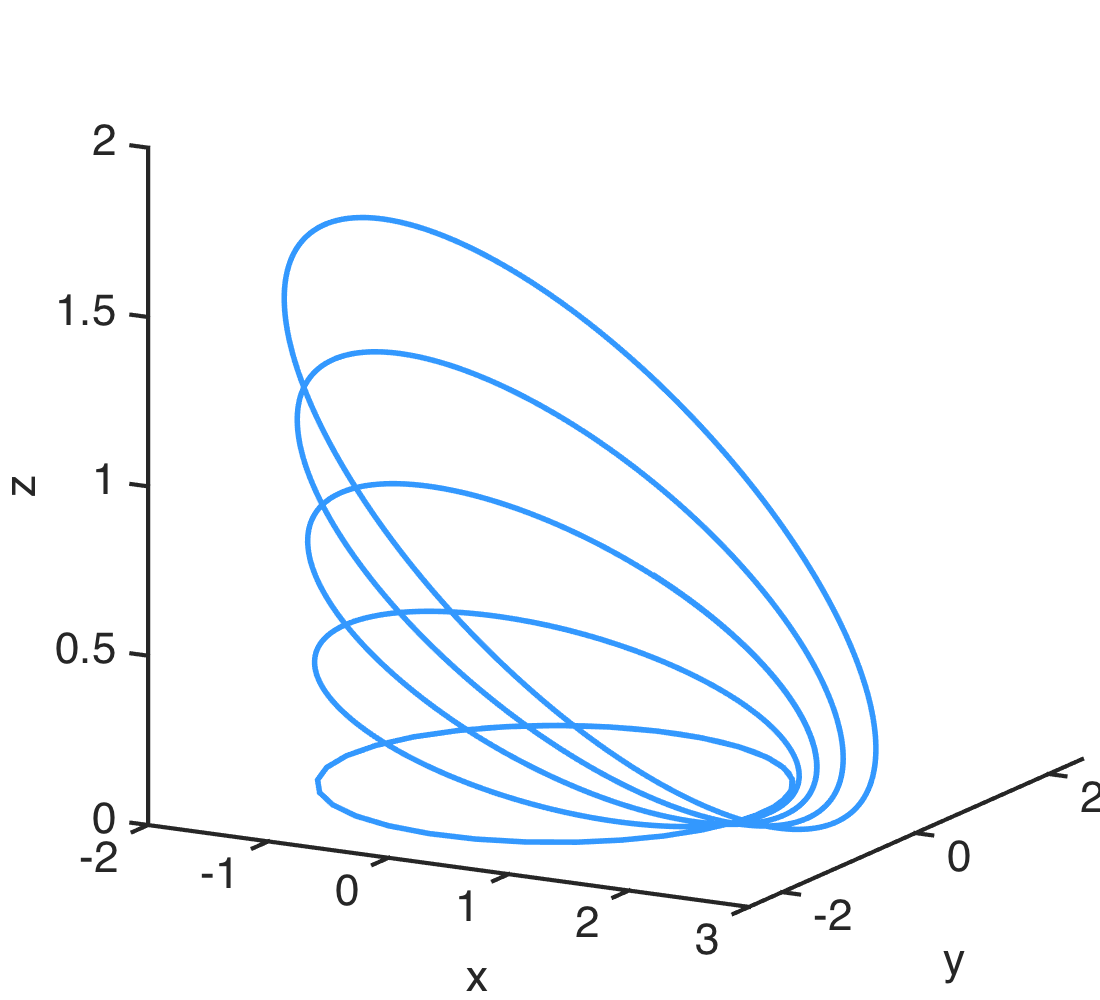} \includegraphics[width=5cm]{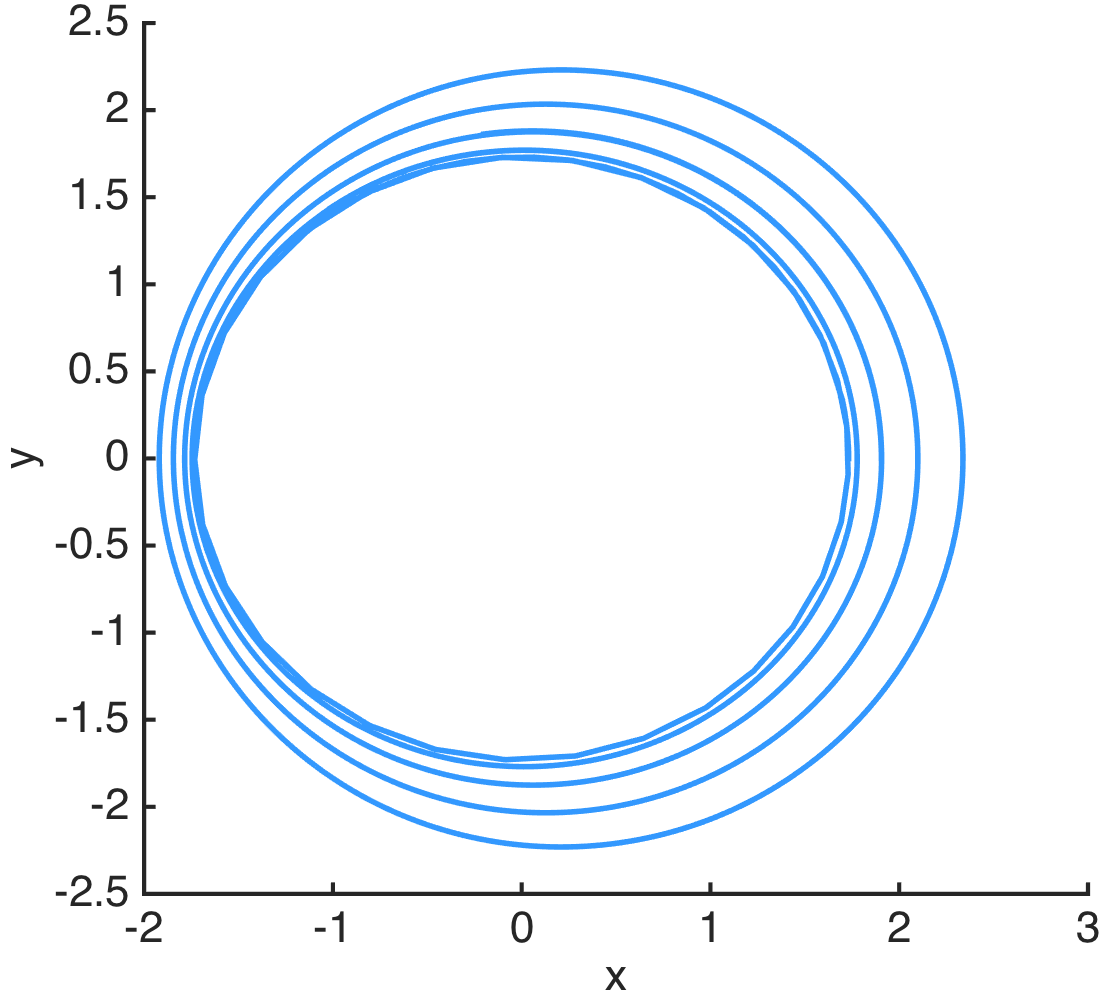}
    \caption{Light rings corresponding to spacetimes with a fixed $m$ 
	and various $\ell$ and their 
	projection onto the $xy$-plane. Both the radius and the elevation of each ring increase with $\ell$.} \label{fig:NUT_rings}
\end{figure}

\begin{figure}[H]
    \centering
    \includegraphics[width=6cm]{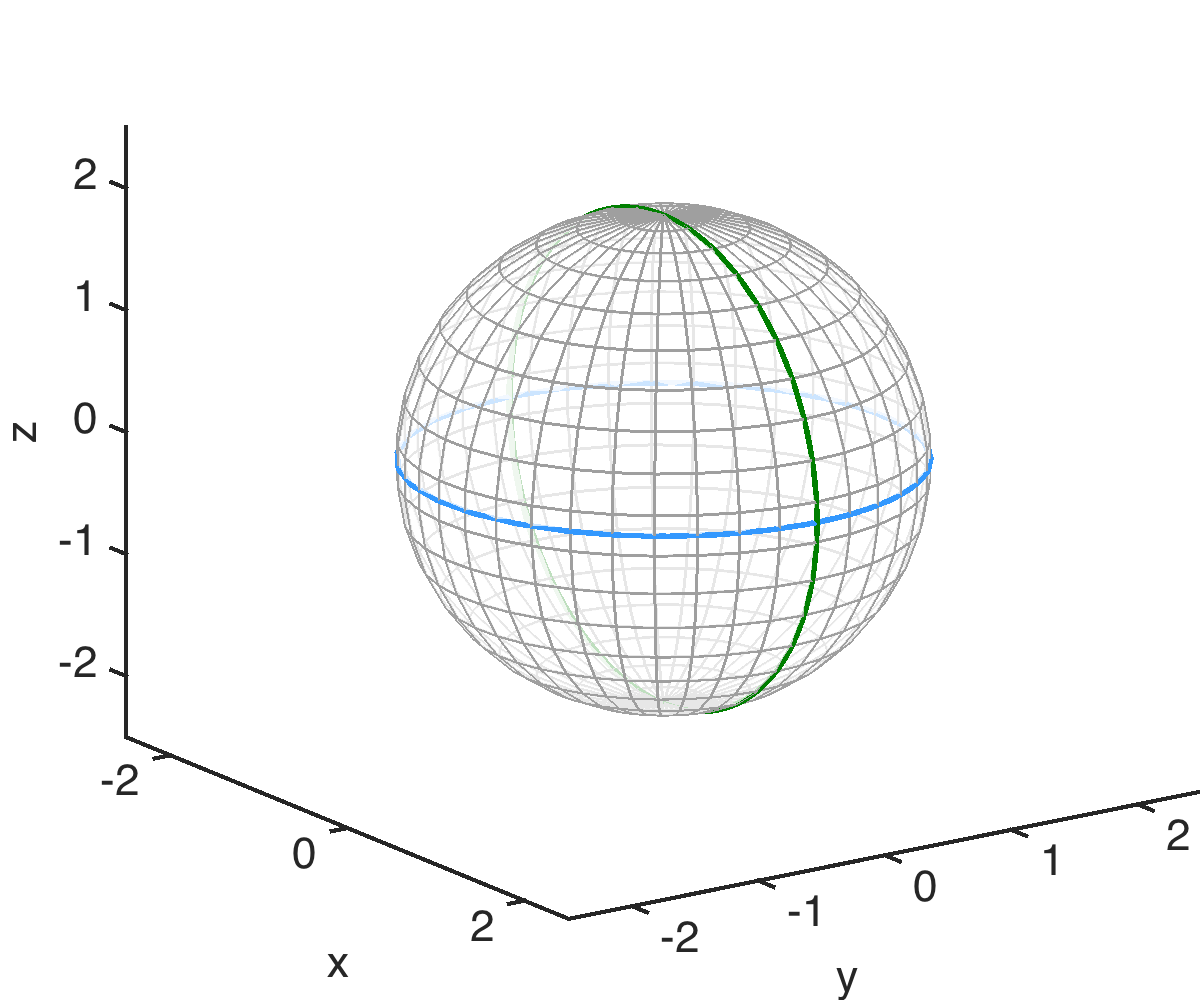} \includegraphics[width=6cm]{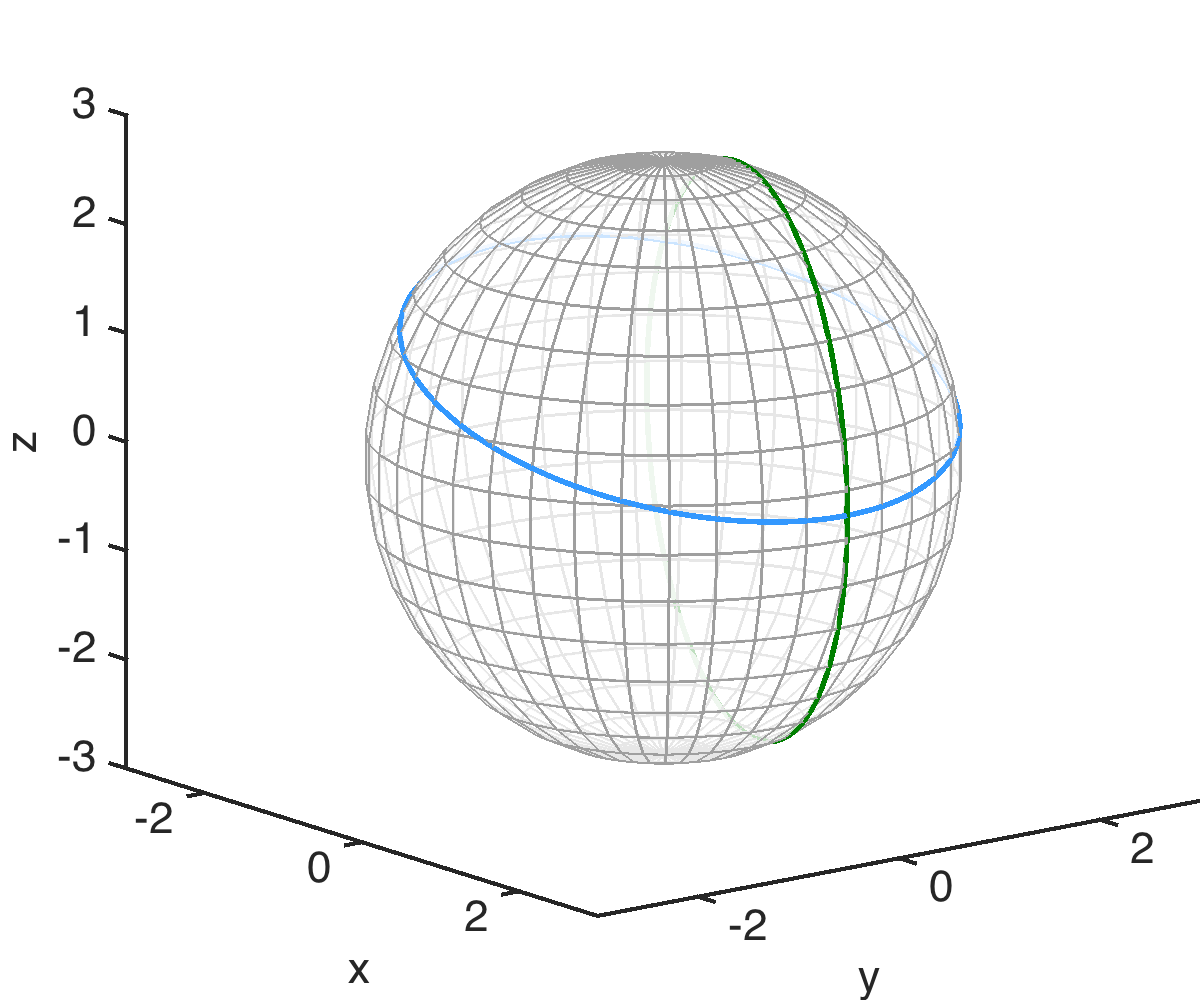}
    \caption{Light rings on the photon spheres in a Schwarzschild and a NUT spacetime.} \label{fig:NUT_ring_shell}
\end{figure}

From the previous examples, we observe that 
one of the effects of the NUT parameter $\ell$ is the 
lifting of the light rays from the equatorial plane.  We expect this to have a consequence on the observed light rings with respect to the ones in Schwarzschild 
spacetime. Let us recall that in that case, the trajectory of the light rays on the photon 
sphere move along the sphere's great circles. Thus the trajectory can
be described as the intersection of the photon sphere and a plane passing 
through the origin. The first effect of a non-vanishing $\ell$ on light rings is that the size of the photon sphere is  bigger, as shown by the formula \eqref{eq:rad_photon_sphere} and Fig. \eqref{fig:NUT_ring_shell}. The second effect is that as $\ell$ is increased, the light rings originally on the equatorial plane is  lifted out of the plane as seen in Fig. \ref{fig:NUT_rings}; in fact, as observed in Fig. \eqref{fig:NUT_ring_shell}, these rings are small circles of the photon sphere.

The chosen initial conditions for the examples in Fig. 
\eqref{fig:NUT_rings} are $t_0=0$, $\rho_0=\sqrt{r_p^2-2mr_p-\ell^2}$, $z_0=1$, $\phi=0$ with $p^t_0=1$, $p^\rho_0=0$, $p^z_0=0$, $p^\phi_0=-1$ where $r_p$ is the radius of the photon sphere in Boyer-Lindquist coordinates. Namely, the photons are emitted from  the equatorial plane with the spatial component of their momentum being tangent to the photon sphere and pointing in clockwise direction.
The light rays represented in Fig. \eqref{fig:NUT_rings} correspond to different spacetimes with a fixed $m=1$ and $\ell\in\{0,0.25,0.5,0.75,1\}$, but they are plotted together in order to visualize the influence of the gravimagnetic mass $\ell$. 
It can be observed that the trajectories of these light rays are still rings, but in this case they are small circles of the photon spheres, meaning that they are described by the intersection of a sphere and a plane not passing through the origin as shown in Figs. \ref{fig:NUT_rings} and \ref{fig:NUT_ring_shell}. Additionally, Fig. \ref{fig:NUT_ring_shell} shows the comparison of light rings in a Schwarzschild and a NUT spacetime, given analogous initial conditions. These results agree with the estimation of \cite{HL} for the deviation with respect to the equatorial plane (denoted therein as \textit{conicity}), which is proportional to the NUT parameter $\ell$. 

\subsection{Shadows and apparent image of the background}

It has been shown in \cite{GPL} that the shadows of NUT black holes are circular. Given a fixed $m$, the size of the shadow increases in dependence of $\ell$, as inferred from formula \eqref{eq:rad_photon_sphere}. With the help of ray tracing techniques, one can additionally study the effects of $\ell$ on the apparent image of the background space. For example, we can place an observer whose line of sight is directed towards the black hole, as shown by the diagram in Fig. \ref{fig:camera_diagram}.

\begin{figure} [H]
    \centering
    \includegraphics[width=11cm]{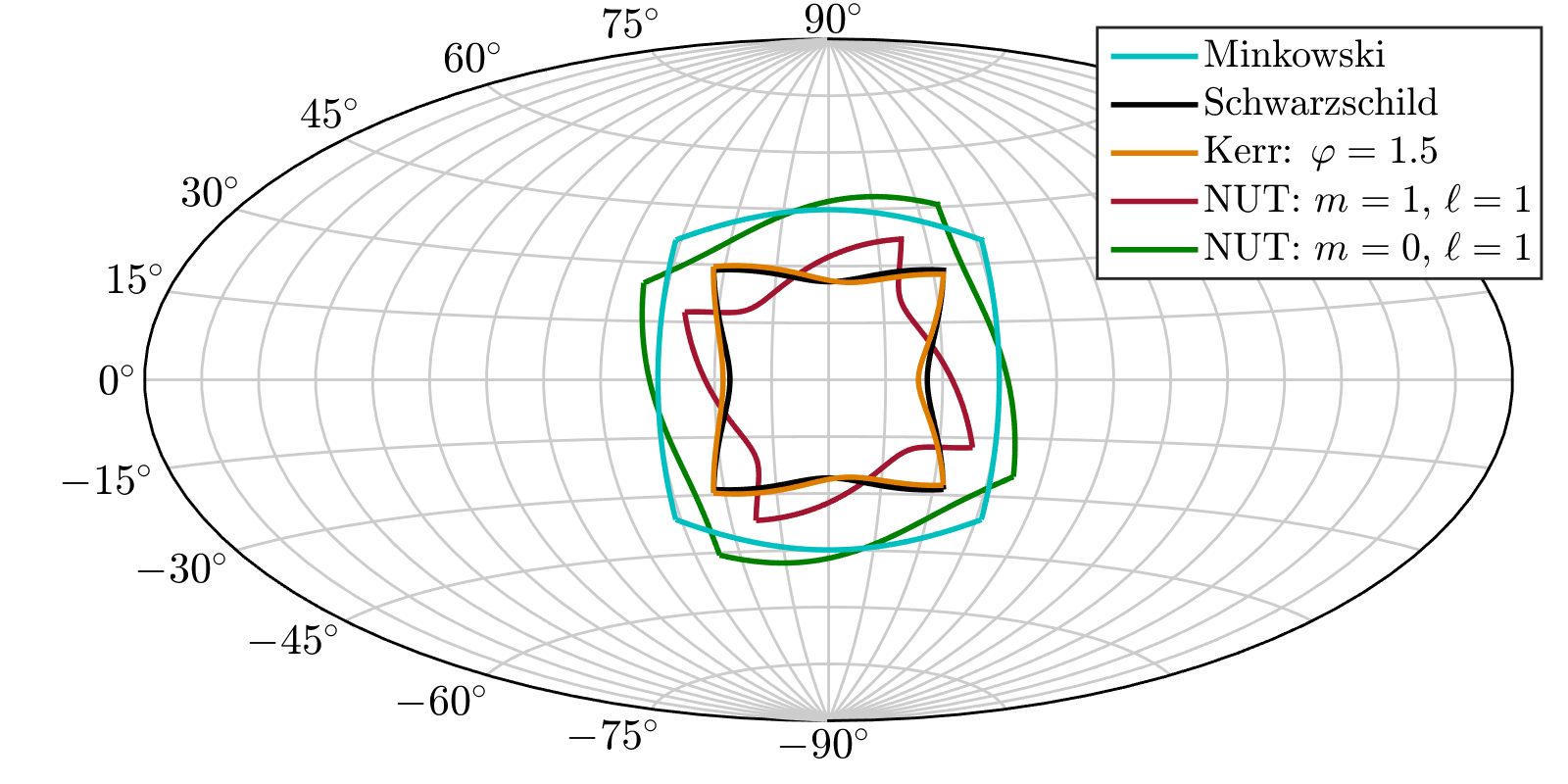}
    \caption{Boundary of the primary image on the celestial sphere in Minkowski, Schwarzschild, Kerr and NUT spacetimes. In all cases, the focal length of the virtual camera is chosen such that the angles of view in Minkowski spacetime are $\delta_H=\delta_V=90^\circ$. } \label{fig:primary_image}
\end{figure}

Regarding the image of the background space, it is well known that if a Schwarzschild black hole is located along the line of sight of the observer, it takes the role of a magnifying lens as well as introducing other deformations (gravitational lensing). In order to illustrate the role of the NUT parameter $\ell$ on the apparent image seen by a distant observer, we begin by drawing the outer border of the primary image for various spacetimes, see Fig. \ref{fig:primary_image}. The observer is placed on the equatorial plane, i.e., its inclination with respect to the symmetry axis is set to $\psi=90^\circ$.
The features of the virtual camera are chosen such that the angles of view in Minkowski spacetime are $\delta_H=\delta_V=90^\circ$. As expected for Schwarzschild's spacetime, we observe that the effective angle of view of the observer is reduced (which means that the image is magnified) due to the bending of light caused by the black hole's mass $m$. The Kerr parameter $\varphi$ induces a frame-dragging effect, but this is only noticeable in the vicinity of the black hole. Thus, for this choice of angular apertures, the difference between the outer boundary of the primary image in an almost-extreme Kerr spacetime with respect to the one in Schwarzschild's is minimal. However, the effect of the gravimagnetic parameter $\ell$ on the primary image's boundary is evident, as observed in the two examples for NUT spacetimes. From the drawing for the case $m=0$ we observe that the main effect of $\ell$ on the apparent image is a twist about the line of sight of the observer. Notice that a magnifying effect is not observed in this case. 

Except for Minkowski, all the apparent images in all these spacetimes will display multiple copies of the background space due to the existence of photon spheres.  
Let us recall that if a photon approaches the gravitating object sufficiently close without crossing the smallest photon sphere in inwards direction, then it can orbit the object several times until it reaches the observer. Thus, photons from the same light source can reach the observer several times from various directions.

We continue the discussion by comparing the primary and secondary images in Schwarzschild and 
NUT spacetimes. To ease the analysis, we add a simple artificial 
coloring; the upper half of the celestial sphere is colored in light 
blue and the lower part in grey. In Schwarzschild and Kerr spacetimes the existence of 
Einstein rings is well known, and they correspond to the light rays 
reaching the observer coming from the same source. 
More precisely, due to light bending about a Schwarzschild black hole, an observer whose line of sight is aligned with both the black hole and a light source (such as a bright star or a nebula) will observe a ring-shaped structure, known as the Einstein ring. Due to the spacetime's symmetry, photons from this light source will reach the observer from various continuous directions, see Fig. \ref{caustic} for a visual reference. The Einstein ring is also present in Kerr spacetimes, but the position of the observer must be shifted (with respect to the alignment with the light source and the gravitating object) in order to account for the frame-dragging effect. In contrast, this structure seems not to be present in NUT spacetimes, at least for an observer placed on the equatorial plane.

\begin{figure}[H]
    \centering
    \begin{tabular}{cc}
    \small Schwarzschild  & \small NUT \\ 
    \includegraphics[width=6cm]{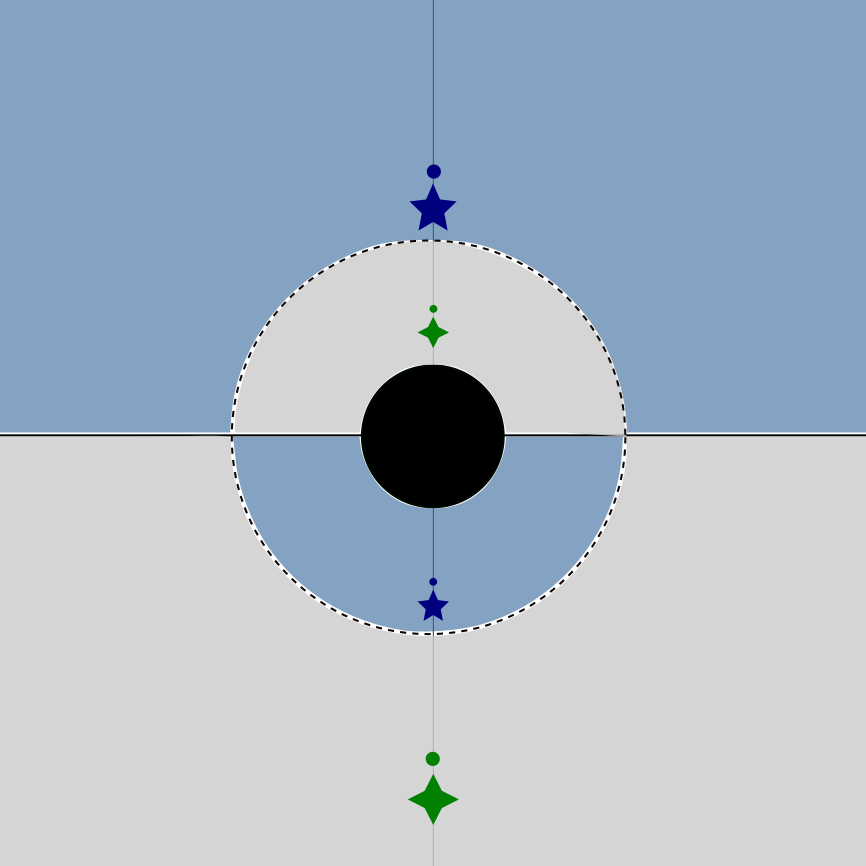} &
    \includegraphics[width=6cm]{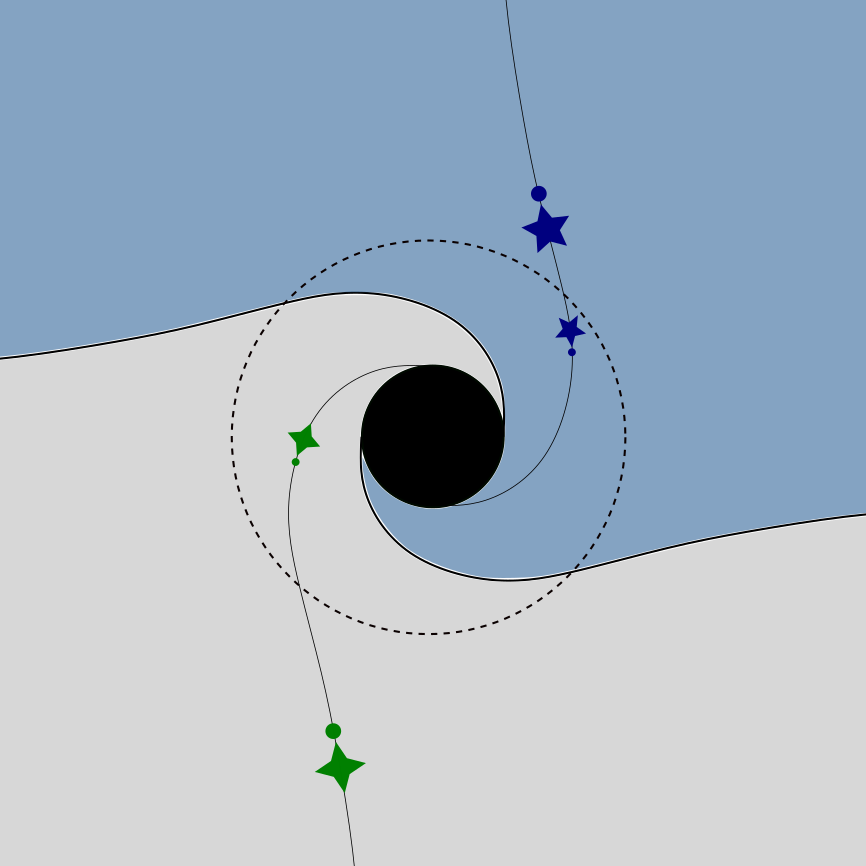} 
    \end{tabular}
    \caption{Primary and secondary images in Schwarzschild and NUT spacetimes (with $m=0$ and $\ell=1$). The dashed lines correspond to the \textit{border} between the primary and secondary images, while the solid ones show the parts of the celestial sphere with constant $z=0$, $\phi=0$ and $\phi=\pi$.} \label{fig:shadow_diagrams}
\end{figure}

The diagram shown in Fig. \ref{fig:shadow_diagrams} corresponds to the Schwarzschild case, but the same phenomenon is observed in Kerr spacetimes as well. The only difference is that the shadow is no longer circular, and it shifts to the right of the screen. Subsequent rings would be observed if the image is magnified near the shadow. The first Einstein ring acts as a \textit{border} between the primary and secondary image of the background space. However, although there is a border between the primary and secondary images in NUT spacetimes, they do not originate from the same source. Moreover, since the light rays are either trapped or repelled by the NUT black hole, as shown in Figs. \ref{fig:NUT_xy} and \ref{fig:NUT_xz}, the region behind the black hole is a \textit{blind spot}, meaning that the light of the objects right behind the black hole will not reach the observer (at least in the primary and secondary images). As the gravimagnetic mass $\ell$ increases, the size of the hidden region will also increase. In contrast, the celestial sphere is visible in its entirety in Schwarzschild and Kerr spacetimes, although the image appears deformed the closer they are to the Einstein ring. The effect of the gravimagnetic mass on the apparent image of the background space is that this image will be twisted as $\ell$ increases, analogous to the observations in the spacetimes studied in \cite{MHC}, in which the metric component $g_{t\phi}$ is skew-symmetric with respect to the equatorial plane, while the remaining components are even. To highlight the purely gravimagnetic effects, we show the extreme case with $m=0$ and $\ell\neq 0$ in Fig. \ref{fig:shadow_diagrams}.

To finish this qualitative comparison, let us place a blue star on the celestial sphere at $\phi=\pi$ and $z>0$ close to the equatorial plane and a green star at $\phi=\pi$ and $z<0$ away from the plane and observe how they appear in the primary and secondary images. In order to simplify this comparison, let us ignore the deformation of the images and look at their apparent locations only. In Kerr spacetimes, the component $p^\phi$ will always have the same sign as the initial $p^\phi_0$ if this value is sufficiently big (i.e., the light ray does not fall directly into the horizon). 
As a consequence, light rays not falling into the horizon will always travel in prograde or retrograde direction (depending on the initial conditions), which implies that the secondary image will show the other half of the celestial sphere but with a left-right flip. 
On the other hand, as the light rays in NUT spacetimes approach the black hole they will be repelled and sent back to infinity. The consequence of this is that the secondary image of the celestial sphere with $z>0$ will appear on the same side and the image of each object will be flipped upside down, as observed in the right-hand side of Fig. \ref{fig:shadow_diagrams}.

In order to visualize the appearance of the phenomenon discussed 
above as the parameter $\ell$ increases in a more quantitative manner, we assign an artificial coloring to the celestial sphere following the colors of the X-ray map of the sky in Fig. \ref{fig:celestial_sphere} obtained by the eROSITA telescope. The line of sight of the observer is pointing towards the center of the galaxy (center of the X-ray map), which would correspond to $(\phi,\theta)=(\pi,0)$ in the celestial sphere coordinates shown in \ref{fig:aitoff}, but with a black hole in between.

\begin{figure}[H]
\centering
\begin{tabular}{cc}
\small Schwarzschild  & \small Kerr with $\varphi=1.5$ \\
    \includegraphics[width=6cm]{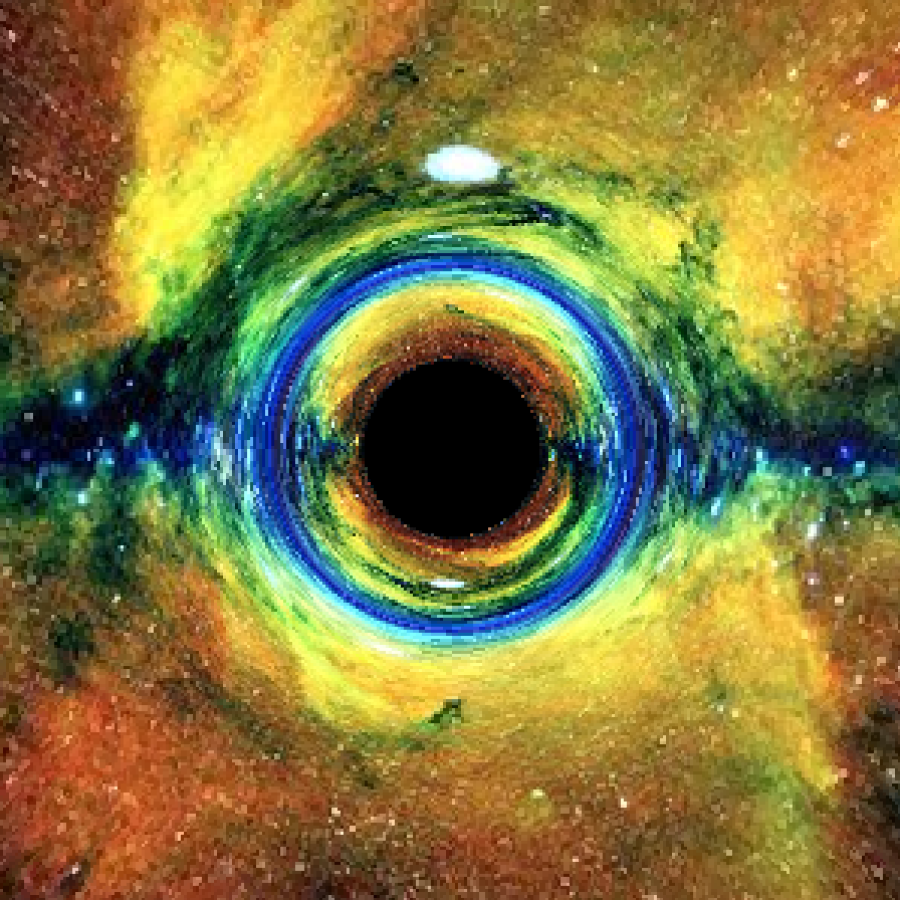} &
    \includegraphics[width=6cm]{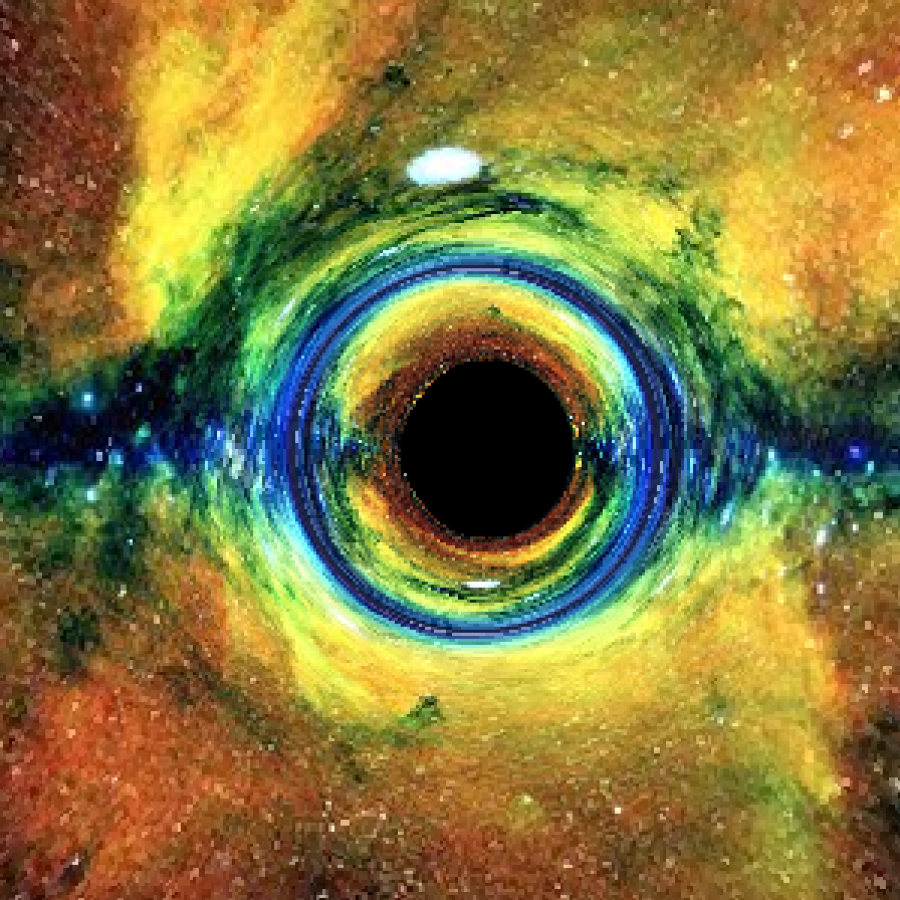} \\
\small NUT with $m=1$, $\ell=0.5$ & \small NUT with $m=0$, $\ell=1$ \\
    \includegraphics[width=6cm]{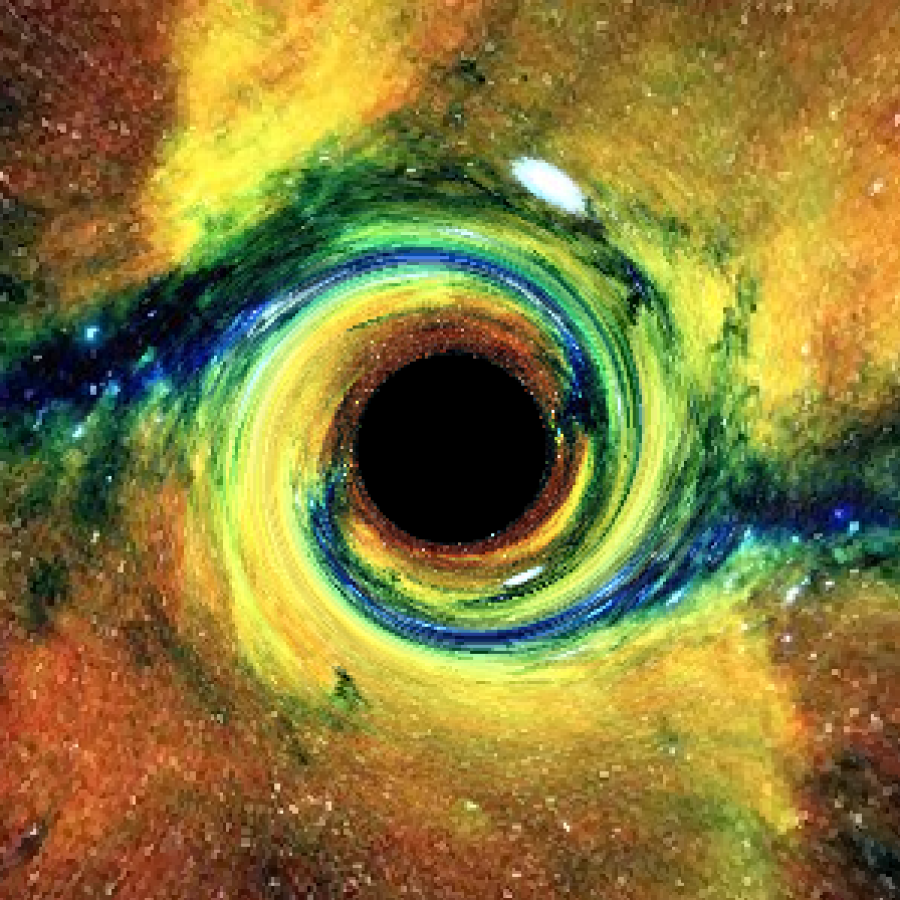}  &
    \includegraphics[width=6cm]{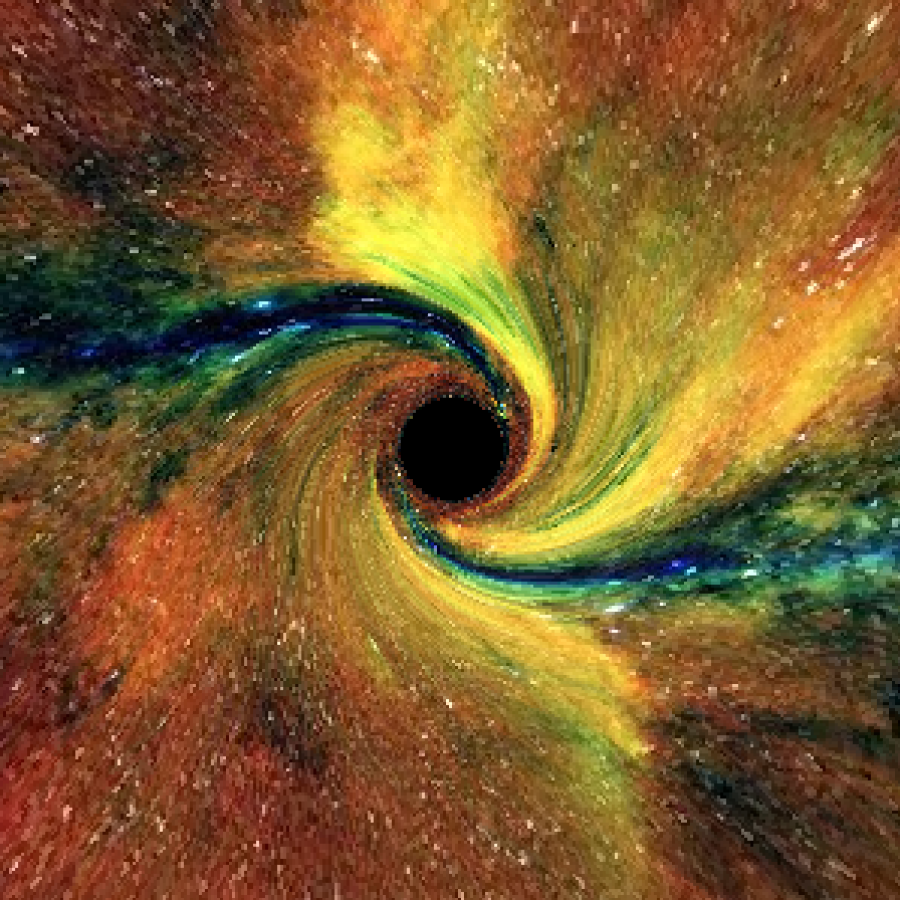} 
\end{tabular}
\caption{Shadows and apparent images of the background space in Schwarzschild, Kerr and NUT spacetimes.} \label{fig:shadow_nut_color}
\end{figure}

\begin{figure}[H]
    \centering
    \includegraphics[width=7cm]{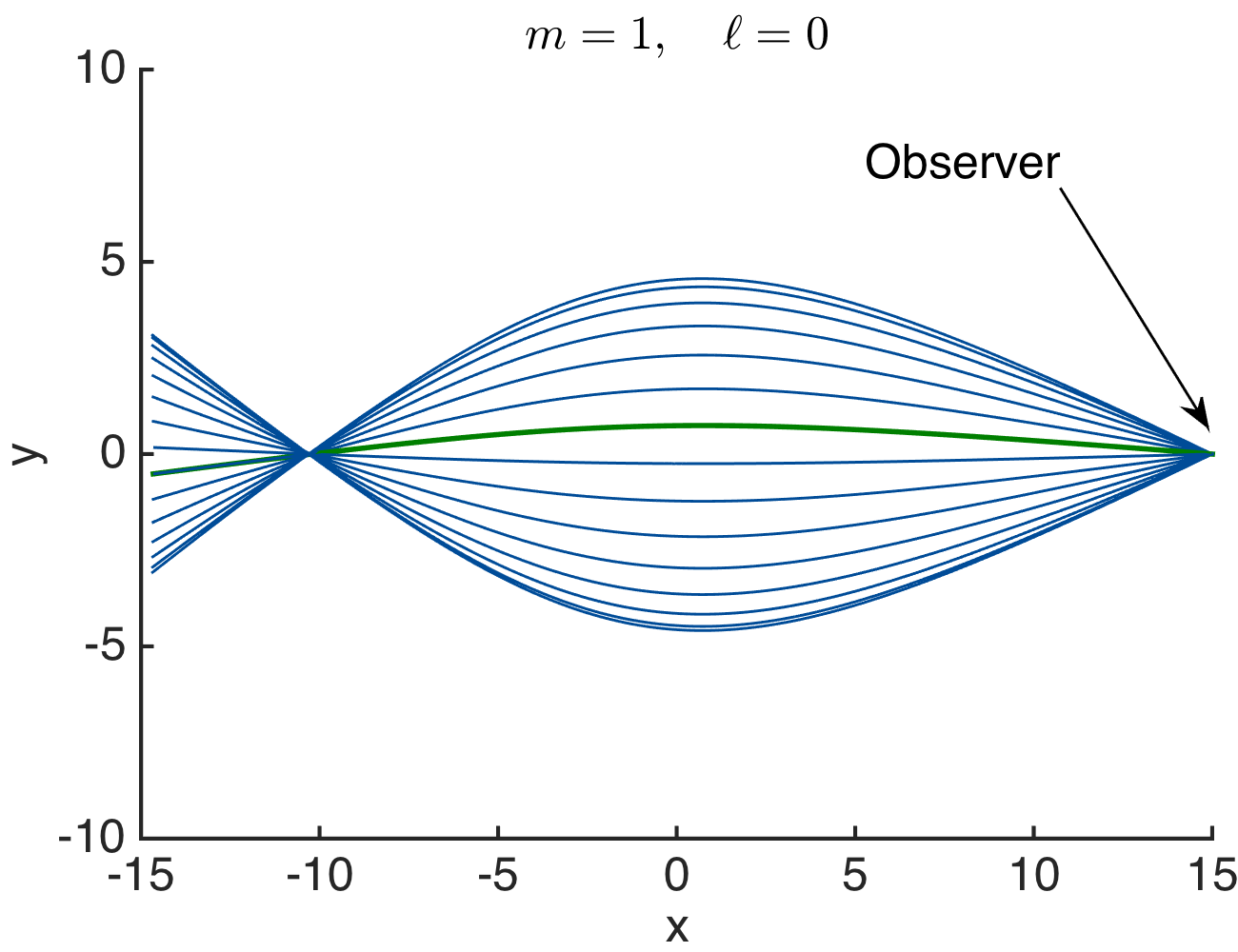} \includegraphics[width=7cm]{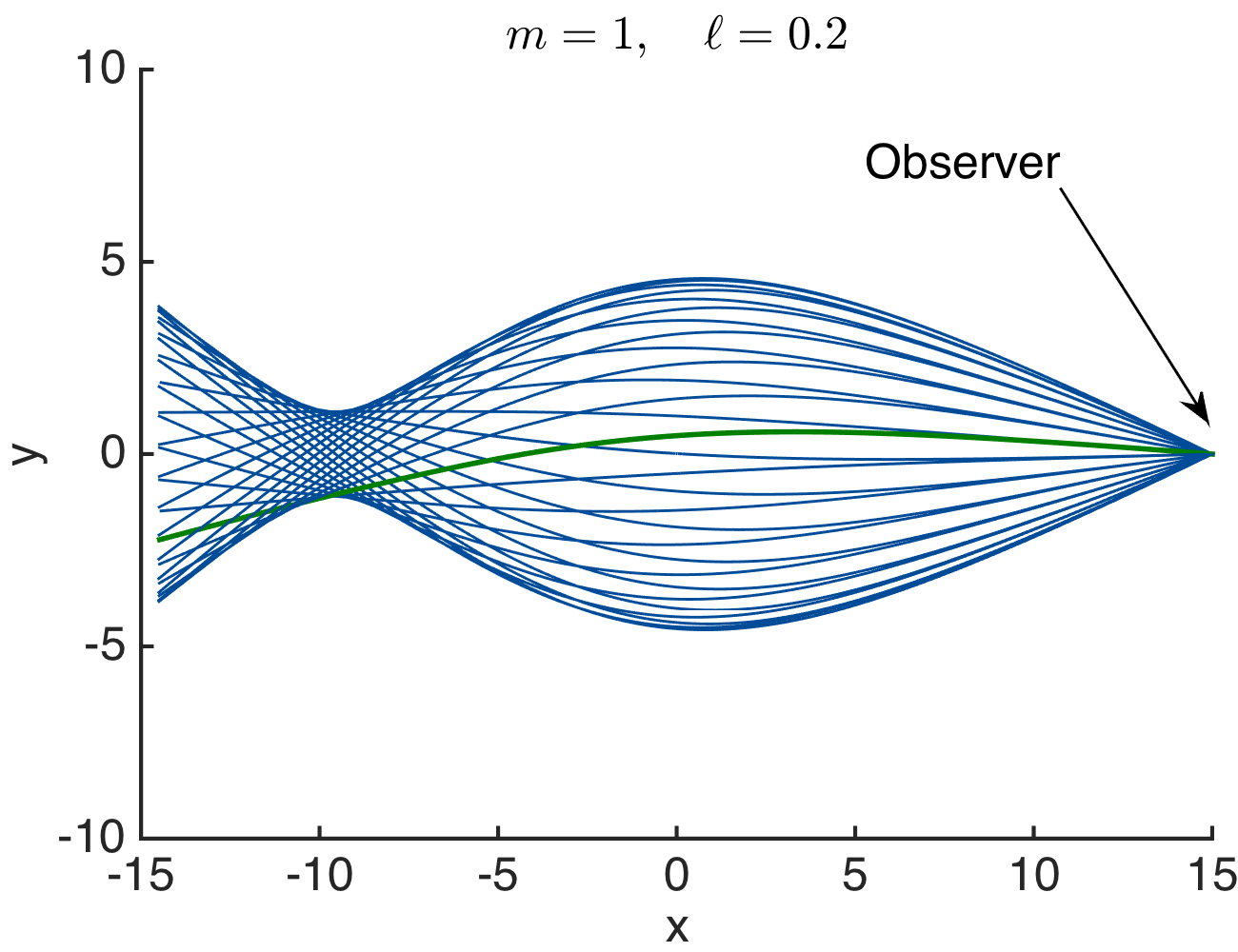}
    \includegraphics[width=7cm]{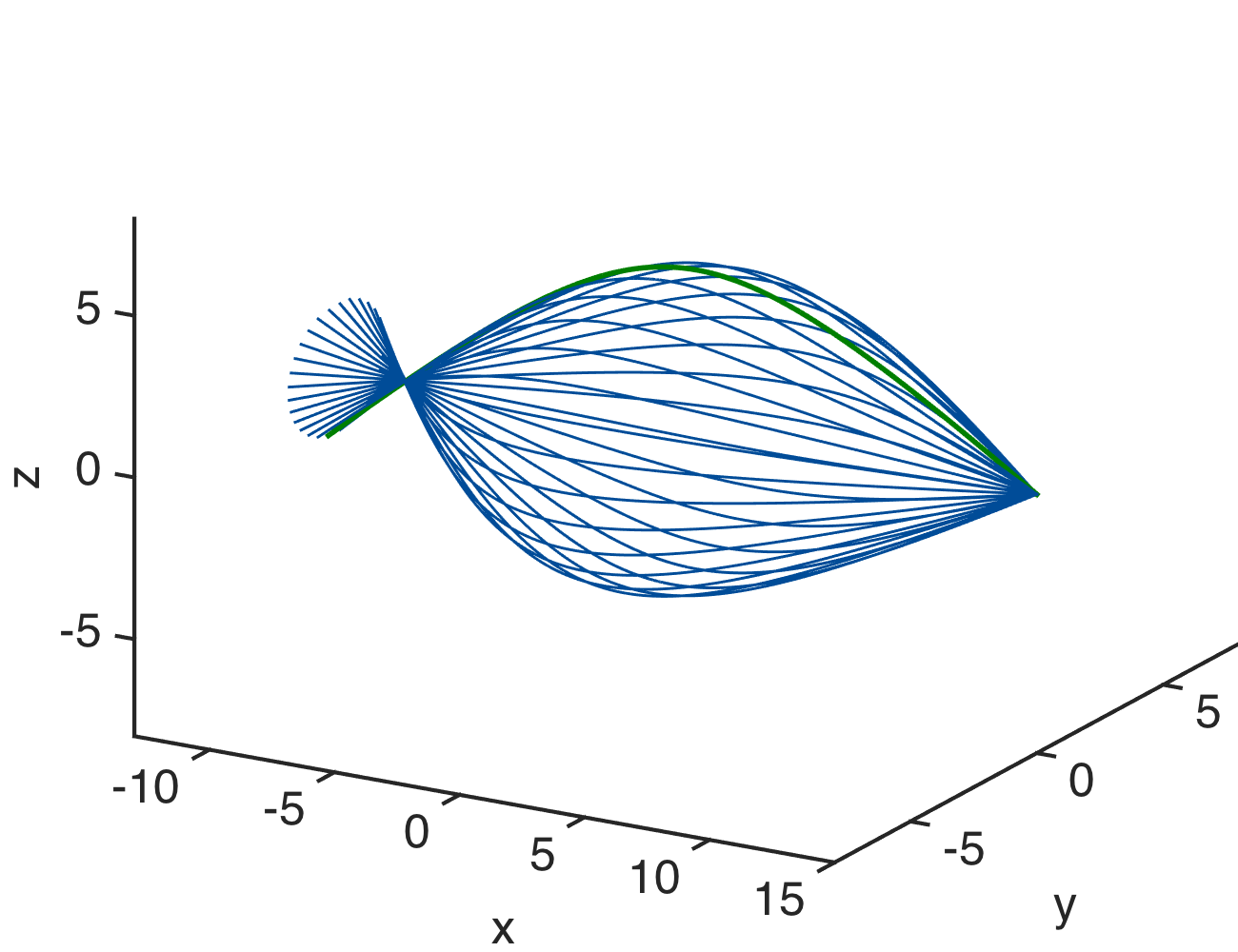} \includegraphics[width=7cm]{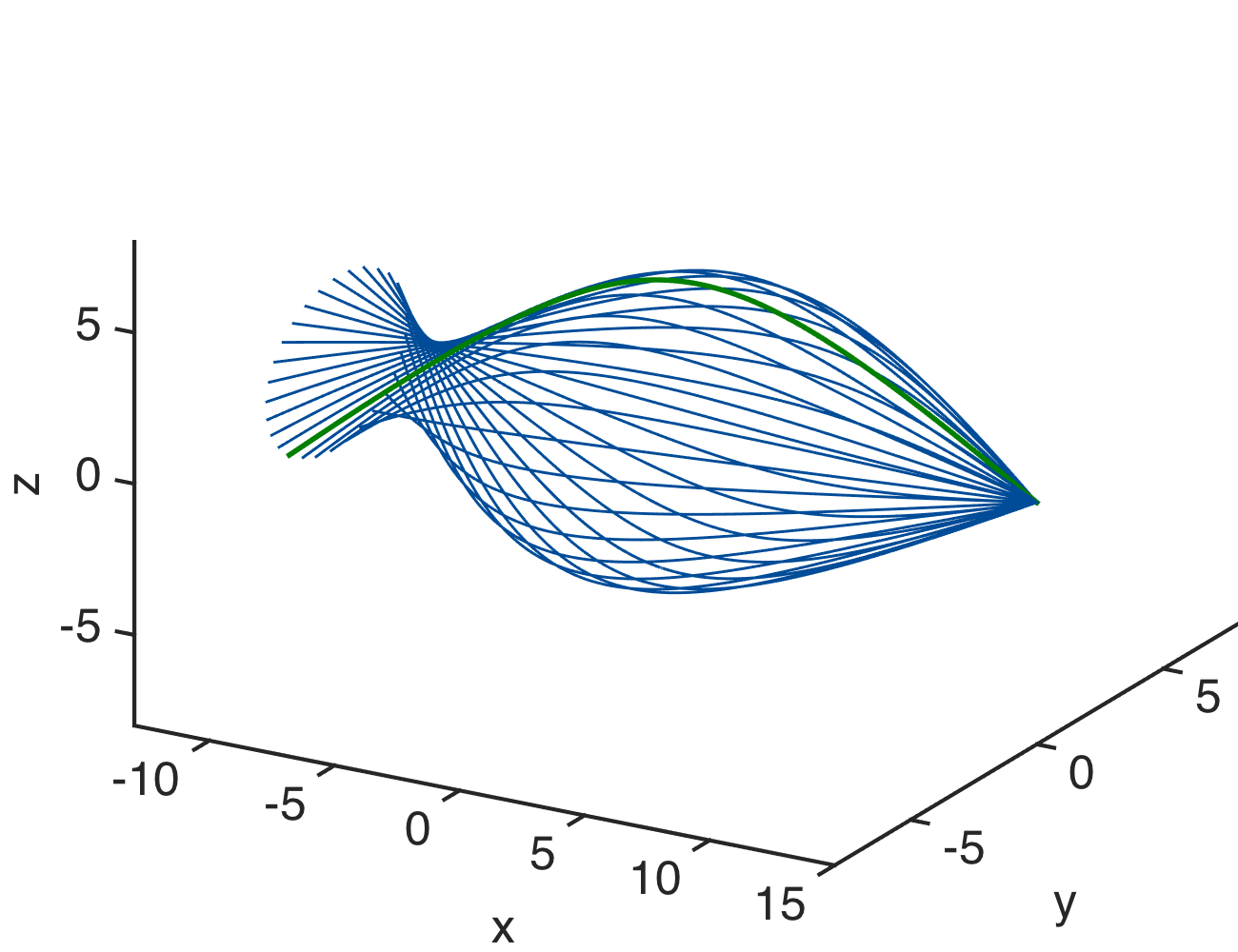}
    \caption{Multiple light rays traced back in time from the position of an observer on the equatorial plane. We show one of these lines in green to highlight the twisting about the $x$-axis.} \label{caustic}
\end{figure}

The apparent images in Schwarzschild, Kerr and NUT spacetimes are shown in Fig. \ref{fig:shadow_nut_color}. To obtain these simulations, the observer is located at a distance $R_c=20$ from the black hole, its line of sight has an inclination $\psi=90^\circ$ with respect to the symmetry axis and the radius of the celestial sphere is set to $R_\infty=100$. The screen has a fixed width and height $d_H=d_V=0.1$, a resolution of $300\times 300$ pixels and a focal length $d_L=0.04$. With these choices, the angles of view in Minkowski spacetime would be approximately $\delta_H=\delta_V=120^\circ$. 

In a Schwarzschild spacetime the region on the celestial sphere located right behind the black hole (the center of the Milky Way) sends light rays reaching the observer from multiple directions, which is why the Einstein ring looks blue. Such a ring is still present in Kerr spacetimes. However, this phenomenon is no longer observed in NUT spacetimes. As the parameter $\ell$ increases, the Einstein ring disappears; moreover, the center of the sky is hidden by the NUT black hole, which becomes more evident in the limiting case $m=0$ and $\ell=1$, in which the center of the galaxy is completely hidden. 

To explain why  a region of the celestial sphere is hidden when $\ell\neq 0$, we show the backwards tracing of some light rays 
reaching the observer in both Schwarzschild and NUT spacetimes in Fig. \ref{caustic}. We 
plot the light rays reaching the screen with a constant angular aperture, i.e., those pixels drawing a perfect circle on the screen of the virtual camera. In Schwarzschild, each of these light rays travels on a plane and will eventually cross again, as observed in the plot. This behavior gives rise to the Einstein ring. However, if $\ell\neq 0$ we observe that these light rays are twisted about the line of sight (the $x$-axis in this example) and they will never cross again, unlike those in Schwarzschild's spacetime. This is the reason for the existence of a blind region behind the NUT black hole in the primary and secondary images. We color one of the light rays in green in order to observe the twist of a single light ray about the $x$-axis.
The light rays shown in the figure would correspond to the pixels showing the secondary image of the background space.

\begin{figure} [H]
    \centering
    \small Kerr: $\varphi=1.5$ \\
    \includegraphics[width=8cm]{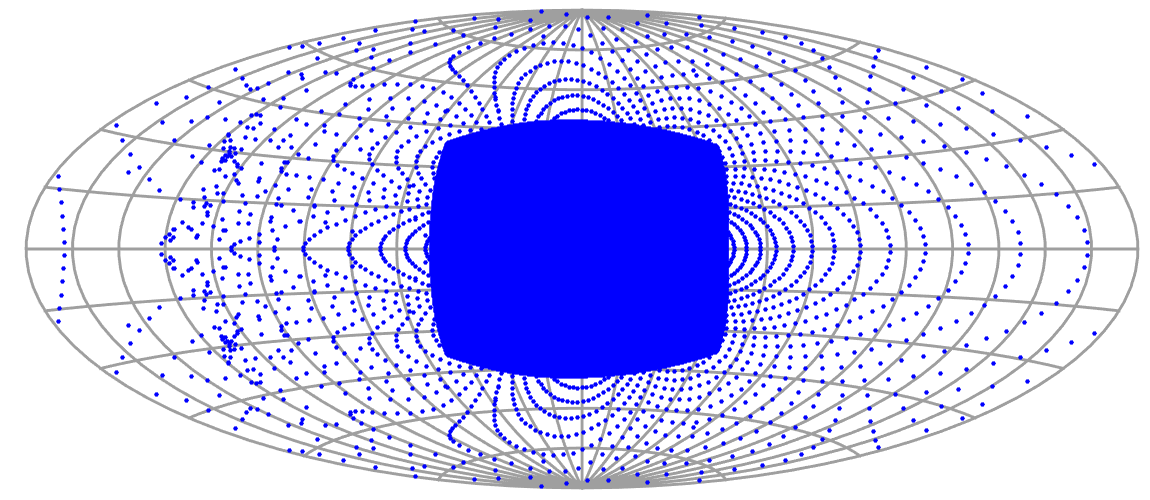} \\
    \small NUT: $m=1$, $\ell=0.5$ \\
    \includegraphics[width=8cm]{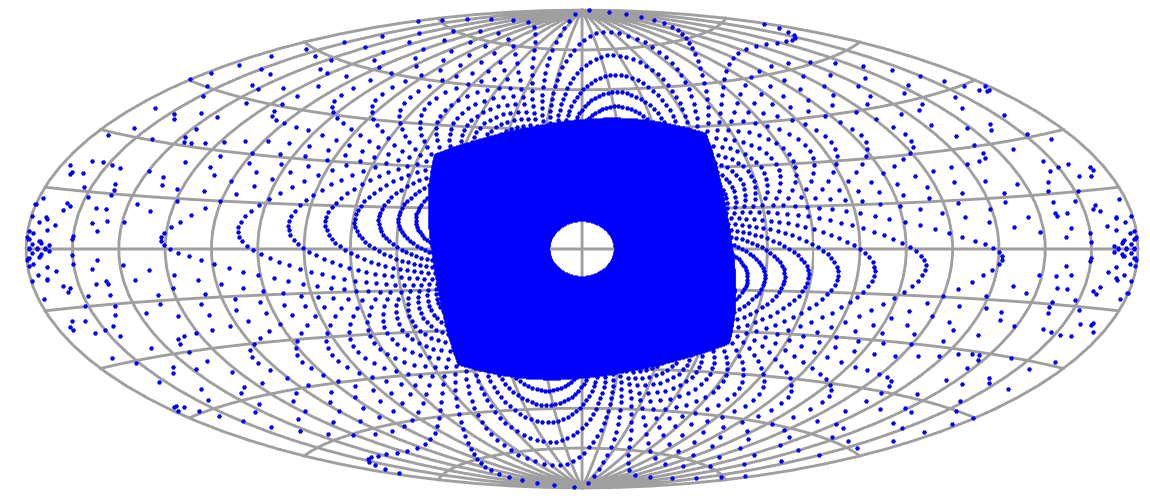}
    \caption{Origin on the celestial sphere of every pixel in Kerr and NUT spacetimes. } \label{fig:origin_Kerr-NUT}
\end{figure}

It can be observed in Fig. \ref{caustic} that light rays in NUT spacetimes with a given constant angular aperture are driven away at the same $x$-component $c<0$, which suggests that there exists a critical value $x_c$ such that regions on the celestial sphere satisfying $x_c/R_{\infty}<-1+\varepsilon_\ell$ are never reached by shooting light rays in this manner, where $\varepsilon_\ell$ is a positive value that increases in dependence of the magnitude of $\ell$ and $x_c$ is the $x$-component of the light ray on the celestial sphere. In other words, light rays originating from this region will never reach the observer. In terms of the celestial coordinates $(\phi,\theta)$, this hidden region is given by the inequality 
$$\cos(\theta)\cos(\pi-\phi)>1-\varepsilon_\ell.$$ 
Evidently, this inequality does not have real solutions if $\varepsilon_\ell=0$ (Schwarzschild). If $\varepsilon_\ell>0$, then this inequality indicates a quasi-circular shape (whose size increase with $\ell$) in the Aitoff projection centered at $(\phi,\theta)=(\pi,0)$.
To visualize this, we can plot the celestial coordinates $(\phi,\theta)$ of each of the light rays originating from the celestial sphere, namely, the coordinates of origin of each colored pixel in Fig. \ref{fig:shadow_nut_color}. These light ray sources are shown in Fig. \ref{fig:origin_Kerr-NUT} for both Kerr and one of the NUT spacetimes.
Notice that, as expected, the entire region behind the Kerr black hole is visible in the primary image, but this is no longer the case for the NUT black hole. In this case, the center of the galaxy is hidden.

%
%
\section{Geodesics in the toron spacetime}
\label{toron_geod}

In this section we present simulations analogous to those presented in the preceding section, 
both for individual geodesics and for the apparent images of extended objects. First, we determine the solutions of the IVP \eqref{eq:ivp} with the same initial 
conditions used for Figs. \ref{fig:NUT_xy} and \ref{fig:NUT_xz} in 
order to illustrate the similarities and differences between toron 
and NUT spacetimes, especially the case with $m=0$ and $\ell\neq 0$, since 
the disk described be the toron solution can be interpreted as a 
massless object, as discussed before. 

Subsequently we show how certain light rays may form the fundamental photon orbits. However, due to the more asymmetric nature of the metric functions, these geodesics 
are more sensitive to small perturbations of the initial conditions; therefore, instead of showing photon rings as in Fig. \ref{fig:NUT_ring_shell}, we only show light rays that approach a particular photon sphere. 
For extended objects, we simulate the apparent image of the  branch disk  of the toron solution, as well as the distorted image of the background space seen by a distant observer. 

\subsection{Initially-parallel light-rays} 

We study the effect of the toron parameter $\alpha$ on a beam of initially-parallel light rays, first on the equatorial plane and then on the $xz$-plane.
In order to compare with well-understood spacetimes, we use the same 
set of initial conditions we used in NUT spacetimes in subsection \ref{sec:parallel_nut}. This will allow us to visualize the effect of the parameter $\alpha$ as its magnitude increases and thus provide a physical explanation based on the observations in NUT spacetimes.

\begin{figure}[H]
    \centering
   \includegraphics[width=6cm]{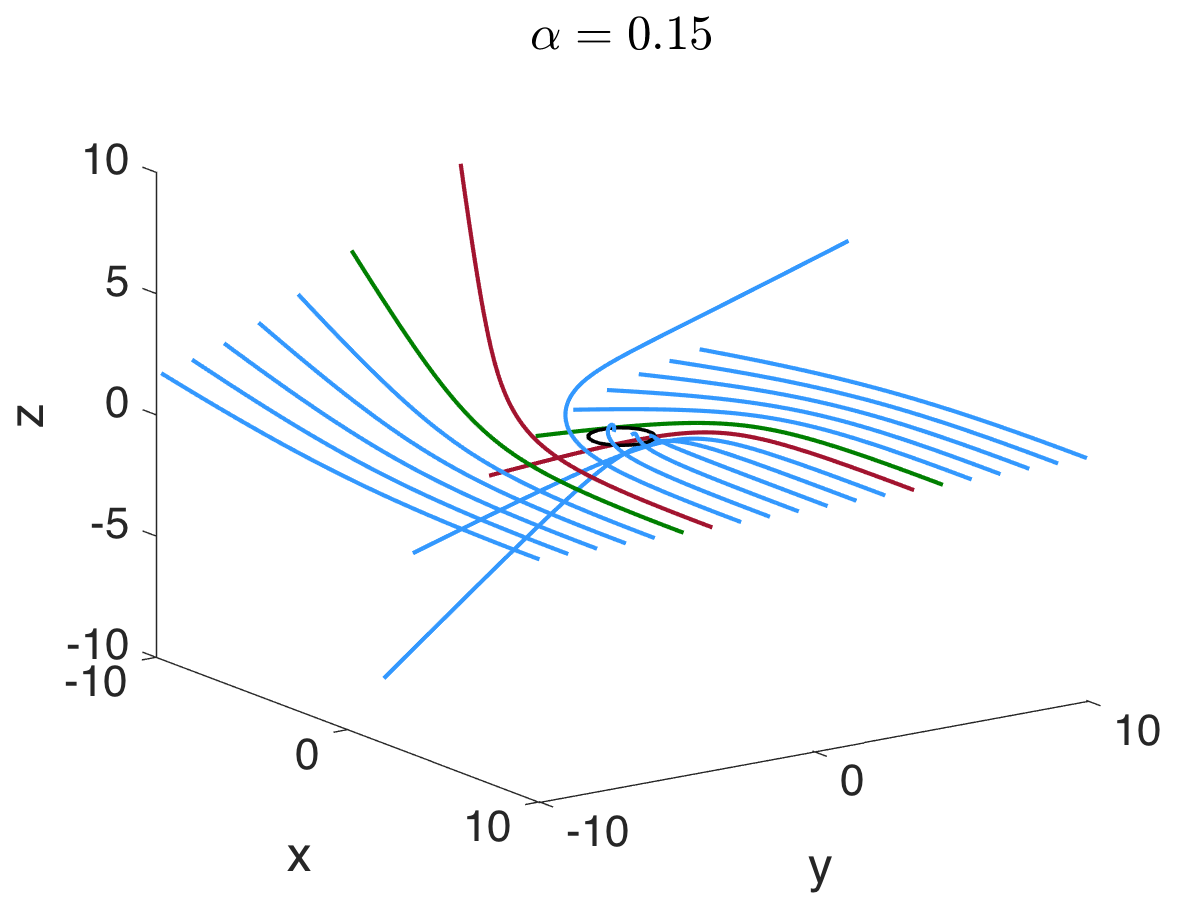}
    \includegraphics[width=6cm]{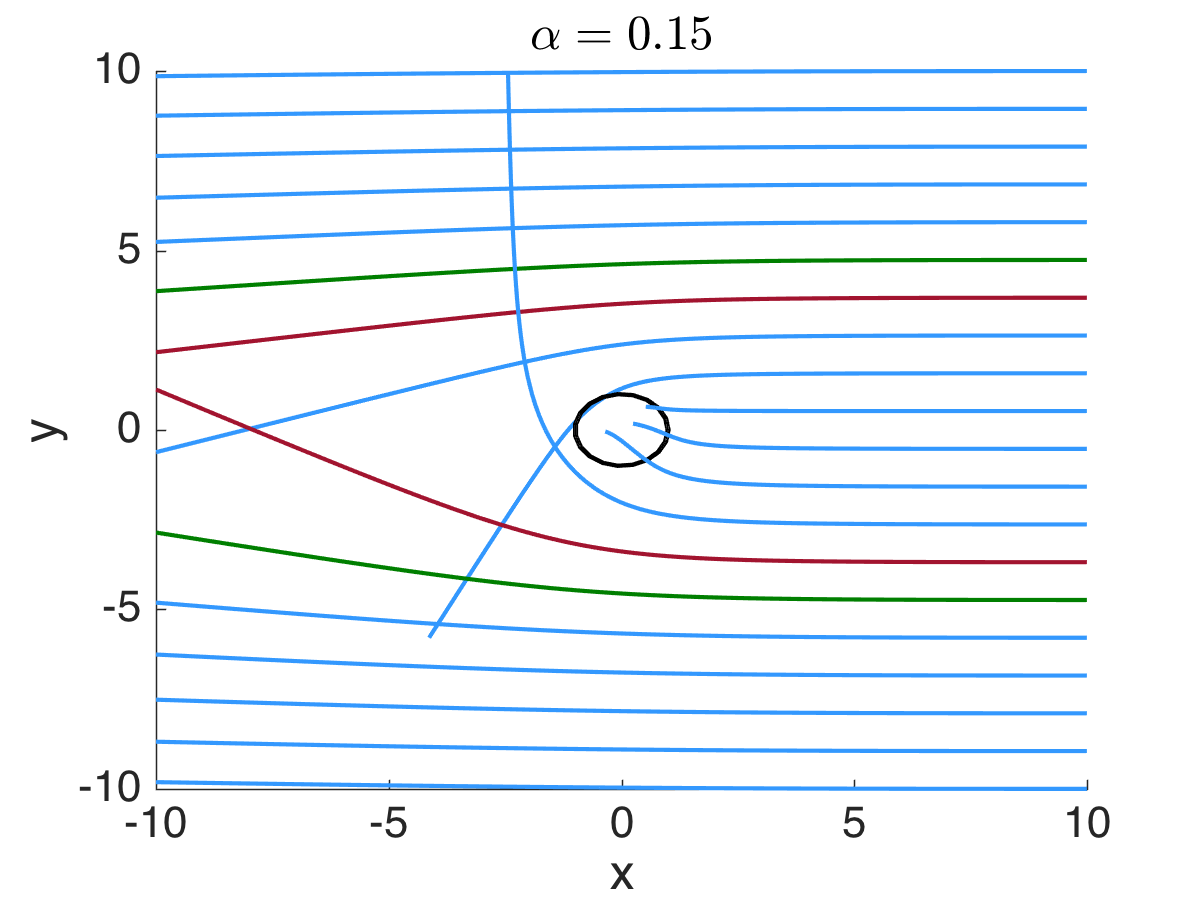}
    \includegraphics[width=6cm]{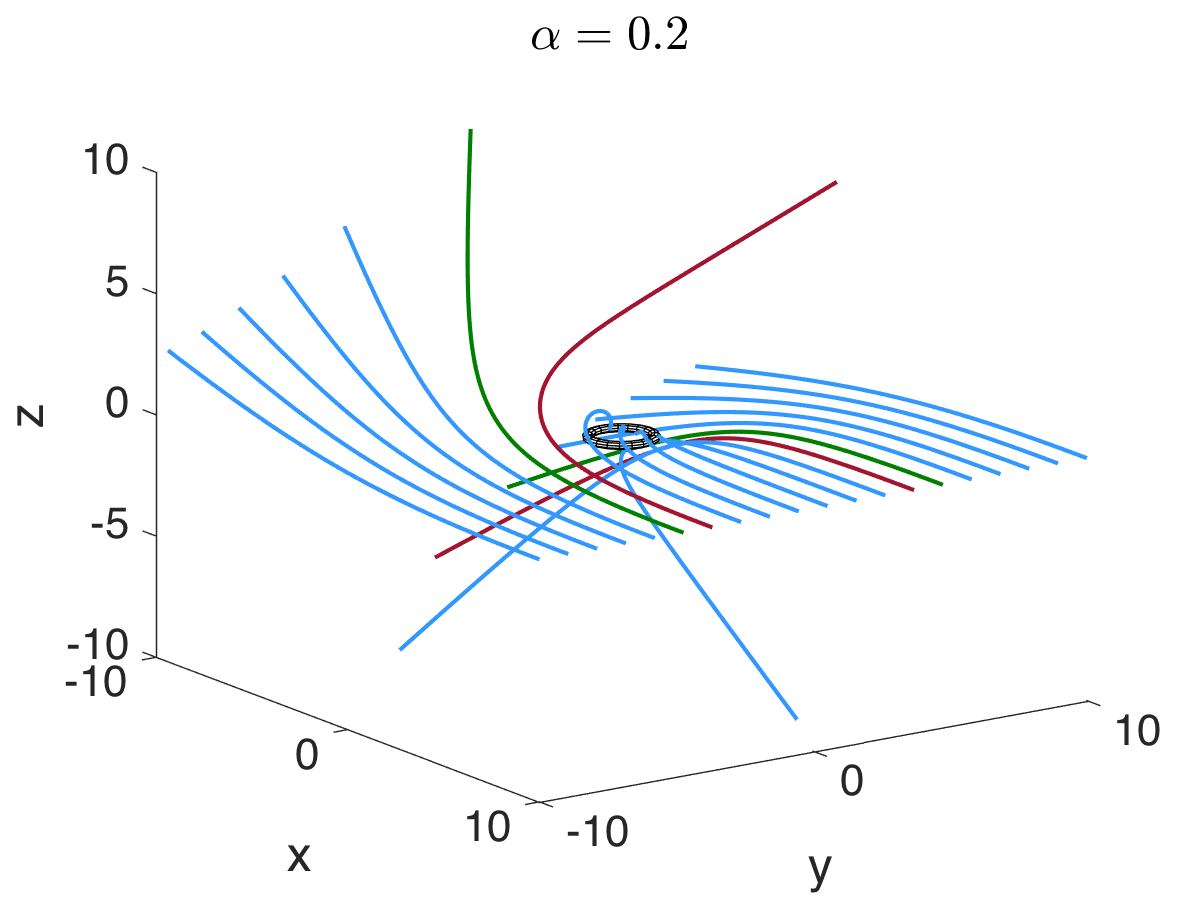}
    \includegraphics[width=6cm]{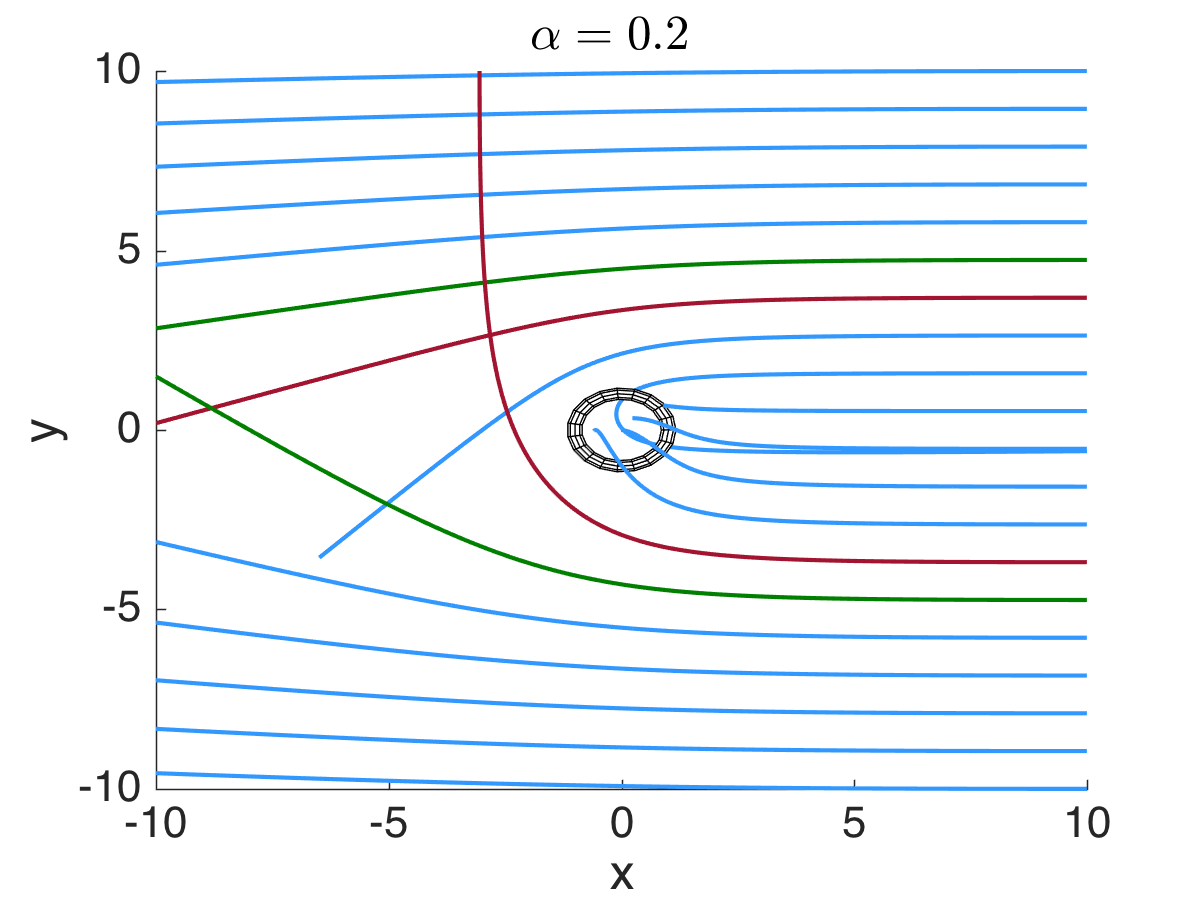}
    \includegraphics[width=6cm]{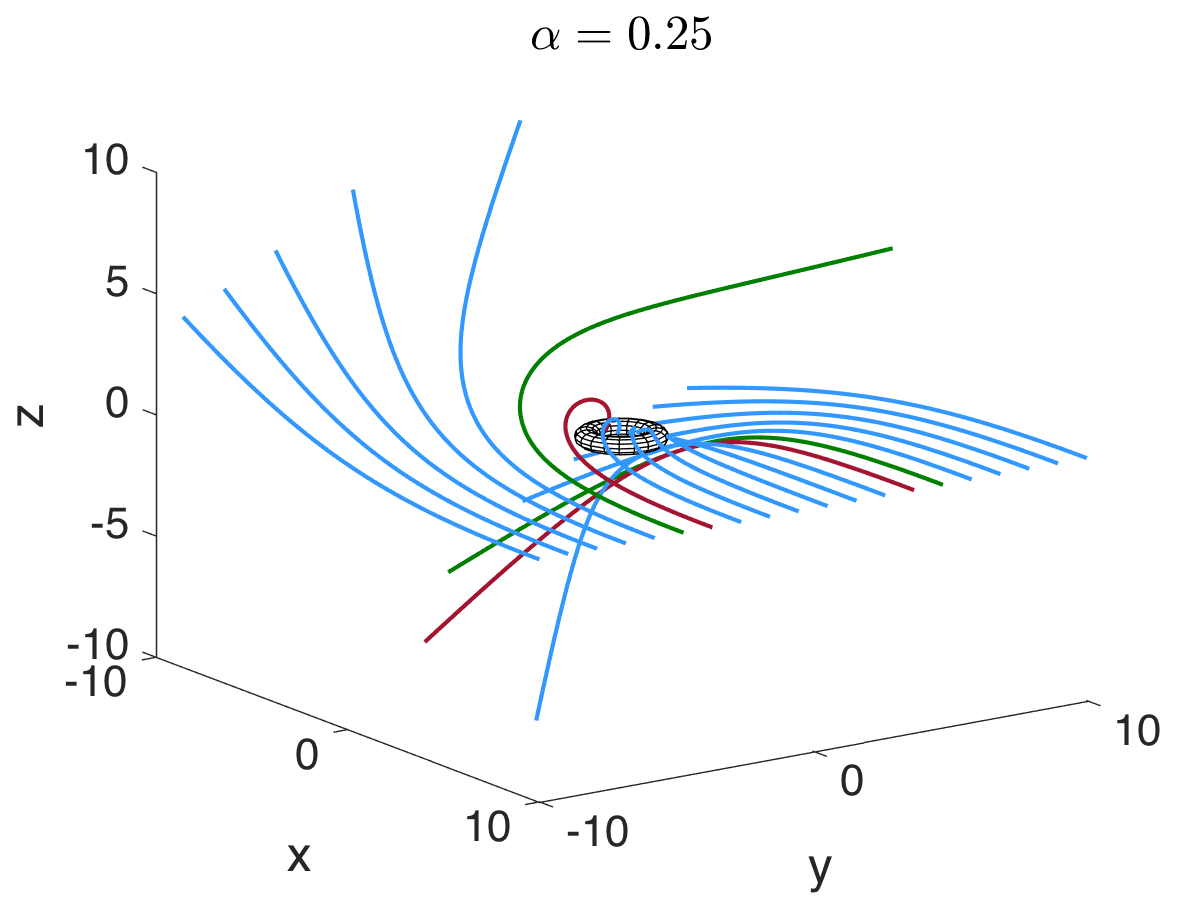}
    \includegraphics[width=6cm]{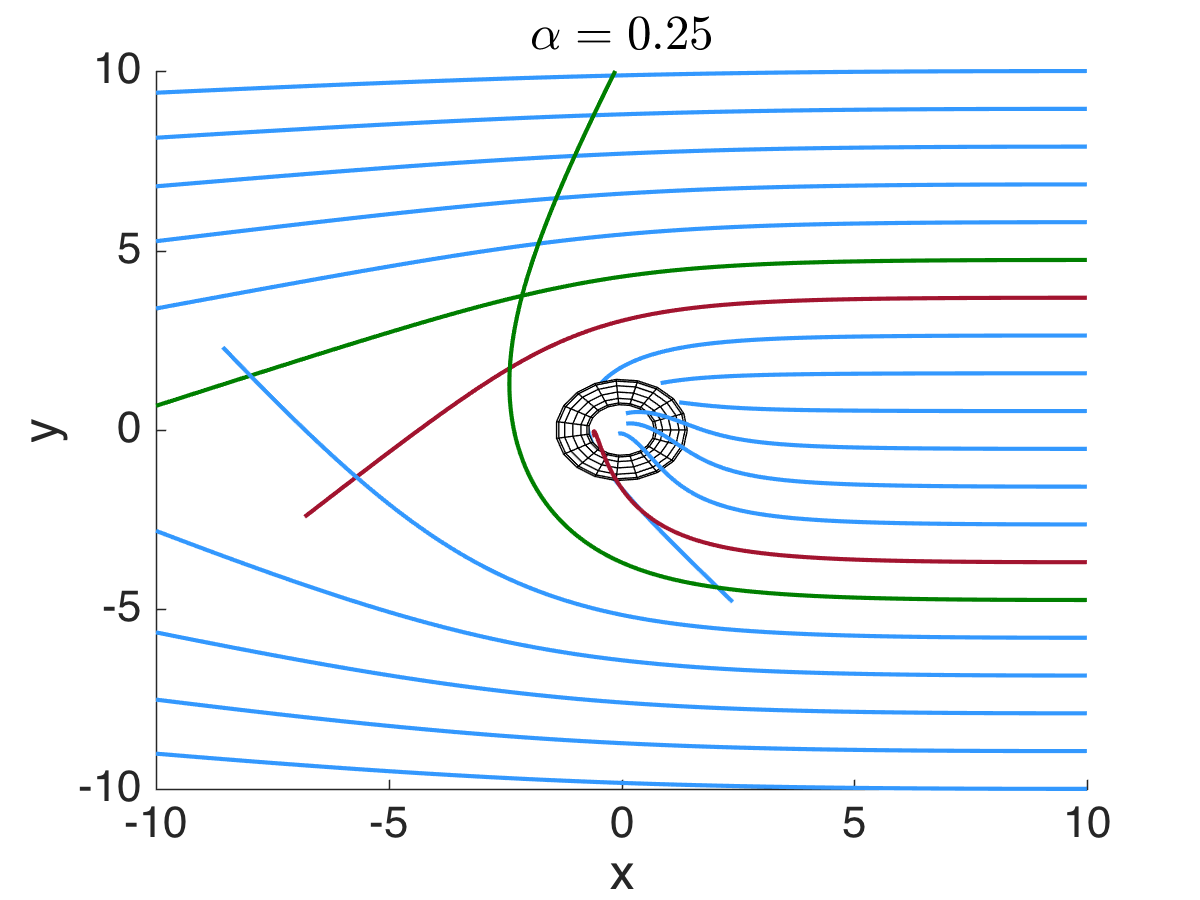}
    \caption{Initially-parallel light rays on the equatorial plane in 
	toron spacetimes for various values of $\alpha$ and their 
	projection onto the $xy$-plane. In each example, two pairs of light rays corresponding to the same initial conditions except for the $y$-component (we chose $y_0>0$ for one of the rays and $y'_0=-y_0$ for the other) are colored in red and green.}\label{fig:LR_xy_elliptic}
\end{figure}

We begin by studying light rays with initial conditions of the type 
(i) given in subsection \ref{sec:parallel_nut}, namely, we consider a beam of initially-parallel light rays on the equatorial plane.
Fig. \ref{fig:LR_xy_elliptic} shows that the gravitational pull 
increases with the magnitude of $\alpha$, even though the toron 
solution describes a massless disk. However, this is also observed in 
the massless NUT solution, see Fig. \ref{fig:NUT_xy}. 
The projection onto the $xy$-plane shows 
that the photons are dragged in a counterclockwise direction, 
suggesting that the disk is a source of angular momentum, analogous to a Kerr black hole with a positive angular momentum.
On the other hand, the three-dimensional representation shows the deviation of some of the light rays in upwards or downwards direction, suggesting that the disk is a source of gravimagnetic momentum as well. Thus, the behavior of photons is similar to what would be expected in Kerr-NUT spacetimes.

The value of $d p^z/ds|_{s=0}$ for a given initial condition will 
indicate whether the light ray will stay on the plane (if this value 
vanishes) or it will move in upwards (resp. downwards) direction if this value is positive (resp. negative). Let us recall that for initial conditions of this type, $d p^z/ds|_{s=0}$ vanishes identically in Kerr spacetimes, while in NUT spacetimes it only depends on the sign of the initial $p^\phi_0$, as indicated by formula \eqref{eq:parallel_xy}.
However, such a formula does not hold in toron spacetimes since the metric functions are not symmetric in $z$ and one would need to numerically compute the full expression for $d p^z/ds|_{s=0}$ to estimate the deviation.
Therefore, numerical computations as shown in Fig. \ref{fig:LR_xy_elliptic} are necessary not only because of this difficulty, but also because of the unique behavior of the toron  metric functions in the vicinity of the gravitating object.

In order to extract additional information from the motion of photons initially on a plane (especially the contribution of the gravimagnetic mass component), we also study the evolution of light rays with initial conditions of type (ii) given in subsection \ref{sec:parallel_nut}. Namely, a beam of initially-parallel light rays on the $xz$-plane.

\begin{figure}[H]
    \centering
    \includegraphics[width=6cm]{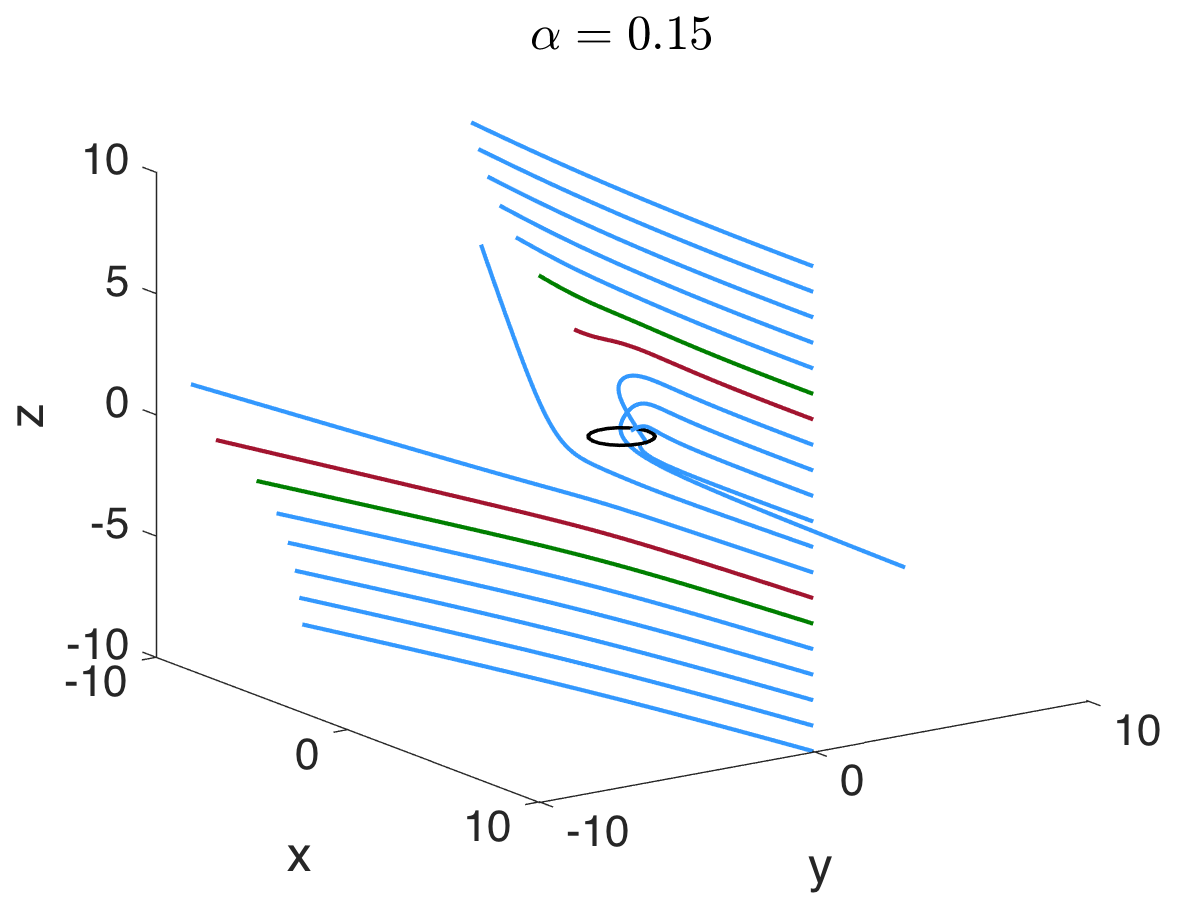}
    \includegraphics[width=6cm]{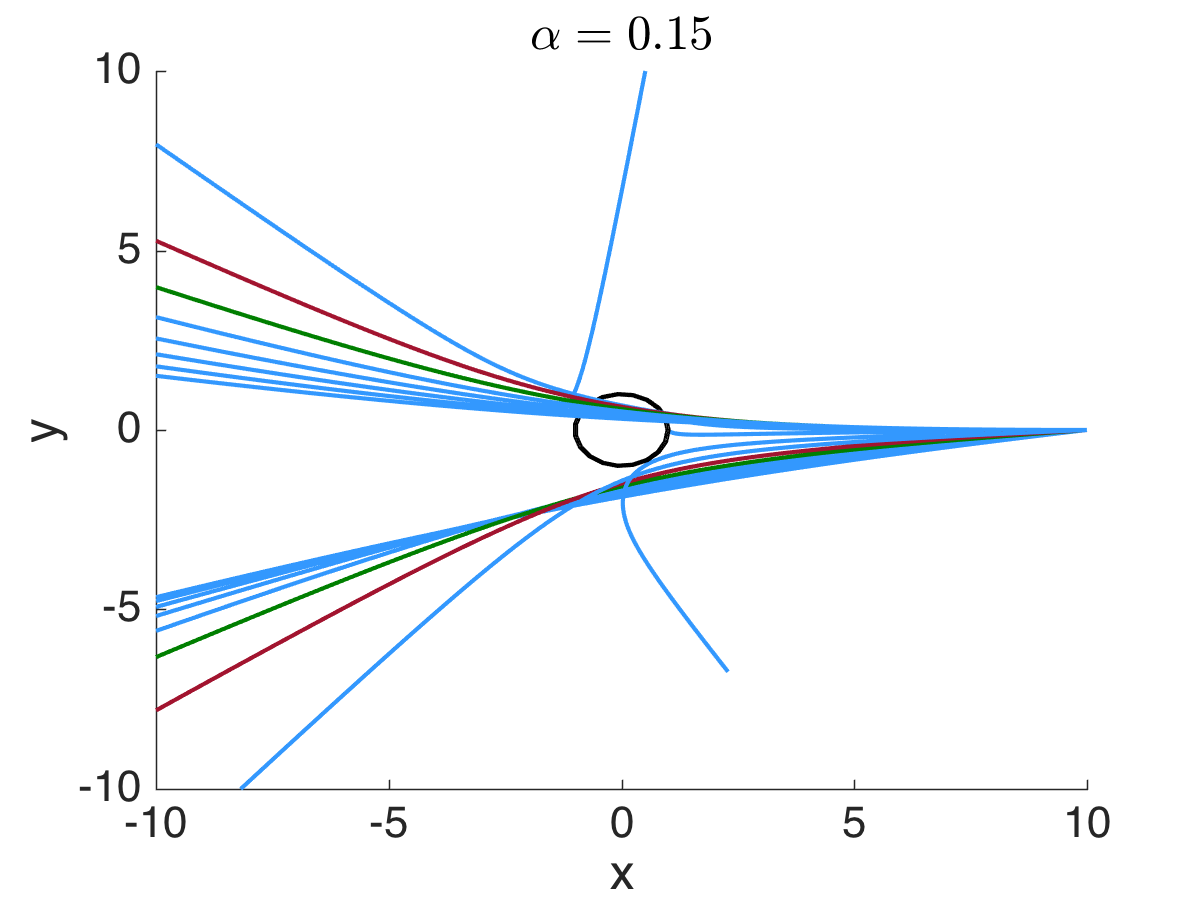} 
    \includegraphics[width=6cm]{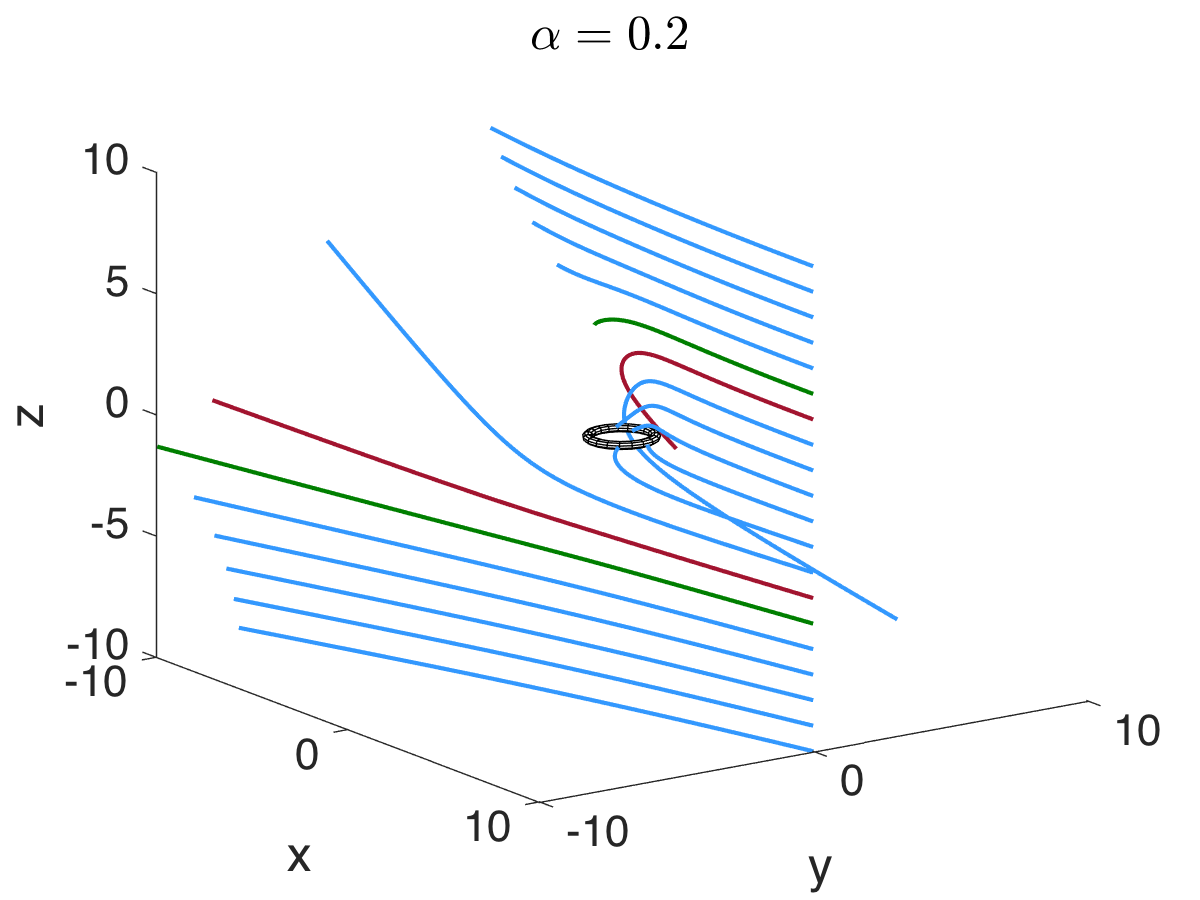}
    \includegraphics[width=6cm]{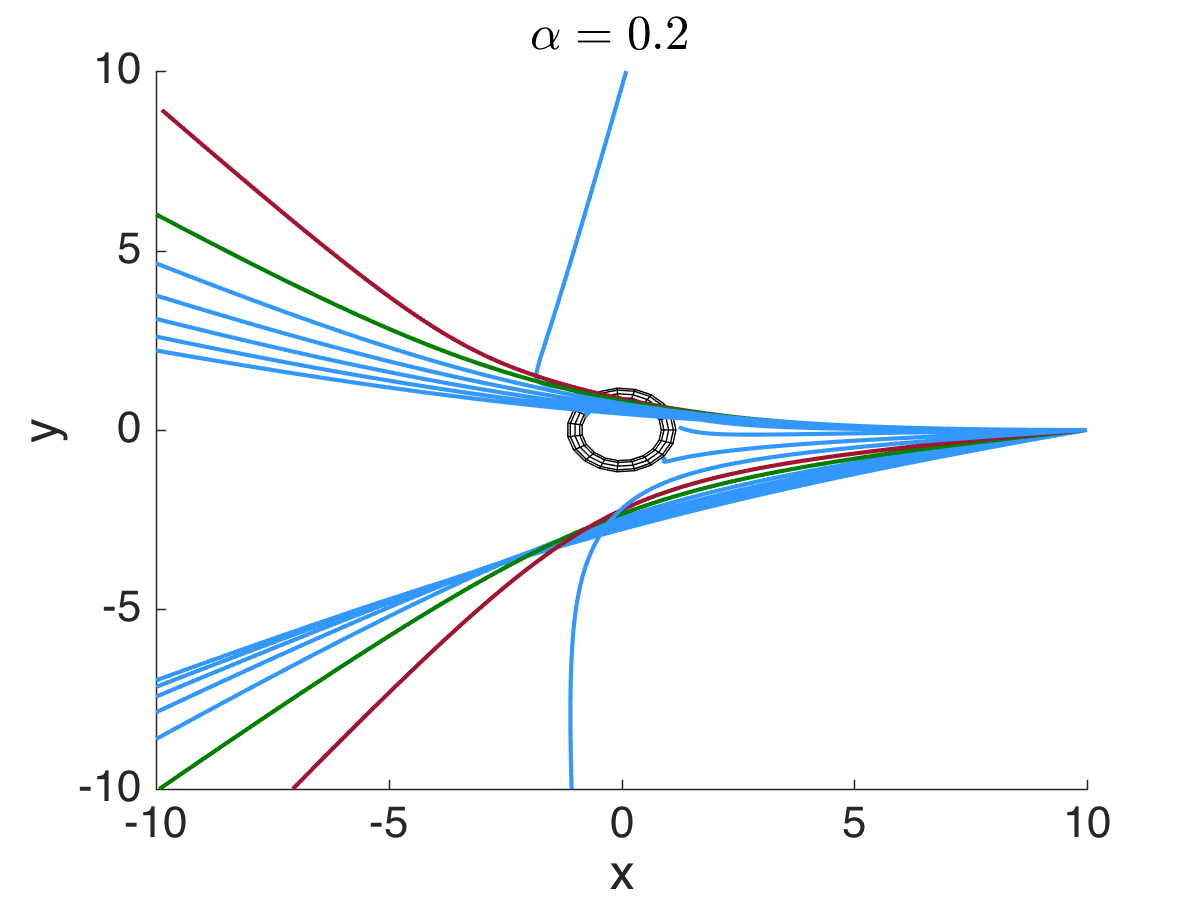}
    \includegraphics[width=6cm]{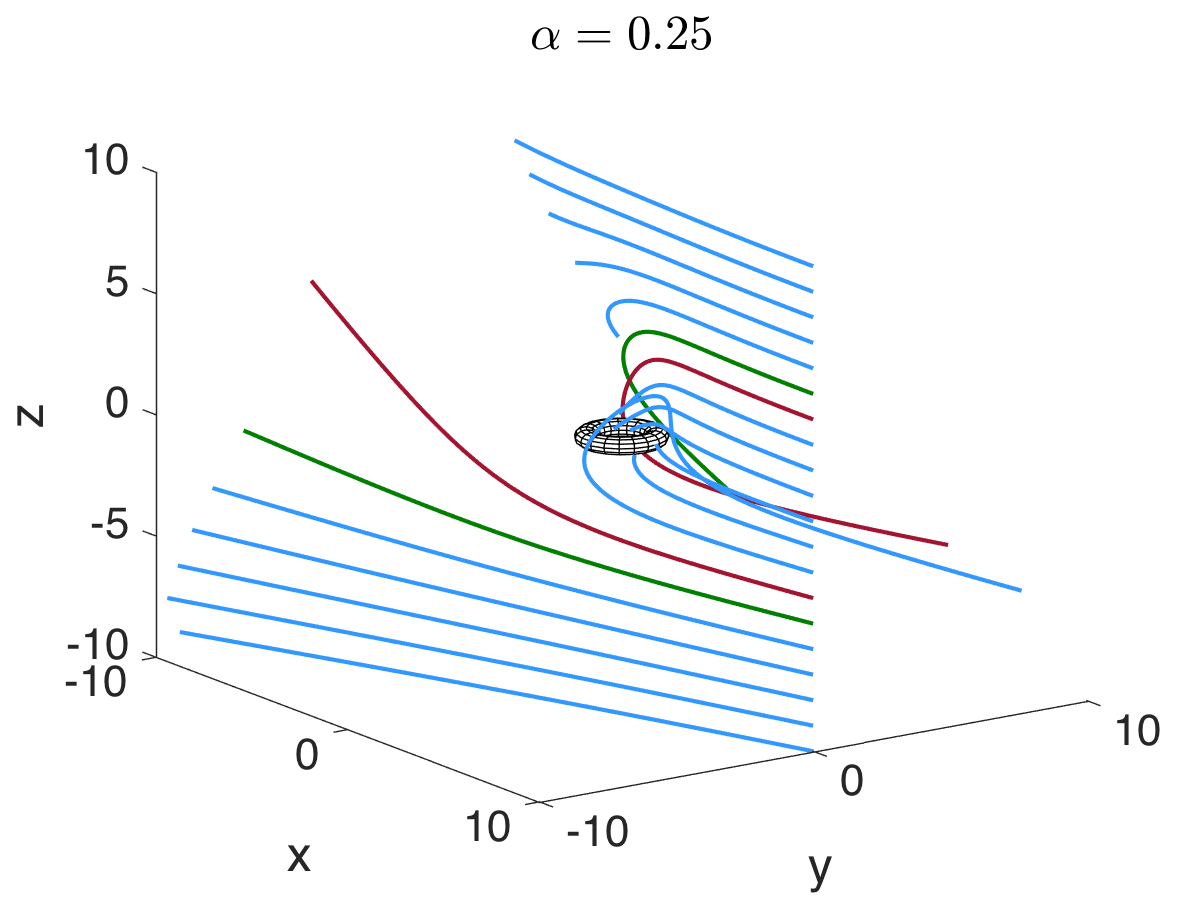}
    \includegraphics[width=6cm]{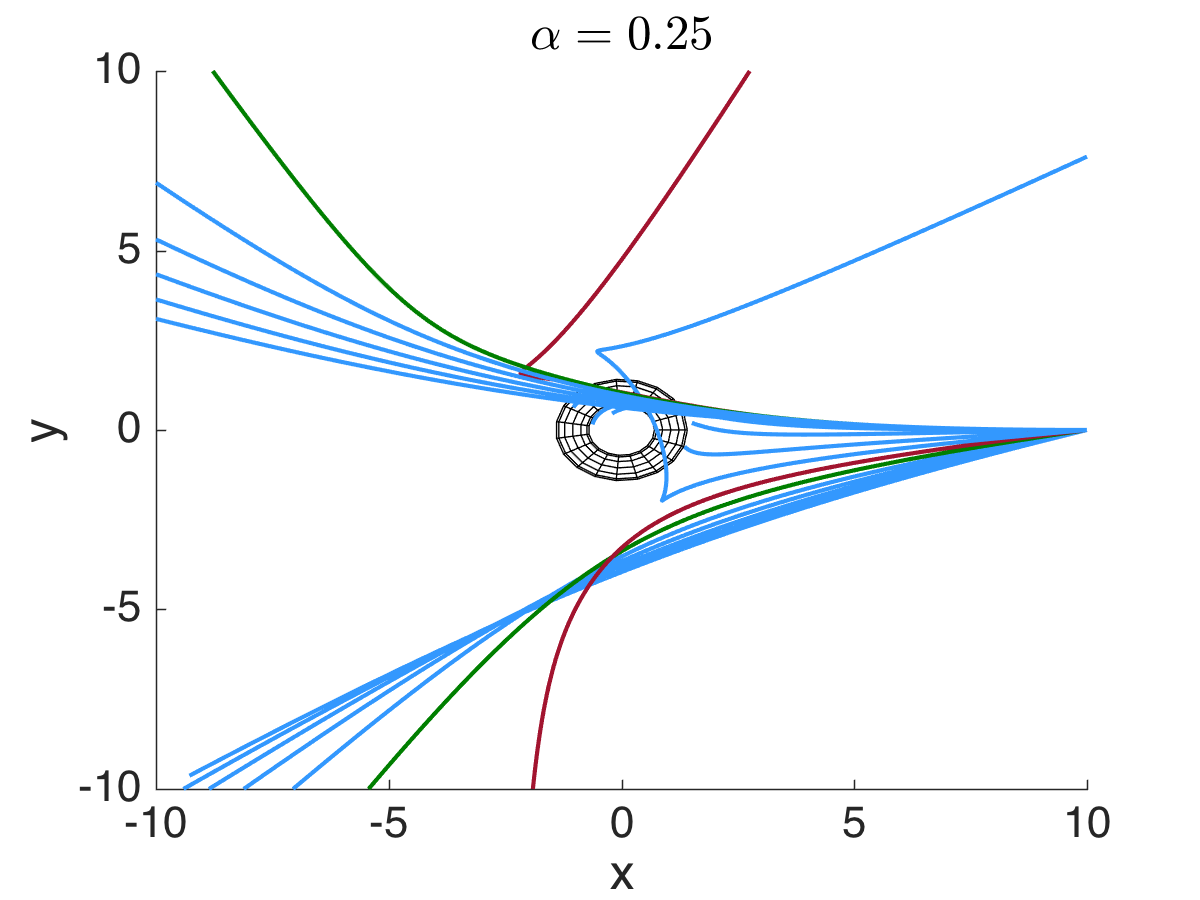}
    \caption{Initially-parallel light rays on the $xz$-plane in a toron spacetimes with  various values of $\alpha$ and their projection on the $xy$-plane. In each example, two pairs of light rays corresponding to the same initial conditions except for the $z$-component (we chose $z_0>0$ for one of the rays and $z'_0=-z_0$ for the other) are colored in red and green.  } \label{fig:LR_xz_elliptic}
\end{figure}

The effect of the angular momentum in addition to the 
gravimagnetic mass is further observed in Fig. 
\ref{fig:LR_xz_elliptic}. 
Although the formula \eqref{eq:pphi_eq} holds in toron spacetimes, determining the deviation of the light rays based solely on $z$ is not possible; one would need to numerically evaluate $\partial_\rho A$ at the initial position. 
The deviation of these light rays can be observed in Fig. \ref{fig:LR_xz_elliptic}. The dragging direction in toron spacetimes is clockwise for some light rays and counterclockwise for others, analogous to NUT spacetimes. The difference is that the dragging effect is stronger when $z_0<0$, which is explained by the fact that the magnitude of $A$ is bigger when $z$ is negative, which can be observed in Fig. \ref{ellipticmetric}. 

Since the disk is a source of angular momentum contributing to the dragging effect in counterclockwise direction, compared to spacetimes with only gravimagnetic mass, the deviation of a light ray that would otherwise be dragged in clockwise direction is slowed down and it is accelerated if it were to be dragged in clockwise direction. This is well observed by comparing the $xy$ projections in Figs. \ref{fig:NUT_xz} and \ref{fig:LR_xz_elliptic}. As expected for toron spacetimes, the symmetry of the $xy$ projection of the geodesics with respect to the $x$-axis is lost and an additional dragging component in counterclockwise direction is observed.

\subsection{Search for fundamental photon orbits} 
\begin{figure}[H]
    \centering
    \includegraphics[width=6cm]{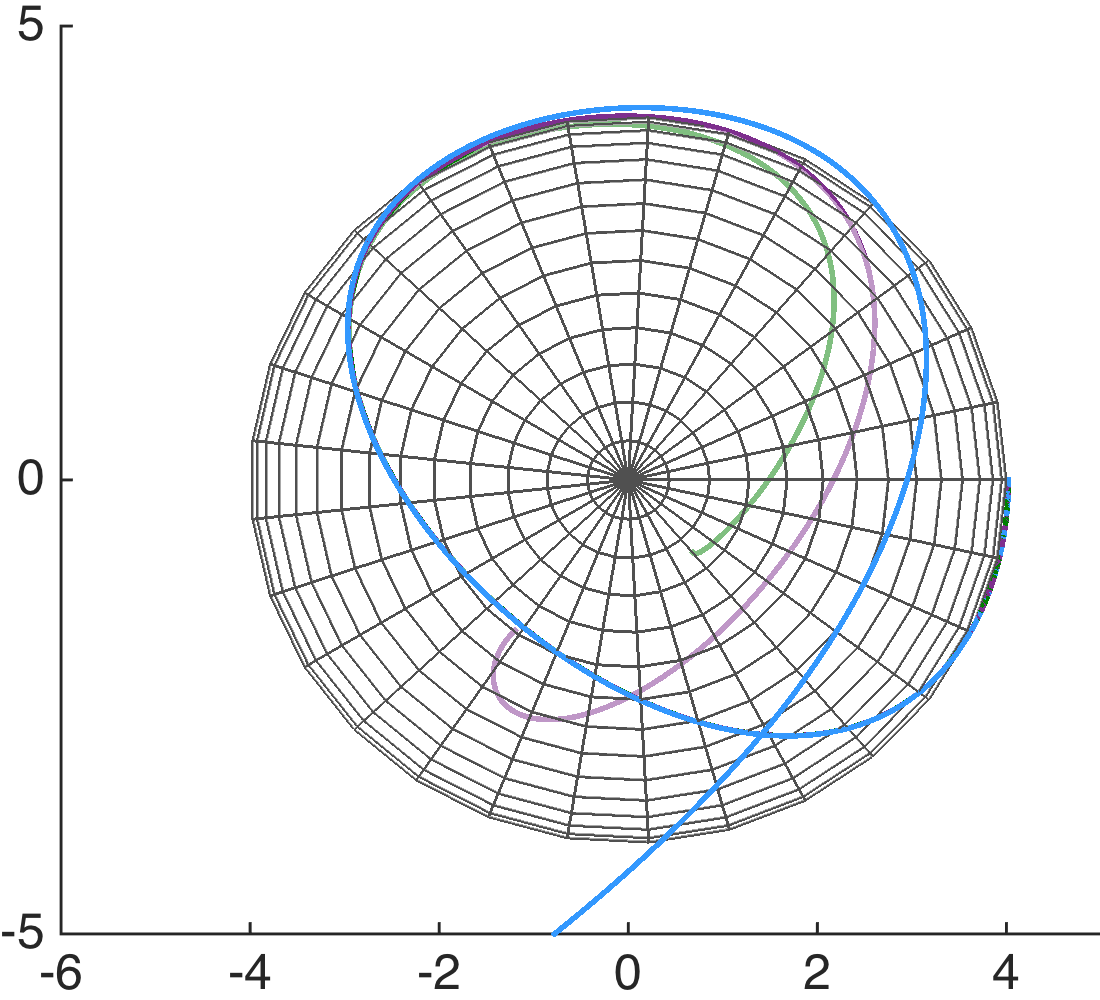} \includegraphics[width=6cm]{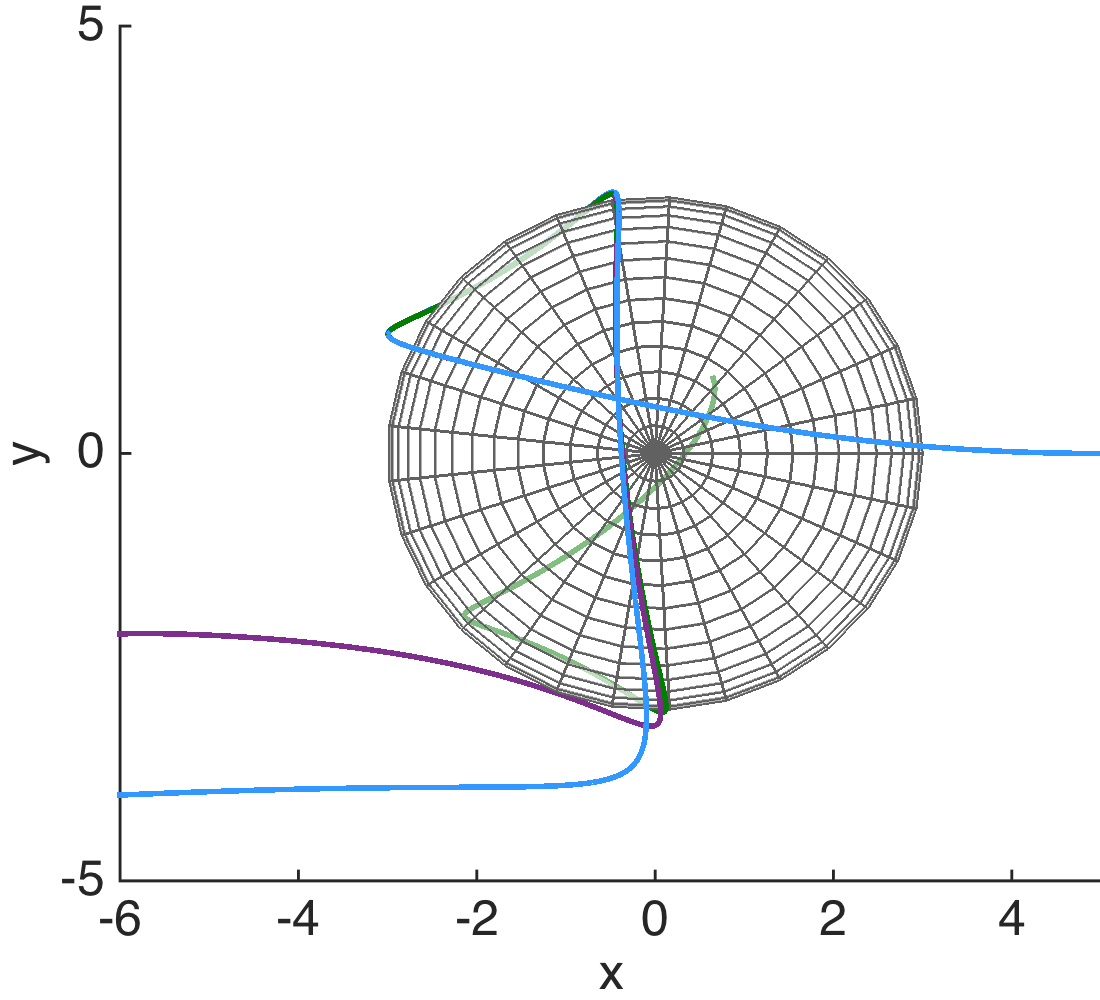} 
    \includegraphics[width=6cm]{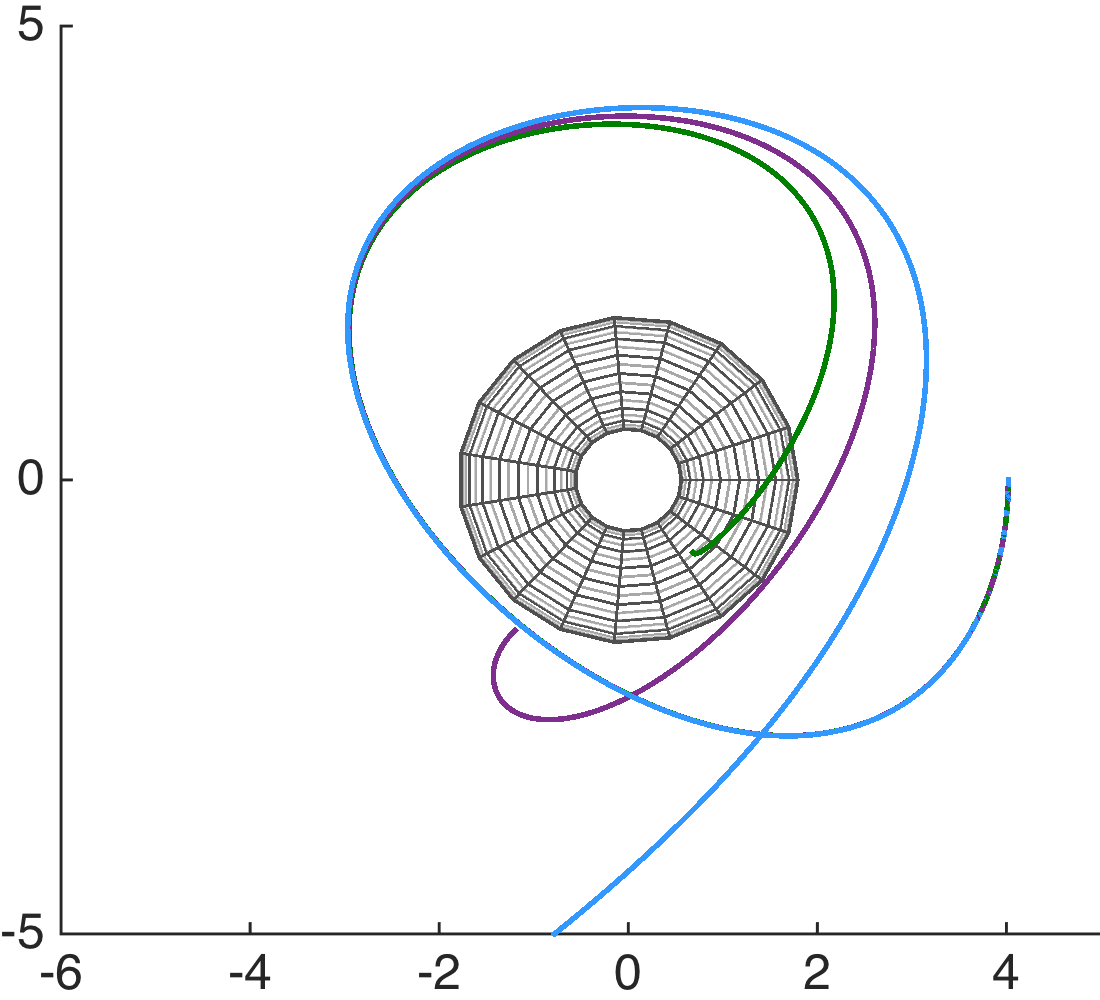} \includegraphics[width=6cm]{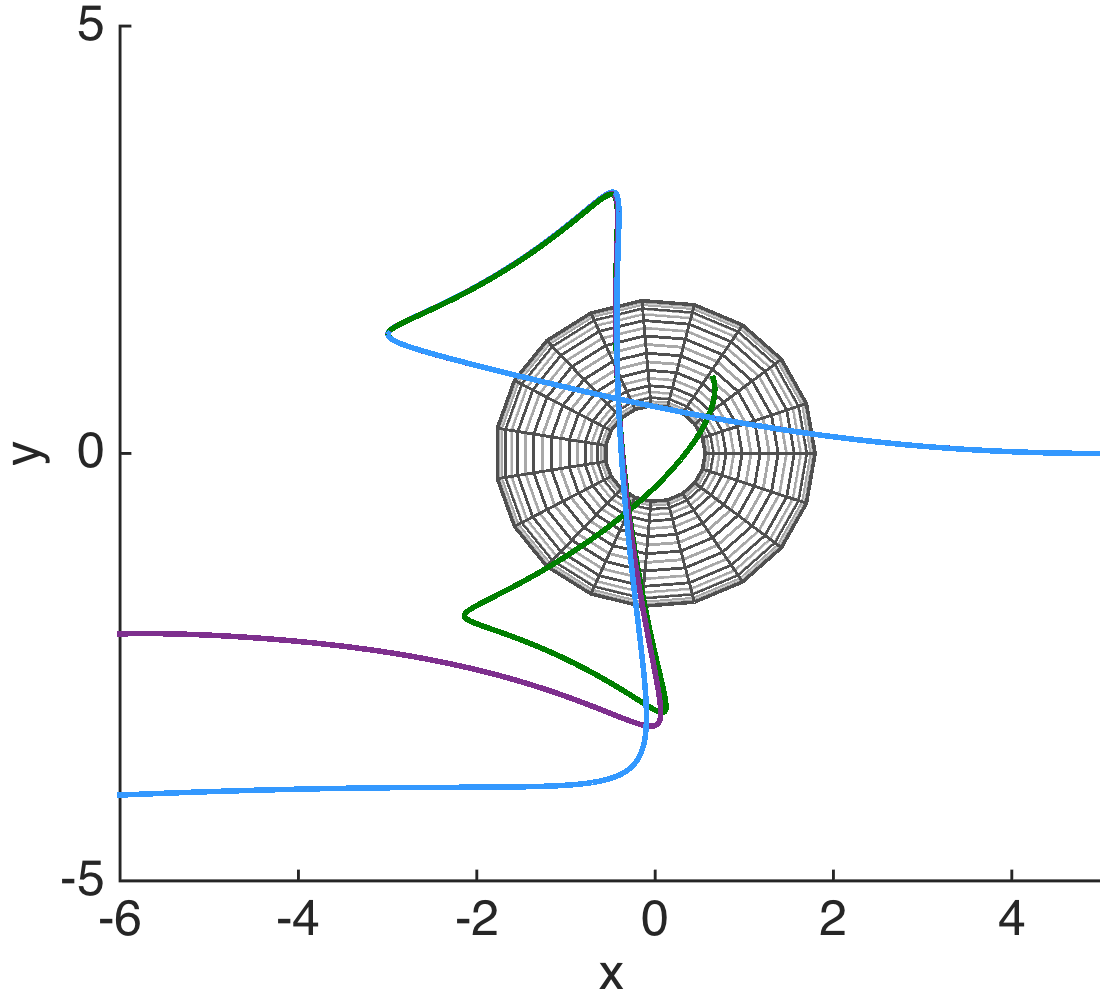} 
    \caption{Light rays approaching photon spheres in a toron spacetime with $\alpha=0.3$. The lower figures show how these light rays either approach the ergosphere or escape to infinity.} \label{fig:photon-sphere_elliptic}  
\end{figure}

Here we are trying to find fundamental photon orbits in
toron spacetime with $\alpha=0.3$. As in Kerr-NUT 
spacetimes, the fundamental photon orbits with  maximal distance from the object  correspond to a retrograde 
photon passing  the equatorial plane with  zero $z$-component of 
velocity. This orbit is  shown on the left-hand side of Fig 
\ref{fig:photon-sphere_elliptic}. However, unlike Kerr-NUT 
spacetimes, it is not possible to decouple the geodesic equations and thus, determining the shape of the fundamental orbit  is a less trivial task. Moreover, light rays  have unstable orbits, meaning that tiny perturbations in the initial conditions of the IVP \eqref{eq:ivp} give rise to quite different outcomes at large times, as seen in Fig. \ref{fig:photon-sphere_elliptic}, which shows two particular photon orbits being approached by three photons whose initial conditions differ only slightly. In order to see how some photons are eventually trapped by the disk, the bottom part of Fig. \ref{fig:photon-sphere_elliptic} shows the same trajectories in a diagram with the ergosphere. The size of the fundamental photon orbit depends on the magnitude of the parameter $\alpha$ of a given toron spacetime.

%

\subsection{Apparent image of a disk} 
We simulate the apparent image of the disk described by the toron solution seen by an observer at a 
distance $R_c$ from its center. For the following simulations, we use a virtual camera 
with an inclination angle $\psi$ with 
respect to the symmetry axis, as described in Fig. \ref{fig:camera_diagram}. 
Since the gravitational pull increases with the magnitude of $\alpha$, 
the size of the apparent image of an object will be bigger as this 
magnitude increases, given a fixed focal length for the virtual camera. 
Analogous to a \textit{camera obscura} in flat space, the apparent 
image becomes bigger as the focal length $d_L$ decreases. Therefore, in order to visualize the entire image of the disk, the focal length needs to be decreased as the gravitational attraction increases. In particular, for the examples in Fig. \ref{fig:shadows_elliptic}, we simulate pictures of resolution $300\times 300$ pixels with the following values for the virtual camera in toron spacetimes with given values of $\alpha$.

\begin{table}[H]
    \centering
    \begin{tabular}{c|c|c|c|c}
   \hline $\alpha$ &  $R_c$ & $d_{H}$ & $d_{V}$ & $d_L$ \\ \hline
   $0.2 $ & $20$ & $0.10$ & $0.10$ & $0.25$ \\ \hline
   $0.3 $ & $20$ & $0.10$ & $0.10$ & $0.12$ \\ \hline
    \end{tabular}
    \caption{Chosen features of the virtual camera in a toron spacetime with parameter $\alpha$.} \label{table:cam}
\end{table}

\begin{figure}[H]
    \centering
    \includegraphics[width=7cm]{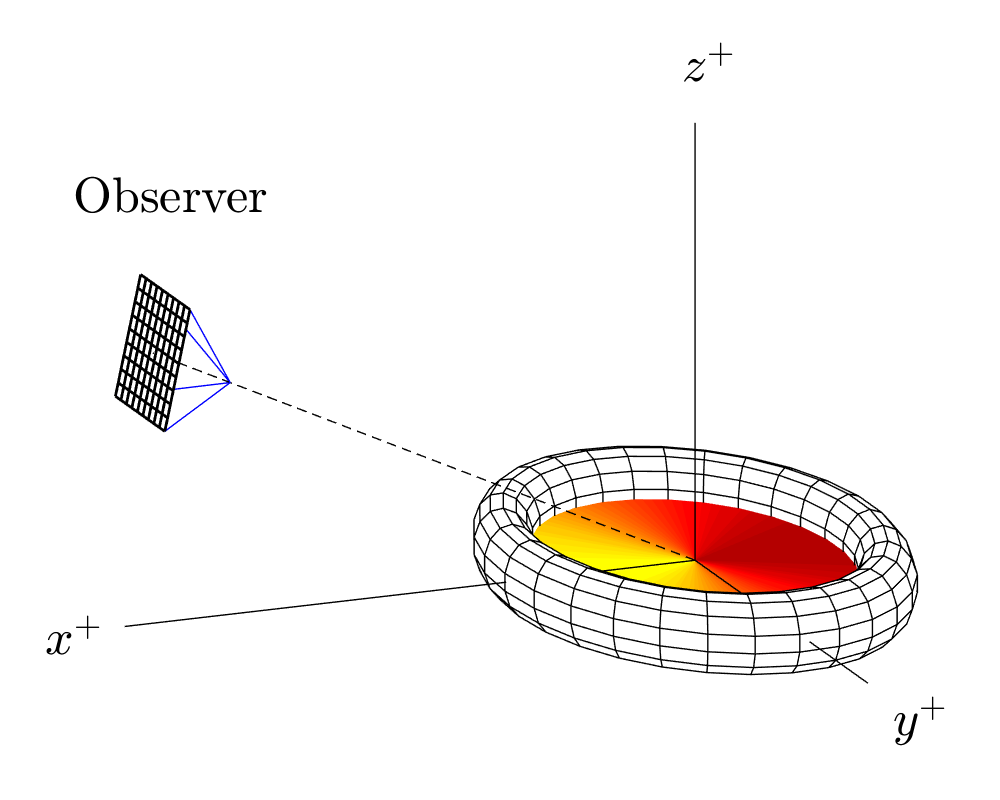}
   \caption{Diagram of the virtual camera, the ergosphere and the disk with its artificial coloring pattern.} \label{fig:camera_elliptic} 
\end{figure}

\begin{figure}[H]
    \centering
    \begin{tabular}{cc}
    \small $\alpha=0.2$  & \small $\alpha=0.3$ \\    
    \fbox{\includegraphics[width=5.5cm]{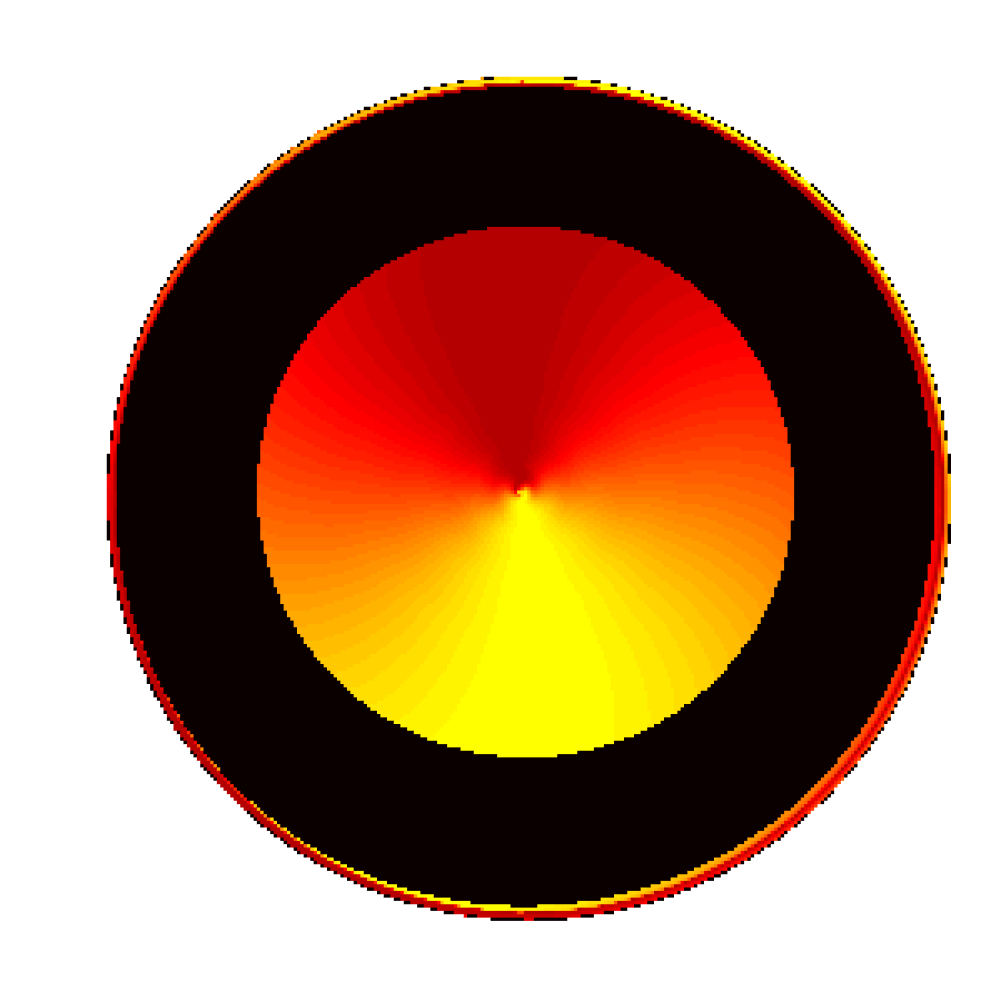}} &
    \fbox{\includegraphics[width=5.5cm]{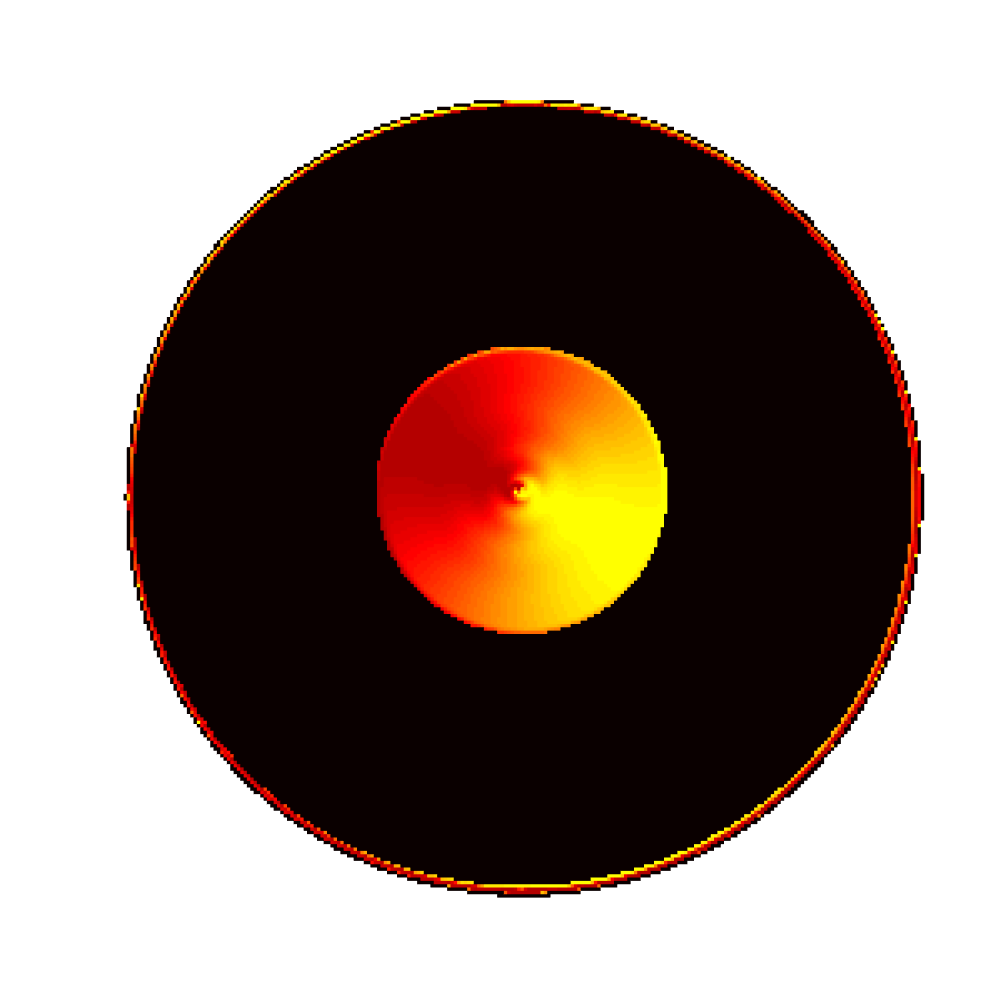}} \\   
     &  \\
    \fbox{\includegraphics[width=5.5cm]{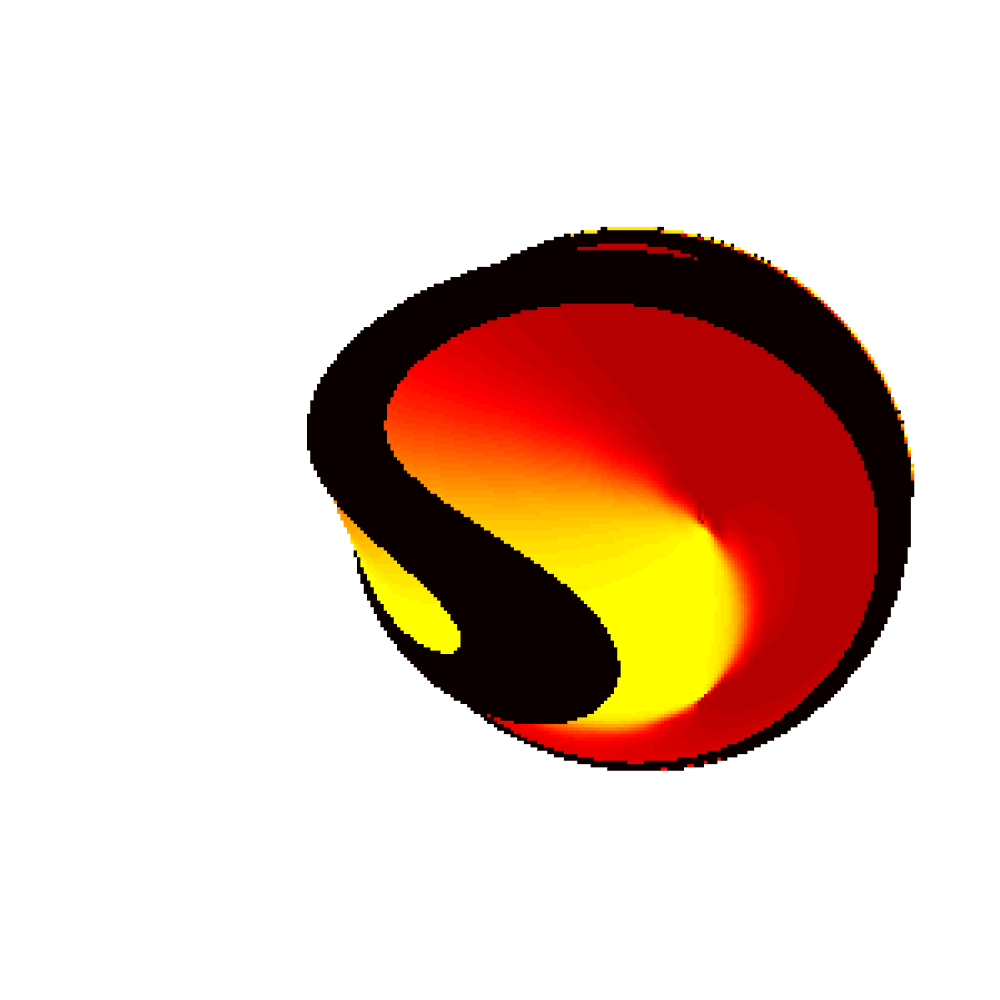}} &
    \fbox{\includegraphics[width=5.5cm]{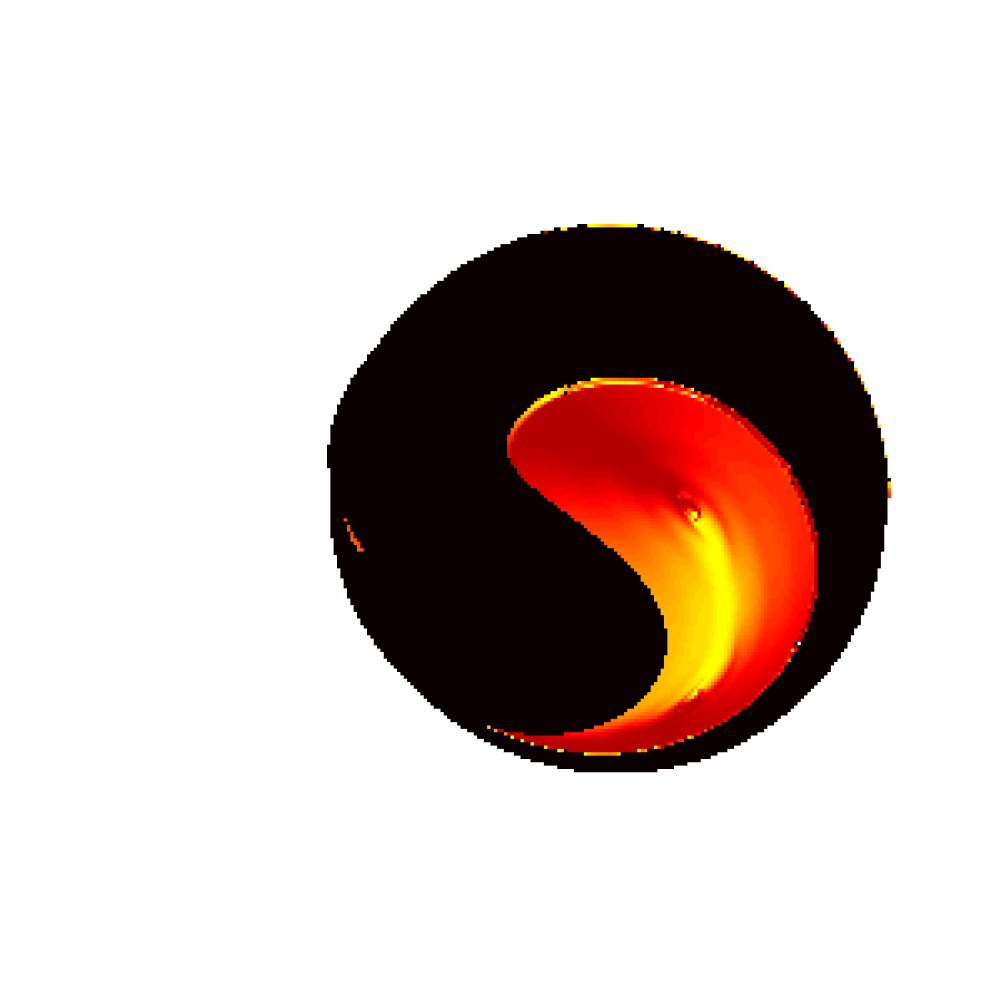}} 
    \end{tabular}
    \caption{Apparent images of disks in two different spacetimes. The images at the top correspond to an observer (virtual camera) with an inclination angle $\psi=10^\circ$ with respect to the symmetry axis and those at the bottom correspond to $\psi=90^\circ$. } \label{fig:shadows_elliptic} 
\end{figure}

For the simulations shown in Fig. \ref{fig:shadows_elliptic}, we 
assume that the disk is a light-emitting source, and in order to 
visualize the effect of the parameter $\alpha$ on light, we add an 
artificial coloring on the disk in dependence of the physical 
coordinate $\phi$ at which the backward integration stops. The chosen coloring of the disk is shown in Fig. \ref{fig:camera_elliptic}. This 
will allow us to infer the part of the disk from which the photons 
were emitted. 

Since we only consider the exterior of the first ergosphere for the numerical computation of the geodesics, we represent the 
ergosphere as a shadow. If the line of sight of the observer is 
almost aligned with the rotation axis, the apparent image of the disk 
will look almost circular, as observed in Fig. 
\ref{fig:shadows_elliptic}, but as the magnitude of the parameter 
$\alpha$ increases, it becomes more evident that light rays swirl around 
the axis, which does not happen in Kerr spacetimes since the 
component $g_{t\phi}$ of the metric vanishes on the whole axis 
outside of the event horizon. As the 
inclination of the observer with respect to the axis increases, the 
shadow is shifted from the center of the screen, as shown in Fig. \ref{fig:shadows_elliptic} for an observer on the equatorial plane. 
The bigger shifting of the apparent image of the disk with $\alpha=0.3$, compared to the one with $\alpha=0.2$, is a visual indication that the angular momentum of the disk described by the toron solution increases with the magnitude of $\alpha$, in analogy to the observed phenomenon in Kerr spacetimes. Furthermore, if the observer is located in the equatorial plane, the apparent image will look highly asymmetrical (as opposed to what would be observed in Kerr spacetimes), which is due to the metric functions being neither odd nor even in $z$. 

In order to visualize further effects of the gravimagnetic mass and the angular momentum in dependence of the parameter $\alpha$, we simulate the 
apparent image of the celestial sphere seen by an observer on the equatorial plane with the disk placed along its line of sight. In analogy to Fig. \ref{fig:primary_image}, we begin by showing the outer boundary of the primary image seen by a distant observer in various toron spacetimes.

\begin{figure} [H]
    \centering
    \includegraphics[width=11cm]{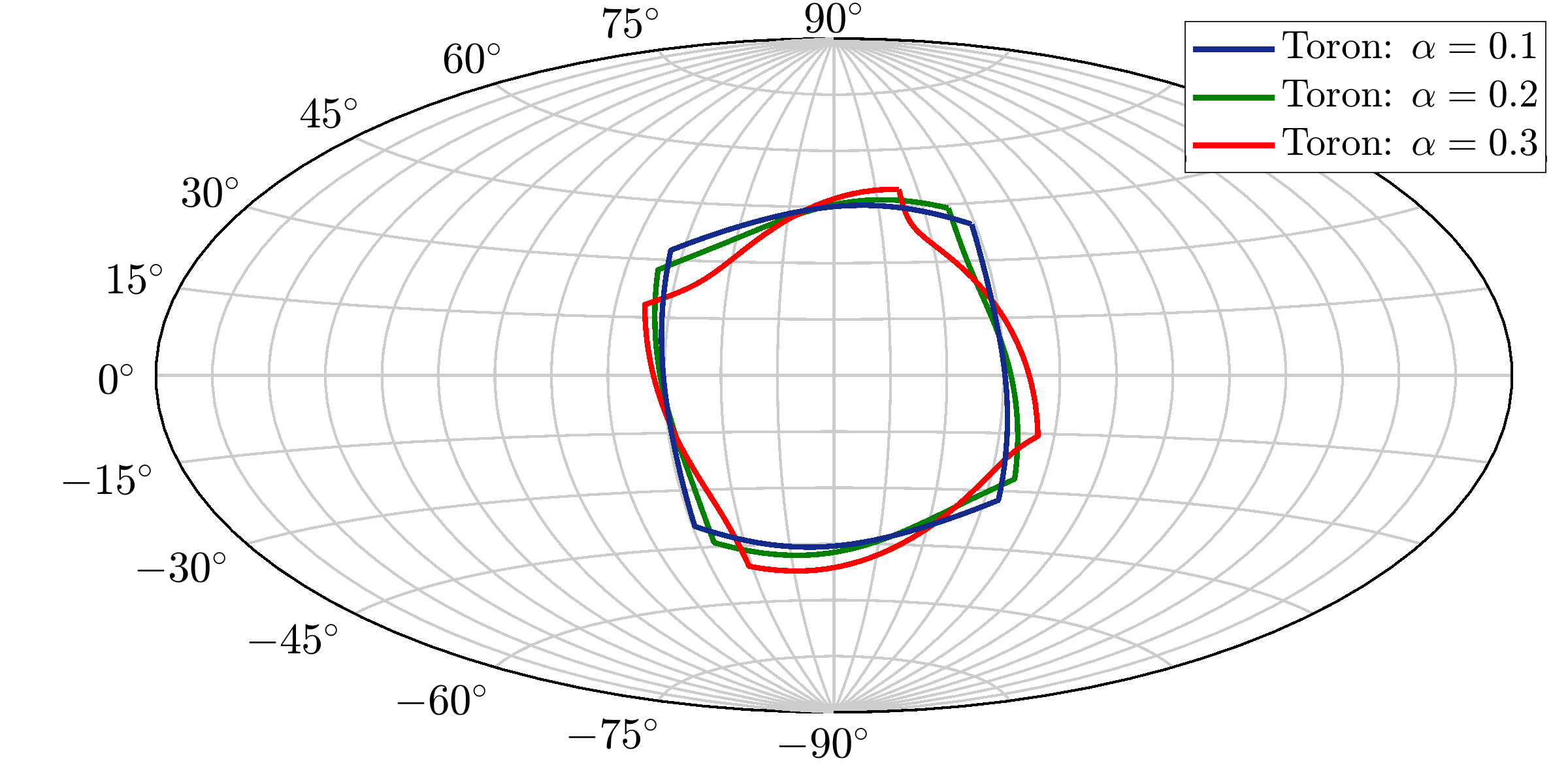}
    \caption{Boundary of the primary image on the celestial sphere in various toron spacetimes. In all cases, the focal length of the virtual camera is chosen such that the angles of view in Minkowski spacetime are $\delta_H=\delta_V=90^\circ$.}
\end{figure}

\begin{figure}[H]
\centering  
\begin{tabular}{cc}
\small $\alpha=0.2$ & \small $\alpha=0.3$ \\ 
\includegraphics[width=6cm]{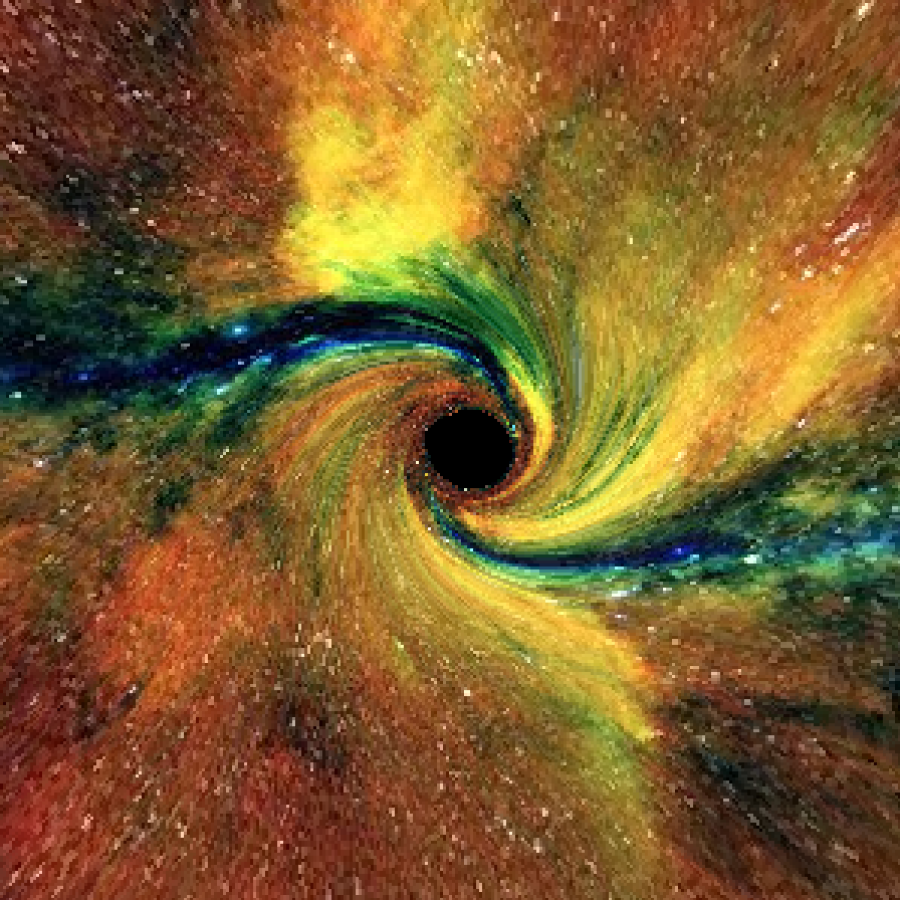} & 
\includegraphics[width=6cm]{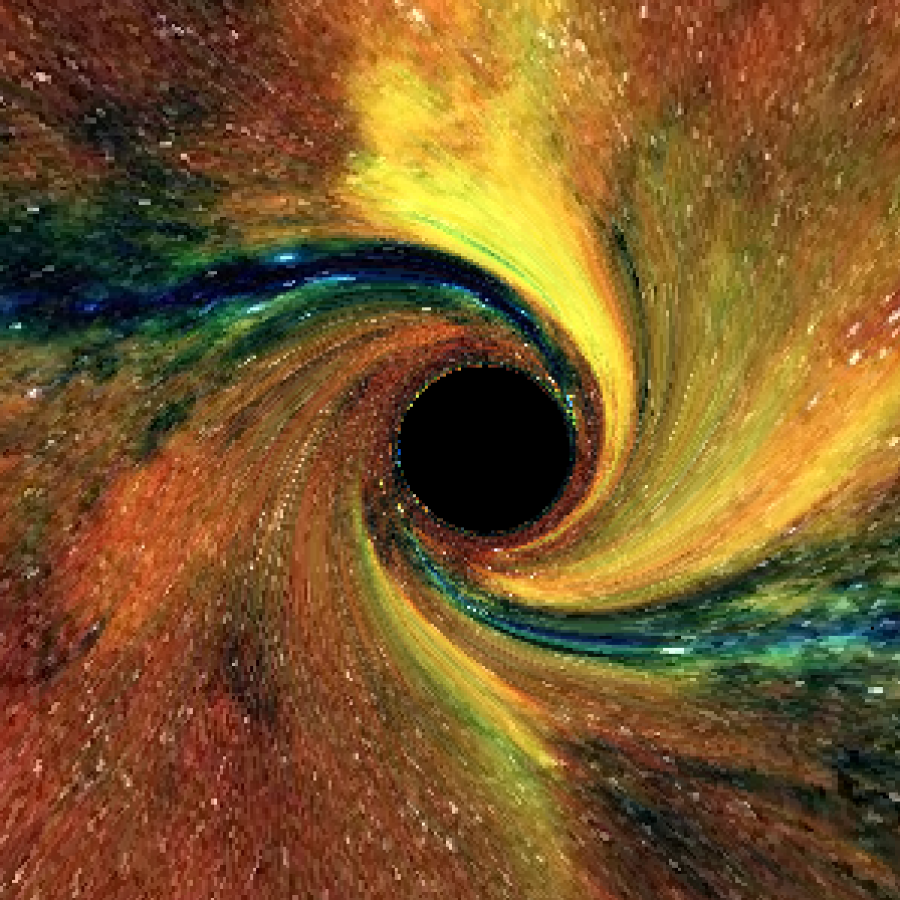}   
\end{tabular}     
\caption{Apparent image of the celestial sphere in toron spacetimes. } \label{fig:shadow_elliptic_color} 
\end{figure}

We continue the discussion by showing simulations analogous to the images in Fig. \ref{fig:shadow_nut_color}, in which 
a black hole is in front of the observer and therefore, it casts 
a shadow. 
To do this, we assume that the 
disk is not emitting light in order to observe a full shadow as in black hole spacetimes.
Since there is an angular momentum component, the shadow is not circular in toron spacetimes. 
Qualitatively, the effect of the $\alpha$ parameter on the apparent 
image of the celestial sphere is similar to the effect of the gravimagnetic mass $\ell$ in NUT spacetimes, as one can observe by comparing Figs. \ref{fig:shadow_nut_color} and \ref{fig:shadow_elliptic_color}; 
in particular, the case with $m=0$ and $\ell\neq 0$. The main consequence is the twisting of the apparent image of the background space, analogous to the observed phenomenon in NUT spacetimes and the spacetimes studied in \cite{MHC}. However, there are differences as well, since the disk is also a source of angular momentum as discussed above. This is mainly observed in the shifting of the shadow to the right. Nevertheless, the dominant effect is that of the gravimagnetic mass component, since the primary and secondary images of the upper and lower halves of the celestial sphere show more resemblance to the image shown in Fig. \ref{fig:shadow_diagrams} corresponding to the NUT spacetime with $m=0$ and $\ell\neq 0$. 
To obtain these simulations, we use the same values chosen for Fig. \ref{fig:shadow_nut_color}, which would produce angles of view of approximately $\delta_H=\delta_V=120^\circ$ in Minkowski spacetime. Namely, the observer is placed at a distance $R_c=20$ from the center of the disk, its line of sight has an inclination $\psi=90^\circ$ with respect to the symmetry axis and the radius of the celestial sphere is set to $R_\infty=100$; the screen has a width and height $d_H=d_V=0.1$, a resolution of $300\times 300$ pixels and a focal length $d_L=0.04$. 

\begin{figure} [H]
    \centering
    \small Toron: $\alpha = 0.2$ \\
    \includegraphics[width=8cm]{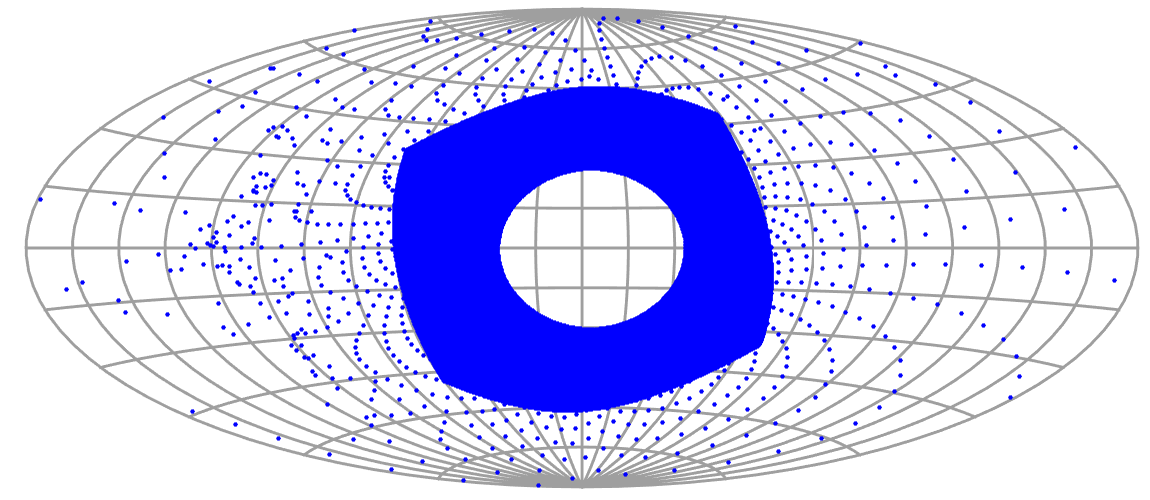}
    \caption{Origin of each light ray that reached the observer in the toron spacetime with $\alpha=0.2$. The quasi-circular region corresponds to the hidden region.  } \label{fig:origin_elliptic}
\end{figure}

To finish the discussion, let us look at the blind region on the celestial sphere in a toron spacetime. Analogous to Fig. \ref{fig:origin_Kerr-NUT}, we plot the celestial coordinates of origin of each of the light rays reaching the observer.
Fig. \ref{fig:origin_elliptic} shows these points on the celestial sphere in the Aitoff projection.
The \textit{blind region phenomenon} observed in NUT spacetimes is also present in toron spacetimes due to the existence of a gravimagnetic mass component. However, due to the clockwise angular momentum arising from a non-zero $\alpha$ (which in turn induces the frame-dragging effect), the blind region is no longer centered at $(\phi,\theta)=(\pi,0)$ and the quasi-circular shape is slightly deformed as well, as observed in Fig. \ref{fig:origin_elliptic}. Analogous to the observed effect on the shadow, the blind region presents a horizontal shift when compared to a zero-angular momentum spacetime with non-zero gravimagnetic mass.

%
%
\section{Conclusion}
\label{conclusion_sec}

In this paper we have presented a discussion of physical properties 
of toron solutions to the stationary axisymmetric 
Einstein equations in vacuum. They have vanishing mass and a non-zero 
NUT parameter. The  parameter $\alpha$ in the solution determines both gravimagnetic mass and angular momentum. 
One key difference with respect to well-known spacetimes is the existence of an infinite number of ergospheres in these spacetimes, all of which have a toroidal shape.

The physical properties of these spacetimes are explored via 
ray tracing. Well-known results of ray tracing in Kerr and NUT 
spacetimes are recalled as a reference for similar effects in 
toron spacetimes. The main findings are as follows:\\
- Role of a NUT parameter: as a consequence of the presence of a NUT parameter 
the light rays initially 
moving in the equatorial plane deviate in upwards or downwards 
directions. This behavior is numerically observed in toron spacetimes 
similarly to NUT spacetimes. This is an indication that the 
so-called gravimagnetic mass is probably not relevant in known
astrophysical situations  since no light ray  deviation  of such 
type was observed so far.\\
- Frame dragging: as in any stationary axisymmetric 
spacetime, light rays initially moving in a purely radial direction 
(i.e. with $p^z_0=p^\phi_0=0$) are dragged in a direction 
depending on the sign of the metric function $A$.  \\
- Fundamental photon orbits:
the existence of fundamental photon orbits and corresponding  photon 
surfaces is inferred numerically based on the observation of 
individual light rays almost orbiting on some  surface of a spherical topology
(then eventually falling into the gravitating object or escaping to 
infinity), as well as the observation of shadows similar to those 
appearing in black hole spacetimes. Due to a lack of symmetry in 
toron spacetimes compared to NUT spacetimes, it is more 
straightforward to identify them in the latter. \\
- Blind regions: there 
are blind regions on the celestial sphere in both NUT and toron
spacetimes, unlike in Kerr spacetimes in which the full celestial sphere 
is visible in the secondary image.   The primary images in NUT and 
toron spacetimes look qualitatively similar to each other, but 
quite different from Kerr. The same observation applies to secondary 
images. This means that ray tracing in toron spacetimes shares  
qualitatively  many features known from NUT spacetimes, but the 
phenomena are richer due to toroidal topology of the ergosphere. 

It is an interesting question whether higher genus toron solutions 
\cite{K91_1} having no NUT parameter will show ray tracing closer to 
the well known Kerr solution. This would allow to generate more 
models for astrophysically  interesting 
objects as, e.g., in \cite{VWA}. This will be the direction of future 
research.


\appendix
\section{Elliptic curves and theta functions} \label{sec:elliptic_curves}
In this appendix we collect some basic mathematical facts on  
toron solutions.
The solutions from \cite{K91} are constructed as follows.
Consider  the family of elliptic  curves  $\L_\xi$ depending on spacetime variables  $\rho$ and $z$ given by the equation
\begin{equation} \label{eq:curves_L_xi}
    \L_\xi=\{ (\lambda,\mu)\in\C^2 \mid \mu^2=(\lambda-\ii\sigma  )(\lambda+\ii\sigma)(\lambda-\xi)(\lambda-\Bar{\xi}) \} ,
\end{equation}
where  $\sigma\in \R$ is a constant (recall that $\xi=z-\I \rho$). 

 Following  \cite{K91}, we introduce the canonical cycles $(a,b)$ on $  \L_\xi$ shown in 
 Fig.~\ref{Axis1fig} such that the $a$-cycle goes around the branch 
 cut $[-\ii\sigma,\ii\sigma]$ while the $b$-cycle goes around the branch points $\xi$ and $\ii\sigma$.
 
Consider the normalized ($\int_a \omega=1$) differential of 
the first kind  
$$\omega=\frac{1}{\mathcal{A}} \frac{d\lambda}{\mu},$$
 where
$\mathcal{A}=\oint_a  \frac{d\lambda}{\mu}=  2\int_{-\ii\sigma}^{\ii\sigma }\frac{d\lambda}{\mu}$. 

In order to express all quantities in terms of standard elliptic 
integrals, we apply a Möbius transform by introducing the new coordinate $x(\lambda)$ via
\begin{equation}
	x(\lambda)=\frac{(\lambda+\ii\sigma)(\ii\sigma-\bar{\xi})}{(\lambda-\bar{\xi})2\ii\sigma},\quad 
	\lambda-\bar{\xi}=\frac{(\bar{\xi}+\ii\sigma )(\ii\sigma-\bar{\xi})}{2 \ii\sigma x-\ii\sigma+\bar{\xi}},
	\label{sing3}
\end{equation}
such that $x(\bar{\xi})=\infty$, $x(-\ii\sigma)=0$, $x(\ii\sigma)=1$ and 
$$x(\xi)=\frac{(\ii\sigma+\xi)(\ii\sigma-\bar{\xi})}{2\ii\sigma(\xi-\bar{\xi})}\;$$
is real,  positive and belongs to the interval $[1,\infty)$.

Then, introducing the elliptic module $k$ via
\begin{equation}
k^2=\frac{1}{x(\xi)}=\frac{4\rho \sigma}{z^{2}+(\rho+\sigma)^{2}},
\label{sing5}
\end{equation}
(which is real and $k\in[0,1]$)
and the new variable $y$ via $x=y^2$ we get 
$$
\frac{d\lambda}{\mu}=\frac{2}{i\sqrt{z^2+(\rho+\sigma)^2}}\frac{dy}{[((y^2-1)(k^2 y^2-1)]^{1/2}}.
$$
We used here that $(i\sigma-\bar{\xi})(\xi+i\sigma)=-(z^2+(\rho+\sigma)^2)$.

Considering the complete elliptic integral of the first kind 
$$
K(k)=\int_{0}^1  \frac{dy}{[((y^2-1)(k^2 y^2-1)]^{1/2}},
$$
we get for the period $\mathcal{A}$:
\begin{equation}
	\mathcal{A}=
	\pm \frac{4K}{
	\sqrt{(\ii\sigma -\bar{\xi})(\xi+\ii\sigma)}}.
	\label{sing7}
\end{equation}

The period of the curve $\tau= \int_b \omega$ 
reads


\begin{equation}
	\tau=-\frac{1
	}{2\mathcal{A} \sqrt{ (\ii\sigma -\bar{\xi})(\xi+\ii\sigma)  }}\int_{0}^{\infty}\frac{dx}{\sqrt{x(1-x)(1-k^2x)}}
	=\frac{\ii K'}{K},
	\label{sing8}
\end{equation}
where $K'(k) = K(\sqrt{1-k^{2}})$. 

For the integral of $\omega$ between $\xi$ and $\infty^+$ 
 we have
\begin{equation}
\int_{\xi}^{	\infty^+}\omega=-\frac{1
	}{2\mathcal{A} \sqrt{ (\ii\sigma -\bar{\xi})(\xi+\ii\sigma)  }}
	\int_{1/k^2}^{\frac{\ii\sigma-\bar{\xi}}{2\ii\sigma}}\frac{dx}{\sqrt{x(1-x)(1-k^2x)}}=
	\frac{1}{2K}(\gamma-K-\ii K')
	\label{sing10}
\end{equation}
where we have defined $\mbox{sn}^2 \gamma
=(\ii\sigma-\bar{\xi})/2\ii\sigma$. Here 
$\mbox{sn}(x)$ is one of the Jacobi elliptic functions. 
The standard definition of the Jacobi sn function is 
via the elliptic integral
$$s:=F(k,\varphi) = 
\int_{0}^{\varphi}\frac{d\theta}{\sqrt{1-k^{2}\sin^{2}\theta}}$$ 
and the elliptic amplitude $\mbox{am}(s,k)=\varphi$. In terms of this 
amplitude, the sn function is defined via 
$\mbox{sn}(s,k)=\sin(\mbox{am}(s,k))$.
\vskip0.7cm

We need to compute the Abelian integral (\ref{sing10}) for all values  
of $\rho>0$ and $z\in\mathbb{R}$. However,  this
integral is a multi-valued  function of $\xi$ since  it picks up a  
term
$2\tau$ when $\xi$ goes around $\ii\sigma$ (see Fig.2 of \cite{K91_1}).

We recall that the theta function with  characteristic $\p\in\R$ and $\q\in\C$ is defined by 
\begin{equation}
    \thetapq(\mathrm{z},\tau) = \sum_{n\in\Z} \exp{(-\pi \I \tau (n+\p)^2+2\pi\I(n+\p)(\mathrm{z}+\q))},
	\label{thetapq}
\end{equation}
where $\mathrm{z}\in\C$ and $\B\in\mathbb{H}$, i.e., $\tau$ is a complex number with $\Im(\B)>0$. The 
theta function is usually defined as a  function with a fixed period
$\B$, but in this paper both arguments depend on  $(\xi,\bar{\xi})$ and 
therefore on the physical coordinates $\rho$ and $z$. Note that 
the theta function (\ref{thetapq}) with characteristics can be 
written in terms of the Jacobi theta function $\vartheta:=\vartheta_{3}=\vartheta_{00}$ 
with zero characteristics,
\begin{equation}
    \thetapq(\mathrm{z},\B) = e^{\ii\pi {\B}\p^{2} +2\pi \ii 
	\p(\mathrm{z}+\q)} \vartheta(\mathrm{z}+2\pi \ii 
	({\B}\p+\q),\B)\;.
	\label{thetaser}
\end{equation}

The  elliptic solutions to the Ernst 
equation found in \cite{K88} can be written in the form
\begin{equation} \label{eq:ernst_potential}
     \ernst(\xi,\bar{\xi}) =e^{-\pi \I \p} \frac{\thetapq(\int^{\infty^+}_\xi 
	 \omega,\B)}{\thetapq(\int^{\infty^-}_\xi \omega,\B)},
 \end{equation}
where $\int^{\infty^\pm}_\xi \omega$ are Abel maps on the elliptic curve \eqref{eq:curves_L_xi} parametrized by $\xi$, which can be expressed in terms of elliptic integrals of the first kind, as shown by \eqref{sing10} in Appendix \ref{sec:elliptic_curves}. Moreover, these two Abel maps satisfy the relation $\int^{\infty^-}_\xi 
	 \omega = - \int^{\infty^+}_\xi \omega$. 
In order for the potential \eqref{eq:ernst_potential} to solve the Ernst equation, the constants $\p$ and $\q$ must satisfy the reality conditions $\p\in\R$ and $\q\in \I \R$ \footnote{In the original paper 
	 \cite{KKS} the necessary  unitary constant factor $e^{-\pi \I 
	 \p}$ in (\ref{eq:ernst_potential}) was omitted; this mistake was 
	 corrected in \cite{dL}.}.

The constants  $\p\in\R$ and $\q\in \I \R$ serve as parametrizers of a class of solutions to the Ernst equation.
For the numerical analysis of the 
solution one can use the approach of \cite{FK}, but 
here we alternatively use the elliptic functions implementation in Matlab. 
Without loss of generality, we 
will assume that $\sigma=1$, i.e., the fixed branch points are placed at $\ii$ and $-\ii$. 
A different choice of $\sigma\neq 0$ corresponds to a constant
rescaling of $\xi$. 

We also put $\p=0$ which simplifies the branching structure and 
the behavior of solution on the axis. 
We also denote
$$
\q= \ii\alpha
$$
for $\alpha\in \R_+$.
Then (\ref{eq:ernst_potential}) coincides with   the 
toron solution of \cite{K91}.

\subsection{Physical properties} \label{sec:physical_prop}

Let us recall that the ergosphere is defined as the hypersurface where the metric component $g_{tt}=-f$ vanishes. From the explicit form of the real part of the Ernst potential, the fact that $\B$ is purely imaginary and the form of the zeros of theta functions (it is well known that $z_*$ is a zero of the function $\theta(z,\B)$ if and only if $z_*=\frac{1}{2}+\frac{\tau}{2} + m + n\tau$ for $m,n\in\Z$), one can deduce that $f$ can only vanish if the theta function $\vartheta(\ii\alpha+1/2,\B)$ vanishes. This condition is equivalent to the existence of $n,m\in\N$ such that

\begin{equation*}
    1/2 + \ii\alpha = (m+1/2)+(n+1/2)\B.
\end{equation*}
Thus $m=0$, and for each period  
\begin{equation}
    \B_n = \frac{\ii\alpha}{n+1/2}\;,\hskip0.7cm n=0,1,\dots
    \label{taun}
\end{equation}
we get a separate component of the ergosphere.
The 
 set of coordinates $\rho,z$ corresponding to the period $\B_n$  can be described  via the equation \eqref{sing5} and by considering the inverse of \eqref{sing8}, see \cite{Chandra}, which is given by
$$
\kappa^2 = \frac{\vartheta_2^4(0,\B)}{\vartheta_3^4(0,\B)}, 
$$
where $\vartheta_2$ and $\vartheta_3$ are Jacobi elliptic functions. 

Denote $\kappa_n=\kappa(\B_n)$ for $n\in\N$. Then the  equation (\ref{taun}) for the $n$th component of the  ergosphere is equivalent to the  equation (see (\ref{sing5}))
$$
\kappa_n^2= \frac{4\rho \sigma}{z^{2}+(\rho+\sigma)^{2}}
$$
which leads to
\begin{equation*}
    (\rho-\rho_n)^2 + z^2 = R_n^2,
\end{equation*}
where 
\begin{equation} \label{eq:E_n}
    \rho_n=\frac{2}{\kappa_n^2}-1\;,\hskip0.7cm R_n =2\frac{\sqrt{1-\kappa_n^2}}{\kappa_n^2}\;.
\end{equation}
 Therefore, for each value of the parameter $\alpha$ (which 
 parametrizes the family of elliptic solutions), there is an infinite family of ergospheres parametrized by $n\in\N$, and since the spacetime is stationary axisymmetric, the spatial projection of each ergosphere is a torus whose major radius is $\rho_n$ and whose minor radius is $R_n$.\\

\subsection{Ernst potential on the axis}

We compute the Ernst potential on the axis when the branch points $\xi$ and $\bar{\xi}$ coincide and the elliptic curve degenerates to the genus 0 curve $\mu^{2}=\lambda^2+1$. This will allow a simple analysis on the asymptotic behavior of the Ernst potential. We only study the case with $\p=0$ in detail, since terms of the type $\rho^{2\p}$ appear in the asymptotic expansion otherwise.

The Ernst potential on the axis can be computed as follows (see (8) of \cite{K91}).

\begin{figure}[htb]
\includegraphics[width=14cm]{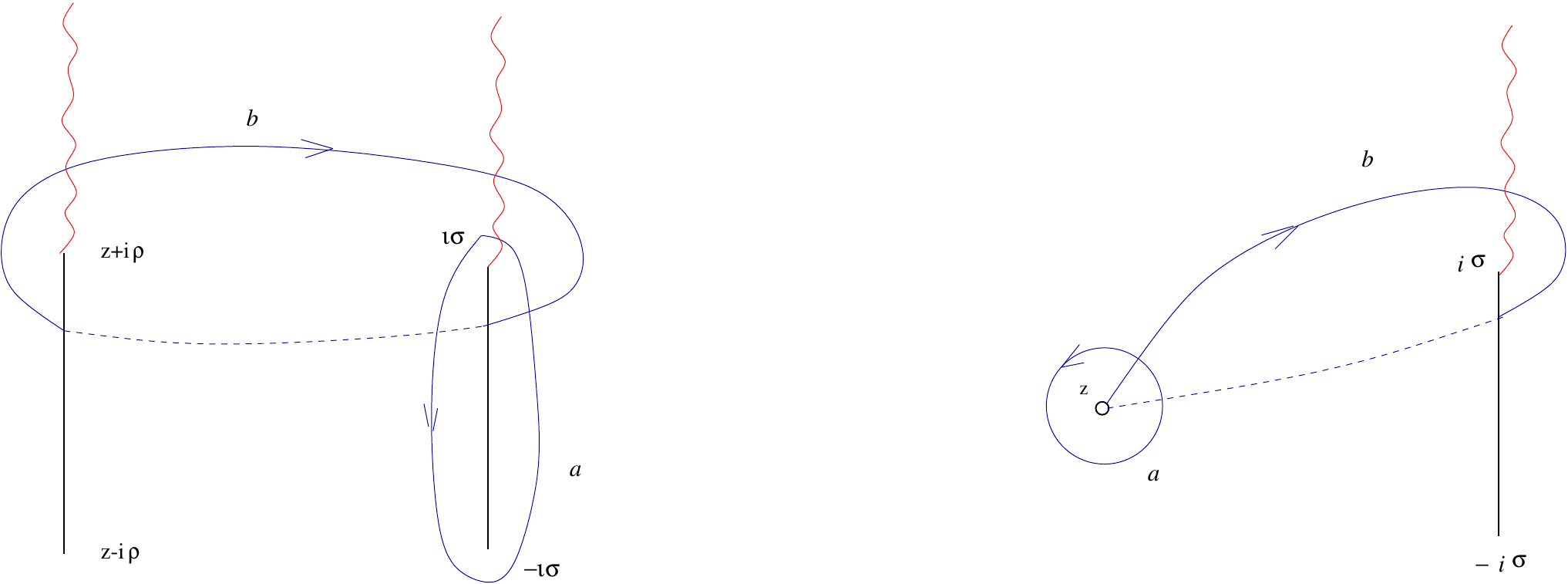}
\caption{Degeneration of the curve ${\mathcal L}_\xi$ into the rational curve in the limit $\rho\to 0$ for $z<0$.}
\label{Axis1fig}
\end{figure}

Let $z<0$ and let $\rho\to 0$. 
Then the curve $\L_\xi$ degenerates into a genus zero curve with double point at $\lambda=z$
(see Fig.\ref{Axis1fig}).  The period of $\L_\xi$ behaves as follows as $\rho\to 0$:
\begin{equation}
\tau=-\pi \ii \ln\rho +\dots
\label{tauas}
\end{equation}

The differential $\omega$ degenerates into the differential of third kind with two simple poles at
$\lambda=z$ on both sheets. On the ``$+$" sheet it has residue $-1/2\pi \ii$, and on the ``$-$" sheet  $ -1/2\pi \ii$
(see Fig. \ref{Axis1fig}).

Therefore,
$$
\omega\to -\frac{1}{2\pi \ii} \frac{\sqrt{(z-\ii\sigma)(z+\ii\sigma)}}{(\lambda-z)\sqrt{(\lambda-\ii\sigma)(\lambda+\ii\sigma)}}d\lambda
$$
and 
$$
\int_{\ii\sigma}^{\infty^+}\omega =
\frac{1}{2\pi \ii}\log\frac{z+\sqrt{z^2+\sigma^2}}{-\ii \sigma}.
$$
Moreover,
$$
\int_{\xi}^{\infty^+}\omega = \int_{\ii\sigma}^{\infty^+}\omega-\frac{\tau}{2}.
$$

As the period of the curve tends to infinity according to (\ref{tauas}) only two terms in the series (\ref{thetaser}) remain non-zero.  Namely, 
\begin{align*}
\vartheta\left( 
\int_{\xi}^{\infty^+}\omega-\ii\alpha\right)&=\vartheta\left(
\int_{\ii\sigma}^{\infty^+}\omega-\frac{\B}{2}-\ii\alpha\right)\\
&=\sum_{m=-\infty}^\infty\exp\left\{\pi \ii \B m^2+
2\pi \ii m \left(-\frac{\B}{2}+\int_{\ii\sigma}^{\infty^+}\omega 
+\ii\alpha\right)\right\}\\
&=\sum_{m=-\infty}^\infty \exp\left\{\pi \ii 
\B m(m-1)+2\pi \ii m 
\left(\int_{\ii\sigma}^{\infty^+}\omega+\ii\alpha\right)\right\}\\
&\to 1+ \exp\{ 2\pi \ii \int_{\ii\sigma}^{\infty^+}\omega+\ii\alpha 
\}=1-e^{2\pi \alpha}\frac{z+\sqrt{z^2+\sigma^2}}{\ii \sigma}.
\end{align*}

Therefore,  we have
\begin{equation}
\ernst(z,\rho=0)=\frac{z+\sqrt{z^2+\sigma^2}-\ii \sigma e^{-2\pi \alpha}}{z+\sqrt{z^2+\sigma^2}-\ii \sigma e^{2\pi \alpha}}\;,
\hskip0.7cm z < 0.
\end{equation}

\begin{figure}[htb]
\includegraphics[width=14cm]{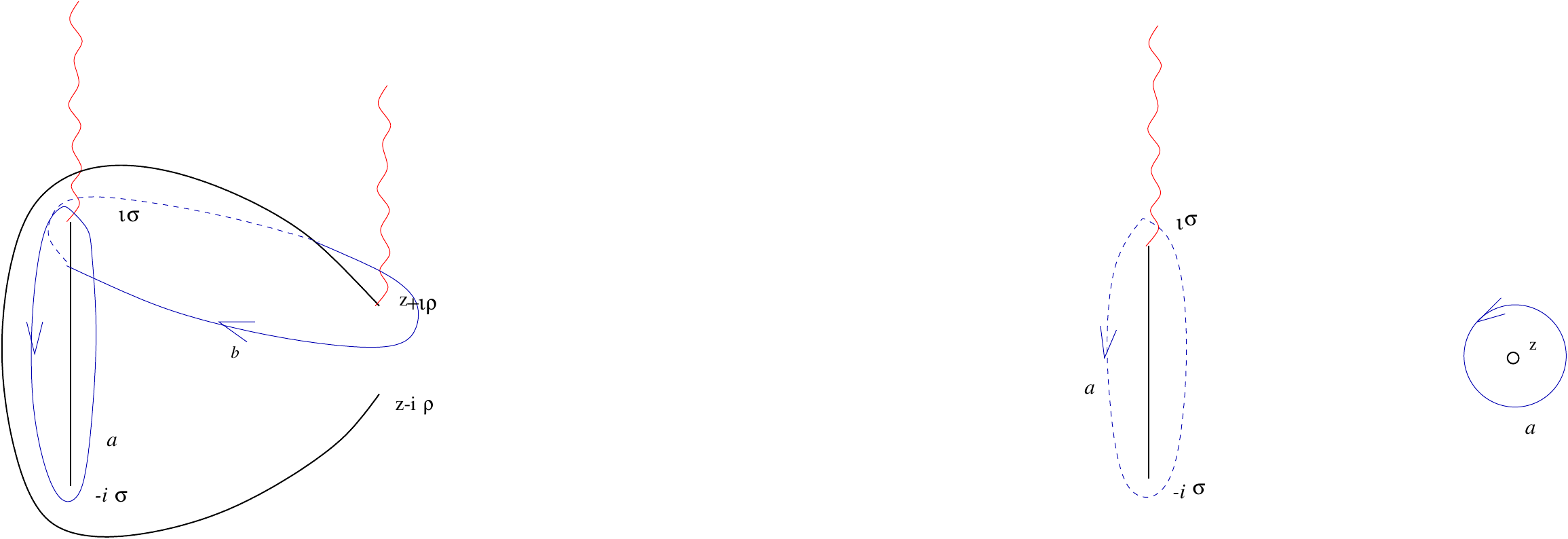}
\caption{{Degeneration of the curve ${\mathcal L}_\xi$ into the rational curve in the limit $\rho\to 0$ for $z>0$.}}
\label{Axis2fig}
\end{figure}

Similarly, for $ z> 0$ the differential $\omega$  has residue $-1/2\pi \ii$ at $z^+$  and the residue  $ -1/2\pi \ii$ 
at $z^-$ (Fig. \ref{Axis2fig}).

Thus, as $\rho\to 0$, 
$$
\omega\to \frac{1}{2\pi \ii} \frac{\sqrt{(z-\ii\sigma)(z+\ii\sigma)}}{(\lambda-z)\sqrt{(\lambda-\ii\sigma)(\lambda+\ii\sigma)}}d\lambda
$$
and 
$$
\int_{\ii\sigma}^{\infty^+}\omega=
-\frac{1}{2\pi \ii}\log\frac{z+\sqrt{z^2+\sigma^2}}{-\ii \sigma}.
$$
Therefore,
\begin{align*}
\vartheta\left( 
\int_{\xi}^{\infty^+}\omega-\ii\alpha\right)&=\vartheta\left(\int_{\ii\sigma}^{\infty^+}w-\frac{\B}{2}-\ii\alpha\right)\\
&\to 1+ \exp\{ 2\pi \ii \int_{\ii\sigma}^{\infty^+}\omega+\ii\alpha 
\}\\
&=1-e^{2\pi \alpha}\frac{\ii \sigma}{z+\sqrt{z^2+\sigma^2}}=1-e^{2\pi \alpha}\frac{z-\sqrt{z^2+\sigma^2}}{\ii \sigma}
\end{align*}
and
\begin{equation}
\ernst(z,\rho=0)=\frac{z-\sqrt{z^2+\sigma^2}-\ii \sigma e^{-2\pi \alpha}}{z-\sqrt{z^2+\sigma^2}-\ii \sigma e^{2\pi \alpha}}\;,
\hskip0.7cm z\geq 0.
\end{equation}

\section{Geodesic equations}\label{appgeodesics}
In this appendix we briefly summarize the approach to 
compute geodesics in 
general stationary axisymmetric spacetimes. 
For further details the reader is referred to \cite{DFK}. 

Due to the existence of two Killing vectors 
$$\eta=\partial_t\;,\hskip0.7cm \tilde{\eta}=\partial_\phi\;,$$
the general geodesic equations 
\begin{equation}
\frac{d^2 x^a}{ds^2}+\Gamma_{bc}^a \frac{d x^b}{ds}\frac{d x^c}{ds}=0,
\label{geodes}
\end{equation}
where $s$ is an affine parameter,
can be simplified. 
Namely, there are two conserved quantities along the geodesics. In analogy to Kerr spacetimes, we denote them as the energy 
and the angular momentum of the test particle:
\begin{equation}
E:=-g_{ab}\eta^a p^b= f p^t +(fA) p^\phi 
\label{energy}
\end{equation}
(here $p^a=d x^a/ds$ are momenta),
\begin{equation}
L:=g_{ab}\tilde{\eta}^a p^b=-fA p^t + \Phi p^\phi
\label{angmom}
\end{equation}
where
$$\Phi:= 
g_{\phi\phi} = \rho^2 f- \frac{A^2}{f} \;.$$
  The Lagrangian 
 $$\mathcal{L} = \frac{1}{2} g_{ab} p^a p^b$$
  is also a conserved quantity, and since we are interested in null geodesics, the Lagrangian is always zero.
However, unlike Kerr spacetimes, the existence of a fourth constant of motion (analogous to the Carter constant) is not known for general stationary axisymmetric spacetimes. 
The momenta $p^t$ and $p^\phi$ can be found from the linear system (\ref{energy}), (\ref{angmom}) once the values of the integrals 
$E$ and $L$ are fixed.

Computing the Christoffel symbols for the metric in  the Weyl-Lewis-Papapetrou form (\ref{eq:wlp}), the  geodesic equations (\ref{geodes}) in Weyl coordinates can be written as  the following  system of ODEs for the remaining 6 variables
which we combine into the vector $\vect{y}=(t,\rho,p^\rho,z,p^z,\phi)\in \R^6$:
\begin{equation} \label{ode_geodesics}
  \begin{split}
    \frac{dt}{ds} &= \frac{1}{\rho^2} \left(E \Phi - (fA) L \right),\\
    \frac{d \rho}{ds} &= p^\rho, \\
    \frac{d p^\rho}{ds} & = \frac{1}{2h} \left[ -f_\rho (p^t)^2 - h_\rho (p^\rho)^2 + h_\rho (p^z)^2 +  \Phi_\rho (p^\phi)^2 - 2 (fA)_\rho p^t p^\phi - 2 h_z p^\rho p^z \right], \\
    \frac{d z}{ds} &= p^z, \\
    \frac{d p^z}{ds} & = \frac{1}{2h} \left[ -f_z (p^t)^2 + h_z (p^\rho)^2 - h_z (p^z)^2 + \Phi_z (p^\phi)^2 - 2(fA)_z p^t p^\phi - 2 h_\rho p^\rho p^z \right], \\
    \frac{d\phi}{ds} &= \frac{1}{\rho^2} ((fA)E+fL),
  \end{split}
\end{equation}
where $h:=g_{\rho\rho}=g_{zz}=e^{2k}/f$  . 
 Then the geodesic 
passing through the initial point $\vect{y}_0$ is  the solution of the initial value problem (IVP)
\begin{equation} \label{eq:ivp}
\left\{ \begin{array}{cc}
       \frac{d \vect{y}}{ds} = F(\vect{y}), \\
   \vect{y}(0) = \vect{y}_0,
\end{array}  \right.
\end{equation}
where $F:\R^6\to\R^6$ is the function described by 
\eqref{ode_geodesics}. Hence, given initial conditions for all 8 variables ($t_0$,  
$p^t_0$, $\rho_0$, $p^\rho_0$, $z_0$, $p^z_0$, $\phi_0$, 
$p^\phi_0$), we first obtain the conserved quantities $E$ and $L$ and find the  momenta $p^t$ and $p^\phi$  from the linear system (\ref{energy}), (\ref{angmom}). For the remaining 6 variables encoded by the vector $\vect{y}$ we
express 
the geodesic equations in the form \eqref{eq:ivp} and solve them numerically with 
the approach presented in \cite{DFK}.
The IVP is solved via the Runge-Kutta  method of fourth order, which 
gives solutions $\vect{y}_1,\vect{y}_2,\ldots,\vect{y}_n$ at steps
$s_{1},s_{2},\ldots, s_{n}$, for some $n\in\N$. Thus, the function 
$F$ must be computed for each stage of the Runge-Kutta method, but this can be computationally expensive as in the case of the metric functions of toron spacetimes, since they are non-elementary functions. However, they are piecewise smooth and the method outlined in \cite{DFK} to compute them on a grid via polynomial interpolation can be used. 
Finally, the third constant of motion $\mathcal{L}=0$ will be used to control the accuracy of the numerical solution.

Using the geodesic equations, we can study the evolution of individual light rays initially moving on a plane (e.g. the equatorial 
plane and a plane containing the symmetry axis), determine the path of light rays bound to a surface (called fundamental photon orbits in the text) and determine the source of light reaching a distant observer. 

%
%
\bibliographystyle{amsplain}

\begin{thebibliography}{}

\bibitem{AAKA} A. Abdujabbarov, F. Atamurotov, Y. Kucukakca, B. Ahmedov and U. Camci. \textit{Shadow of Kerr-Taub-NUT black hole}. Astrophys. Space Sci. 344, 429 (2012)

\bibitem{bardeen}J. M. Bardeen, in  \textit{Black Holes} (Les Astres Occlus), 
edited by C. DeWitt and B. S. DeWitt (Gordon and Breach, New York, 
1973) p. 215   

\bibitem{BZ} V.A. Belinskii, V.E. Zakharov, \textit{Integration of the
  Einstein equations by the methods of inverse scattering theory and
  construction of explicit multisoliton solutions}, Soviet Phys. JETP
  48 (1978) 985–994.

\bibitem{BB} E.D. Belokolos, A.I. Bobenko, V.Z. Enol'skii, A.R. Its and V.B. Matveev. \textit{Algebro-geometric approach to nonlinear integrable equations}. Springer, Berlin (1994)

\bibitem{bonor} W.B. Bonnor. \textit{A new interpretation of the NUT metric in general relativity}. Math. Proc. Cambridge Philos. Soc. 66, 145
(1969)

\bibitem{Chandra} K. Chandrasekharan. \textit{Elliptic Functions}. Volume 281, Grundlehren der mathematischen Wissenschaften (1985)

\bibitem{CH} P.V.P.Cunha, C.A.R.Herdeiro, {\it Shadows and strong gravitational lensing: a brief review}, 
Gen Relativ Gravit 50, 42 (2018). https://doi.org/10.1007/s10714-018-2361-9

\bibitem{DFK} E.B. de Leon, J. Frauendiener and C. Klein. \textit{Visualisation of counter-rotating dust disks using ray tracing methods}. Class. Quantum Grav. 41 155005 (2024)

\bibitem{dL} E.B. de Leon. \textit{On a class of algebro-geometric solutions to the Ernst equation},  arXiv:2310.19095v1 (2023)

\bibitem{einstein}A. Einstein, \textit{Lens-Like Action of a Star by the Deviation of Light in the Gravitational Field,} Science 84, 506 (1936).

\bibitem{ernst1968a}F.J. Ernst, \textit{New formulation of the axially symmetric 
gravitational field problem I}, Phys. Rev. {\bf D 167}, 1175 

\bibitem{ernst1968b}F.J. Ernst, \textit{New formulation of the axially symmetric gravitational field 
problem II},  Phys. Rev. {\bf D 168}, 1415   

\bibitem{FK} J. Frauendiener and C. Klein, \textit{Computational approach to hyperelliptic Riemann surfaces}, Lett. Math. Phys. 105(3), 379-400  

\bibitem{GPL} A. Grenzebach, V. Perlick and C. Lämmerzahl. \textit{Photon regions and shadows of Kerr-Newman-NUT black holes with a cosmological constant}. Phys. Rev. D 89, 124004 (2014)

\bibitem{HL} E. Hackmann and C. Lämmerzahl. \textit{Observables for bound orbital motion in axially symmetric space-times}. Phys. Rev. D 85, 044049 (2012)

\bibitem{KKS} C. Klein, D. Korotkin and V. Shramchenko. \textit{Ernst equation, Fay identities and variational formulas on hyperelliptic curves}. Mathematical Research Letters (2004) 27--45

\bibitem{KRPRL}C. Klein, O. Richter, \textit{Exact relativistic gravitational treatment of a stationary counter-
rotating dust disk}, Phys. Rev. Lett. 83 (1999) 2884.

\bibitem{KR} C. Klein and O. Richter. \textit{Ernst Equation and Riemann Surfaces: Analytical and Numerical Methods}. Lecture Notes in Physics, Vol. 685. Springer (2005)

\bibitem{K88} D. Korotkin. \textit{Finite-gap solutions of the stationary axisymmetric Einstein equation in vacuum}. Theor. Math. Phys. 77, 1018 (1988)

\bibitem{K91}D. Korotkin, \textit{Solutions of the vacuum Einstein equation having toroidal infinite red-shift
surface}, Class. Quantum Grav. {\bf 8} L219 - L222 (1991)

\bibitem{K91_1} D. Korotkin, \textit{Algebraic Geometric Solutions
of Einstein's Equations: Some Physical Properties}, Commun. Math. Phys. {\bf 37}, 383-398 (1991)

\bibitem{Mai}D. Maison, \textit{ Are the stationary axially symmetric Einstein
  equations completely integrable?}  Phys. Rev. Lett. {\bf 41} (1978)
  521–524.

\bibitem{MR} V.S. Manko and E. Ruiz. \textit{Physical interpretation of NUT solution}. Class. Quantum Grav. {\bf 22} 3555-60 (2005)

\bibitem{MHC} Z.S. Moreira, C.A. Herdeiro and L.C. Crispino. \textit{Twisting shadows: Light rings, lensing, and shadows of black holes in swirling universes}. Phys. Rev. D 109, 104020 (2024)

\bibitem{Neu} G. Neugebauer, \textit{Backlund transformations of axially symmetric stationary gravitational fields} J. Phys. {\bf A 12}, L67 (1979).

\bibitem{NM}G. Neugebauer, R.Meinel, \textit{ General relativistic gravitational field of a rigidly rotating disk of dust: Solution in terms of ultraelliptic functions}, Phys. Rev. Lett. {\bf 75} 3046 (1995) 

\bibitem{PT}V. Perlick, O.Y. Tsupko, \textit{ Calculating black hole shadows: Review of analytical studies}, Physics Reports, 2022, arxiv 2105.07101


\bibitem{Silv} J.H. Silverman. \textit{The arithmetic of elliptic curves}. Graduate texts in mathematics (1986)

\bibitem{Exact} H. Stephani, D. Kramer, M. MacCallum, C. Hoenselaers and E. Herlt. \textit{Exact Solutions of Einstein’s Field Equations}. 2nd ed. Cambridge University Press (2003)

\bibitem{synge}J. L. Synge, \textit{The Escape of Photons from Gravitationally Intense Stars}, Mon. Not. R. Astron. Soc. {\bf 131}, 463 (1966)

\bibitem{Teo} E. Teo,  \textit{Spherical Photon Orbits Around a Kerr Black Hole} ,General Relativity and Gravitation. {\bf 35} (11): 1909–1926  (2003)

\bibitem{VWA}F. H. Vincent, M. Wielgus, M. A. Abramowicz,, E. 
Gourgoulhon, J.-P. Lasota, T. Paumard and G. Perrin, Geometric 
modeling of M87* as a Kerr black hole or a non-Kerr compact object, 
A\& A, 646, A37
(2021)

\bibitem{Wilkins} Wilkins, D. C. \textit{ Bound geodesics in the Kerr metric}, Phys. Rev. {\bf D 5}, 814–822 (1972)

\end{thebibliography}

\end{document}